\documentclass[%
 reprint,
superscriptaddress,
 amsmath,amssymb,
prd,
twocolumns
]{revtex4-1}

\usepackage{graphicx}
\usepackage{dcolumn}
\usepackage{bm}

\usepackage{hyperref}
\usepackage{mathrsfs}
\usepackage[caption=false]{subfig}
\usepackage[dvipsnames]{xcolor}
\definecolor{myred}{rgb}{0.89,0.259,0.204}
\def\red#1{\textcolor{red}{#1}}

\def\black#1{\textcolor{black}{#1}}

\newcommand{\etal}{$et\, al.$}
\usepackage{txfonts}
\newcommand{\comment}[1]{}

\begin{document}


\title{Gravitational-wave echoes from spinning exotic compact objects: \\numerical waveforms from the Teukolsky equation}

\author{Shuo Xin}
\affiliation{Tongji University, Shanghai 200092, China}
\author{Baoyi Chen}
\affiliation{Burke Institute for Theoretical Physics, California Institute of Technology, Pasadena, CA 91125, USA}
\author{Rico K.~L.~Lo}
\affiliation{LIGO Laboratory, California Institute of Technology, Pasadena, CA 91125, USA}
\author{Ling Sun}
\affiliation{LIGO Laboratory, California Institute of Technology, Pasadena, CA 91125, USA}
\affiliation{OzGrav-ANU, Centre for Gravitational Astrophysics, College of Science, The Australian National University, ACT 2601, Australia}
\author{Wen-Biao Han}
\affiliation{Shanghai Astronomical Observatory, Shanghai, 200030, China}
\author{Xingyu Zhong}
\affiliation{Shanghai Astronomical Observatory, Shanghai, 200030, China}
\author{Manu Srivastava}
\affiliation{Department of Physics, Indian Institute of Technology Bombay, Mumbai 400076, India}
\author{Sizheng Ma}
\affiliation{Burke Institute for Theoretical Physics, California Institute of Technology, Pasadena, CA 91125, USA}
\author{Qingwen Wang}
\affiliation{Perimeter Institute and the University of Waterloo, Canada}
\author{Yanbei Chen}
\affiliation{TAPIR, California Institute of Technology, Pasadena, CA 91125, USA}

\date{\today}

\begin{abstract}
We present numerical waveforms of gravitational-wave echoes from spinning exotic compact objects (ECOs) that result from binary black hole coalescence.  We obtain these echoes by solving the Teukolsky equation for the $\psi_4$ associated with gravitational waves that propagate toward the horizon of a Kerr spacetime, and process the subsequent reflections of the horizon-going wave by the surface of the ECO, which lies right above the Kerr horizon.  The trajectories of the infalling objects are modified from Kerr geodesics, such that the gravitational waves propagating toward future null infinity match those from merging black holes with comparable masses. 
In this way, the corresponding echoes approximate to those from comparable-mass mergers. 
For boundary conditions at the ECO surface, we adopt recent work using the membrane paradigm, which relates $\psi_0$ associated with the horizon-going wave and $\psi_4$ of the wave that leaves the ECO surface. We obtain $\psi_0$ of the horizon-going wave from $\psi_4$ using the Teukolsky-Starobinsky relation. The echoes we obtain turn out to be significantly weaker than those from previous studies that generate echo waveforms by modeling the ringdown part of binary black hole coalescence waveforms as originating from the past horizon.
\end{abstract}

\maketitle

\section{\label{Section: Introduction}Introduction}
The detection of binary black hole (BBH) mergers, e.g., GW150914~\cite{GW150914}, and binary neutron star (BNS) collisions, e.g., GW170817~\cite{GW170817}, has opened the era of gravitational wave (GW) astronomy. In the third observing run of Advanced LIGO~\cite{LIGO2014} and Virgo~\cite{Virgo2014}, completed in March 2020, the upgraded detectors reached further in space and observed a significantly increased number of events. The LIGO-Virgo Collaboration has confirmed 50 GW events in total during the first, second, and the first half of the third observing runs (as of October 1st, 2019)~\cite{catalog, catalog2}. With this new messenger to observe signals from strong gravity regime, we are now able to test theories of gravity in ways that was inaccessible before~\cite{abbott2019tests,abbott2020tests,berti2018extremea,berti2018extremeb}.

In General Relativity (GR), black holes (BHs) are the standard models for compact stellar remnants of the gravitational collapse of massive stars at the end of their lives and for massive objects at centers of galaxies.  However, exotic matter equations of states~\cite{palenzuela2017gravitational}, phase transitions~\cite{mazur2004gravitational,pani2009gravitational}, or effects of quantum gravity~\cite{giddings2018event,bianchi2020distinguishing,bena2020multipole,mukherjee2020multipole} allow the existence of {\it Exotic Compact Objects} (ECOs), whose external spacetime has the same geometry as a BH --- except in a small region near the horizon (i.e., with size $\ll M$, where $M$ is the BH mass)~\cite{ecology,qc,spinechoSearch1}.  Searching for and detecting ECOs would push the frontier of fundamental physics. Without detection, upper limits on ECO properties derived from the search can be used to quantitatively test ``how black are BHs'', thereby confirming the existence of the event horizon, the boundary of a region within which signals cannot be sent to distant observers. 


Since the ECOs have nearly the same external spacetime geometry as BHs, geodesic motions around them are identical to those around BHs, therefore one way to probe ECOs is through tidal interactions during a binary inspiral process~\cite{li2008generalization,PhysRevD.72.124016,Datta:2020rvo,datta2020tidal}. An additional approach is via GW echoes~\cite{nature17,abedi2017echoes, ashton2016comments}, namely the GWs reflected from the near-horizon region of the final ECO formed during the merger process.  
One prominent feature of these echoes is that the lag of echoes behind the main wave corresponds to the exponentially small distance between the ECO surface and the location of the horizon.  Thus, GW echoes can be used to probe even Planck-scale structures near the horizon~\cite{nature17}. Even though there is still no statistically significant evidence for echoes in existing GW data~\cite{echoADA,echoLowSig,PhysRevD.99.104012,  echoSearch1,echoSearch2,echoSearch3,echoSearch4,Tsang:2019zra, Uchikata:2019frs, Abbott:2020jks}, echo signals would be a promising candidate for probing physics beyond GR.

Many studies about GW echoes have been carried out for spherically symmetric, non-spinning ECOs, whose external spacetime is Schwarzschild. For instance, Mark \etal{} studied echo modes of scalar waves in some ECO models by solving the scalar perturbation equations with reflecting boundaries \cite{echoSchw}. Du \etal{} solved GW echo modes based on the Sasaki-Nakamura formalism and studied its contribution to stochastic background~\cite{PRL18}. Huang \etal{} developed the Fredholm method and a diagrammatic representation of echo solution for general wave equations \cite{huang2019fredhom}.  Maggio \etal{} studied the ringdown of non-spinning ECOs~\cite{maggio2020does}.

Astrophysically, we expect that merging BHs are spinning, at least to some extent.  Indeed, the LIGO-Virgo Collaboration has detected mergers of spinning binary BHs \cite{catalog2}. Inferring these spins has led to better understandings of these BHs' formation history and their stellar environments \cite{LIGOspin}.  More importantly, even non-spinning merging BHs will generally result in significantly spinning remnants.  For example, an equal-mass, non-spinning BH binary in a quasi-circular merger leads to a final BH with $J/M^2 \sim 0.7$ (with units $G = c = 1$), where $J$ is the remnant angular momentum. This motivates the study of GW echoes from spinning ECOs. 

In this paper, we assume that the external spacetime geometry of a spinning ECO is Kerr, except in a small region with distance $\ll M$ near the horizon, and pose a boundary condition at the boundary of that region. Here we need to point out that, while in the spherically symmetric case, one can use the Birkhoff's theorem to argue that the exterior spacetime of a spherical ECO should be Schwarzschild in GR, a similar argument does not exist for spinning objects.  For spinning ECOs, one eventually needs to consider the composite effects of spacetime deviation and near-horizon boundary condition. Note that mapping exterior spacetime geometry of compact objects (or ``bumpy BHs'') has been a subject of extensive studies~\cite{ryan1997accuracy,vigeland2010spacetime,PhysRevD.78.102001}.
Recent studies have further obtained non-spherically symmetric fuzzball geometries that arise from spacetime microstates~\cite{bianchi2020distinguishing,PhysRevLett.125.221602}.

Let us now get to effects of the boundary condition at the ECO surface.   Nakano \etal{} \cite{tanaka} have constructed a model for echoes from spinning ECOs, where the asymptotic behavior of solutions to the Teukolsky equation is used to analyze the reflectivity and echo modes, but the incident wave toward the horizon is phenomenological. Micchi and Chirenti studied effects of the rotation using a scalar charge falling into a Kerr spacetime \cite{micchi2020spicing}. Wang \etal{} \cite{wang2020echoes}, Maggio \etal{} \cite{Maggio} and Micchi \etal{} \cite{micchi2021loud} obtained echo waveforms from spinning ECOs by first deducing waves that propagate toward the horizon from waves that propagate toward infinity, and then imposing a reflectivity for Sasaki-Nakamura (SN) functions. Sago \etal{} \cite{sago2020gravitational} studied echoes from a particle radially falling into a spinning ECO.  


One caveat when studying echoes from ECOs comes from the instability of ECOs, either from the structure of the ECO itself, including superradiance for spinning ECOs~\cite{maggio2017exotic,maggio2019ergoregion} and the existence of stable photon orbits~\cite{cunha2017light}, or from the energy content of GWs that propagate toward the ECO, which may induce the ECO to collapse~\cite{addazi2020gravitational,chen2019instability}.  Despite all such instabilities, we advocate keeping an open mind about echoes. The superradiant instability can either be quenched or lead to non-spinning ECOs. We might also argue that instabilities that cause gravitational collapse can simply cause the event horizon to grow, while non-local effects of quantum gravity would keep appearing right outside the new location of the growing event horizon.  

In this article, we construct echoes from spinning ECOs using the Teukolsky formalism. We evolve a point particle in quasi-circular orbits, until it finally plunges into a Kerr BH.  Our approach improves from previous work, notably Refs.~\cite{wang2020echoes,Maggio,micchi2021loud}, by: (i) using gravitational-wave waveform towards the horizon directly computed in the Teukolsky formalism, and (ii) using a boundary condition on the ECO surface that is connected to tidal tensors measured locally by zero-angular-momentum observers floating right above the horizon (fiducial observers used in the membrane paradigm). Since the current focus of GW astronomy is on binaries with nearly comparable masses, we tune our point-particle trajectory in such a way that the waveform at infinity matches the Numerical Relativity (NR) surrogate waveform from non-precessing binaries with comparable masses; this is similar to the approach taken by Ref.~\cite{micchi2021loud}. 

The paper is organized as follows.  In Sec.~\ref{p_formulation}, we briefly review the Teukolsky formalism and obtain trajectories of  particles in Kerr spacetime whose gravitational waves (in terms of the Newman-Penrose $\psi_4$ scalar) at infinity match those from comparable-mass binaries that merge from quasi-circular inspirals.  In Sec.~\ref{p_waveforms}, we discuss how to obtain echoes at infinity from $\psi_4$ waveforms that go down the horizon, in particular deducing a conversion factor from the echoes obtained in the SN formalism to those obtained by imposing more physical boundary conditions on curvature perturbations. In Sec.~\ref{Section:Results}, we discuss features of the GW echoes, demonstrating the effect of the conversion factor obtained in the previous section, and highlight a subtlety that leads to discrepancy between $\psi_4$ directly obtained using the Teukolsky formalism and $\psi_4$ obtained using approximations imposed by Refs.~\cite{wang2020echoes,Maggio,micchi2021loud}. The main conclusions are summarized in Sec.~\ref{sec:conclusions}.

\section{Particle falling into a BH} 
\label{p_formulation}

In this section, we briefly review the computation of waveforms of a particle falling into a Kerr BH, both at infinity and near the horizon.  We follow the prescription of the effective one-body (EOB) and Teukolsky formalism~\cite{han2014gravitational} and construct particle trajectories in the Kerr spacetime that decay due to radiation reaction, in such a way that the waveforms at infinity match those from {\it comparable-mass binaries} obtained from NR surrogate models~\cite{blackman2015fast}.  Here the study is restricted to non-spinning binaries.  The mass $M$ and the angular momentum per unit mass $a$ of the Kerr spacetime correspond to those of the remnant formed by the binary merger, which can be obtained from the initial total mass and mass ratio of the binary via NR surrogate models~\cite{vijay_varma_2018_1435832,PhysRevLett.122.011101}.

\subsection{GWs emitted by a particle falling into a Kerr BH}
Let us first consider GWs emitted by a particle falling into a Kerr BH. We use the Newman-Penrose scalar curvature $\psi_4$, which can be decomposed into frequency and angular components in the Boyer-Lindquist coordinates 
\begin{align}
\label{eq:psi4}
    &\psi_4(t,r,\theta,\phi) \nonumber\\ 
    &=   \rho^4 \int^{+\infty}_{-\infty}{{d\omega}\sum_{\ell m}{R_{\ell m\omega}(r)_{~-2}S^{a\omega}_{\ell m}(\theta)e^{im\phi}e^{-i\omega t}}},
\end{align}
with $\rho =   \left({r-ia\cos \theta}\right)^{-1}$. 
Here  $_{~-2}S^{a\omega}_{\ell m}$ is the spin-weighted spheroidal harmonic with eigenvalue $E_{\ell m}$ \footnote{See, e.g., Sec.~III of Ref.~\cite{press1973perturbations}.}, while $R_{\ell m\omega}$ is the solution to the radial Teukolsky equation,
\begin{equation}
\Delta^2\frac{d}{dr}\left(\frac{1}{\Delta}\frac{d
	R_{\ell m\omega}}{dr}\right)-V(r)R_{\ell m\omega}=-\mathscr{T}_{\ell m\omega}(r),
\label{Teukolsky}
\end{equation}
with the potential
\begin{equation}
V(r)=-\frac{K^2+4i(r-M)K}{\Delta}+8i\omega r+\lambda,
\end{equation}
where $K=(r^2+a^2)\omega-ma, ~\lambda=E_{\ell m}+a^2\omega^2-2a m\omega -2$, and $\Delta = r^2-2Mr+a^2$. The source term $\mathscr{T}_{\ell m\omega}(r)$ is determined by the mass and trajectory of the particle (discussed in Sec.~\ref{subsec:Traj}).  


Homogeneous solutions for the radial Teukolsky equation have two types of asymptotic behaviors each, near the horizon and at infinity, respectively.  We are particularly interested in two combinations of such solutions, namely the one that is purely in-going near the horizon,
\begin{align}
\label{eq_RH}
R^{{\rm H}}_{\ell m\omega}(r)
=
\left\{
\begin{array}{ll}
B^{{\rm hole}}_{\ell m\omega}\Delta^2 e^{-ipr_*} \quad
 & r\rightarrow r_+  \\ 
 \\
B^{{\rm out}}_{\ell m \omega}r^3 e^{i\omega
	r_*}+r^{-1}B^{{\rm in}}_{\ell m \omega} e^{-i\omega r_*} &  r\rightarrow
\infty,
\end{array}
\right.
\end{align}
and the one that is purely out-going at infinity
\begin{align}
\label{eq_Rinf}
R^{\infty}_{\ell m \omega}(r)= \left\{
\begin{array}{ll}
D^{\rm{out}}_{\ell m \omega} e^{ip r_{*}}+\Delta^2
D^{{\rm in}}_{\ell m \omega} e^{-ip r_{*}}  & r\rightarrow r_+ \\
\\
 D^{\infty}_{\ell m \omega} r^3 e^{i\omega
	r_{*}} &  r\rightarrow \infty.
\end{array}
\right.
\end{align}
Here we define
\begin{equation}
    p=\omega-m\Omega_+\,,\quad \Omega_+ = a/(r_+^2+a^2),
\end{equation}
with $\Omega_+$ being the horizon's rotation angular frequency, and the tortoise coordinate $r_*$ as
\begin{equation}
\label{rs}
    r_* = r+\frac{2Mr_+}{r_+-r_-}\log\frac{r-r_+}{2M}-\frac{2Mr_-}{r_+-r_-}\log\frac{r-r_-}{2M}\,.
\end{equation}
The quantities $B^{\rm in,\,out,\,hole,\,\infty}$ and $D^{\rm in,\,out,\,hole,\,\infty}$ are related to the transmissivity and reflectivity of the compact object; their values here are subject to a choice of conventions, for which we follow the convention of Hughes~\cite{Hughes}. We compute these homogeneous solutions numerically with the help of the SN formalism based on the codes developed in \cite{han2010prd,han2011prd} (see review in the Appendix).
		
When the central object is a BH, we look for solutions that are only in-going at the horizon and only out-going at infinity, by imposing
 \begin{align}
 \label{eq_asym}
  R^{{\rm BH}}_{\ell m \omega}
  =\left\{
  \begin{array}{ll}
Z^{\infty\,{\rm BH}}_{\ell m \omega}r^3 e^{i\omega
 	r_*} & r\rightarrow \infty \\
 	\\
 	Z^{{\rm H\,BH}}_{\ell m \omega}\Delta^2 e^{-ip
 	r_*} & r\rightarrow r_+.
  \end{array}\right.
 \end{align}
This uniquely determines a solution that can be obtained from the Green's function approach,
\begin{align}
    \label{BHsolution}
R^{{\rm BH}}_{\ell m \omega}(r)&=\frac{R^{\infty}_{\ell m \omega}(r)}{2i\omega
	B^{{\rm in}}_{\ell m \omega}D^{\infty}_{\ell m \omega}}\int^{r}_{r_+}{dr'\frac{R^{{\rm H}}_{\ell m \omega}(r')\mathscr{T}_{\ell m \omega}(r')}{\Delta(r')^2}}\nonumber\\
	&+\frac{R^{{\rm H}}_{\ell m \omega}(r)}{2i\omega
	B^{{\rm in}}_{\ell m \omega}D^{\infty}_{\ell m \omega}}\int^{\infty}_{r}{dr'\frac{R^\infty_{\ell m \omega}(r')\mathscr{T}_{\ell m \omega}(r')}{\Delta(r')^2}},
\end{align}
from which we can read off
\begin{equation}
\label{eq_ZH}
Z^{{\rm \infty}\,\rm BH}_{\ell m \omega} = \frac{1}{2i\omega B^{{\rm in}}_{\ell m \omega}} \int_{r_+}^{\infty} dr' \frac{R^{{\rm H}}_{\ell m \omega}(r') \mathscr T _{\ell m \omega}(r') }{\Delta(r')^2 },
\end{equation}
and 
\begin{equation}
\label{eq_Zinf}
Z^{\rm H\,\rm BH}_{\ell m \omega} = \frac{B^{{\rm hole}}_{\ell m \omega}}{2i\omega B^{{\rm in}}_{\ell m \omega} D^{\infty }_{\ell m \omega}} \int_{r_+}^{\infty} dr' \frac{R^\infty_{\ell m \omega}(r') \mathscr T _{\ell m \omega}(r') }{\Delta(r')^2 }.
\end{equation}
In particular, at $r\rightarrow +\infty$, $\psi_4$ is related to GW polarizations $h_+$ and $h_\times$ by 
\begin{equation} 
\psi_4(r\rightarrow \infty)= \frac 12 (\ddot{h}_+ - i\ddot{h}_\times).
\end{equation}
The GWs we observe at a distance $r$, latitude angle $\Theta$ and azimuthal angle $\Phi$, is given by:
 \begin{align}
 \label{BHwaveform}
 &h^{{\rm BH}}_+-ih^{{\rm BH}}_\times|_{(r,\Theta,\Phi,t)}\nonumber\\
 =-&\frac{2}{r} \sum_{\ell m}\int_{-\infty}^{+\infty}{ d\omega \frac{Z^{{\infty}}_{\ell m\omega}}{\omega^2} \,_{-2}S^{a\omega}_{\ell m}(\Theta,\Phi)e^{-i \omega (t -r_*)}}.
 \end{align}

Here we note that the spin-weighted spheroidal harmonics differ from the spin-weighted spherical harmonics, and are frequency dependent. 
In NR waveform catalogs, as well as in surrogate models, GW strains at infinity are decomposed into spin-weighted spherical harmonics
\begin{equation}
r \left[    h_{+}^{\rm NR} - ih_\times^{\rm NR} \right] \Big|_{(\Theta,\Phi,t)} =\sum_{\ell m} h_{\ell m}^{\rm NR} (t) \,{}_{-2}Y_{\ell m}(\Theta,\Phi)\,.
\end{equation}
In this paper, we treat the mode-mixing in the spheroidal harmonics as negligible, writing ${}_{-2}S_{\ell m}^{a\omega} \approx {}_{-2}Y_{\ell m}$. This allows us to make a simple connections between $Z_{\ell m \omega}^{\infty}$ and $h_{\ell m}^{\rm NR}$:
\begin{equation}
    -2\frac{Z_{\ell m \omega}^{\infty}}{\omega^2} \leftrightarrow h_{\ell  m\omega}^{\rm NR}\,.
\end{equation}
We also focus on the $\ell =2$ contributions, with $m = \pm 2$. Note that ${}_{-2}Y_{22} (\Theta,\Phi)$ and ${}_{-2}Y_{2-2}(\Theta,\Phi)$ predominantly emit toward the northern hemisphere ($\Theta < \pi/2$) and southern hemisphere ($\Theta >\pi/2$), respectively.  Both $m=+2$ and $-2$ contributions are equally important to describe the ``(2,2)'' waveform.  Furthermore, for non-precessing binaries with angular momentum along the $z$ axis ($\Theta=0$), we have $h_{\ell m} (f) = h_{\ell -m}^*(-f)$.  The studies in this paper are based on this scenario.



\subsection{Trajectory of particles in Kerr and the Teukolsky source terms}
\label{subsec:Traj}

 
We aim to use Teukolsky waveforms to approximate those from coalescence of BHs with comparable masses.  First of all, in order for the ringdown frequencies to match up, we set the mass and spin of the Kerr background spacetime equal to those of the remnant BH of a comparable-mass binary merger.  We then evolve particle trajectories by modifying the Kerr geodesic equation, adding generalized forces that implement the effect of radiation reaction, in such a way that the late inspiral, merger, and ringdown parts of the waveforms match those from comparable-mass binaries, obtained from surrogate models. 

For a trajectory in the Boyer-Lindquist system, parametrized as $x^\mu(\tau) = (t(\tau), r(\tau),\theta(\tau), \phi(\tau))$, our modified equations are written as
\begin{align}
    \frac{dx^\mu}{d\tau}  & =u^\mu, \\
        \frac{d{u^\mu}}{d\tau} &=-\Gamma^\mu_{\rho\sigma}u^\rho u^\sigma + \mathcal{F}^\mu \,.
\end{align}
The radiation reaction force $\mathcal{F}^\mu$ can be obtained from the GW energy flux $\dot E$, angular momentum flux $\dot L_z$ and rate of change of Carter constant $\dot Q$, by solving the following equations:
\begin{align}
&\dot{E} u^t = - g_{tt} \mathcal{F}^t - g_{t\phi} \mathcal{F}^\phi, \\
&\dot{L_z} u^t = g_{t\phi} \mathcal{F}^t + g_{\phi\phi} \mathcal{F}^\phi,\\
&\dot{Q} u^t=2g_{\theta\theta}^2 u^\theta \mathcal{F}^\theta + 2\cos^2 \theta a^2 E \dot{E} + 2 \cos^2\theta\frac{L_z\dot{L_z}}{\sin^2 \theta}, \\
&g_{\mu\nu} u^\mu\mathcal{F}^\nu=0.
\label{eq_radiationF}
\end{align}
In this paper, we focus on non-precessing binaries.  Hence we consider equatorial circular orbits and impose $\dot Q=0$; $\dot E$ and $\dot L_z$ are determined phenomenologically such that the Teukolsky waveforms (in the BH case) match those obtained from NR. We also assume that the ECO does not modify these forces. 

From the numerical trajectory $x^\mu(\tau)$, or alternatively the 3-dimensional trajectory as a function of time  $r(t),\theta(t), \phi(t)$, the source term in the Teukolsky equation is given by (see Appendix~\ref{app_source} for details) 
\begin{align}
\label{eq:T:detail}
	\mathscr{T}_{\ell m \omega} (r') 
	= \int_{-\infty}^\infty dt & e^{i[\omega t - m \phi(t)]}  \Delta^2(r') \nonumber\\
	& 
	\Big\{ [A_{nn0} + A_{n\bar m 0} + A_{\bar{m}\bar m 0} ] \delta(r'-r(t)) \nonumber \\
	&+\partial_{r'} ([A_{n\bar m 1} + A_{\bar m \bar m 1} ]\delta (r'-r(t)) ) \nonumber\\
	&+\partial_{r'}^2 [ A_{\bar m \bar m 2} \delta(r'-r(t))] \Big\} \,.
\end{align}
Plugging this source term into Eqs.~(\ref{eq_ZH}) and (\ref{eq_Zinf}), we obtain amplitudes of GWs toward infinity,
\begin{align}
\label{source_term_H}
Z^{\infty}_{\ell m \omega} = \frac{1}{2i\omega B^{in}_{\ell m \omega}} \int_{-\infty}^{+\infty} dt \; & e^{i[\omega t - m\phi(t)]} \nonumber\\
\Bigg\{& R^H_{\ell m \omega}(r(t)) [A_{nn0} + A_{n\bar m0}+A_{\bar m \bar m 0}]  \nonumber\\
-&  \frac{d R^H_{\ell m \omega}}{dr}\bigg|_{r(t)} [A_{n\bar m 1}  
+ A_{\bar m \bar m 1} ]\nonumber\\
+&\frac{d^2 R^H_{\ell m \omega}}{dr^2}\bigg|_{r(t)} A_{\bar m\bar m 2} \Bigg\},
\end{align}
and toward the horizon,
\begin{align}
\label{source_term_inf}
Z^{\rm H}_{\ell m \omega} = \frac{B^{\rm hole}_{\ell m \omega}}{2i\omega D^{\infty}_{\ell m \omega} B^{in}_{\ell m \omega}} \int_{-\infty}^{+\infty} dt \; & e^{i[\omega t - m\phi(t)]} \nonumber\\
 \Bigg\{ &  R^\infty_{\ell m \omega}(r(t)) [A_{nn0} + A_{n\bar m0}+A_{\bar m \bar m 0}] \nonumber \\
-& \frac{d R^\infty_{\ell m \omega}}{dr}\bigg|_{r(t)} [A_{n\bar m 1} + A_{\bar m \bar m 1} ] \nonumber\\
+&\frac{d^2 R^\infty_{\ell m \omega}}{dr^2}\bigg|_{r(t)} A_{\bar m\bar m 2} \Bigg\}.
\end{align}
In the above integrals, the integration variable $t$ parametrizes locations on the trajectory of the in-falling particle, with \mbox{$t\rightarrow -\infty$} corresponding to the beginning of the infall, and \mbox{$t\rightarrow +\infty$} corresponding to when the particle approaches $r_+$. The integrand approaches zero for \mbox{$t\rightarrow +\infty$}. For \mbox{$t\rightarrow -\infty$}, we apply a window which selects part of the trajectory within a finite distance from the 
BH/ECO. We note that as \mbox{$t\rightarrow +\infty$}, although the individual terms, e.g., $A_{\bar m \bar m 0}$ in Eq.~\eqref{source_term_inf} can diverge, the sum of those terms approaches zero due to the cancellation between the terms.  

\subsection{Waveform at infinity: calibration with surrogate models}

\begin{figure*}[t]
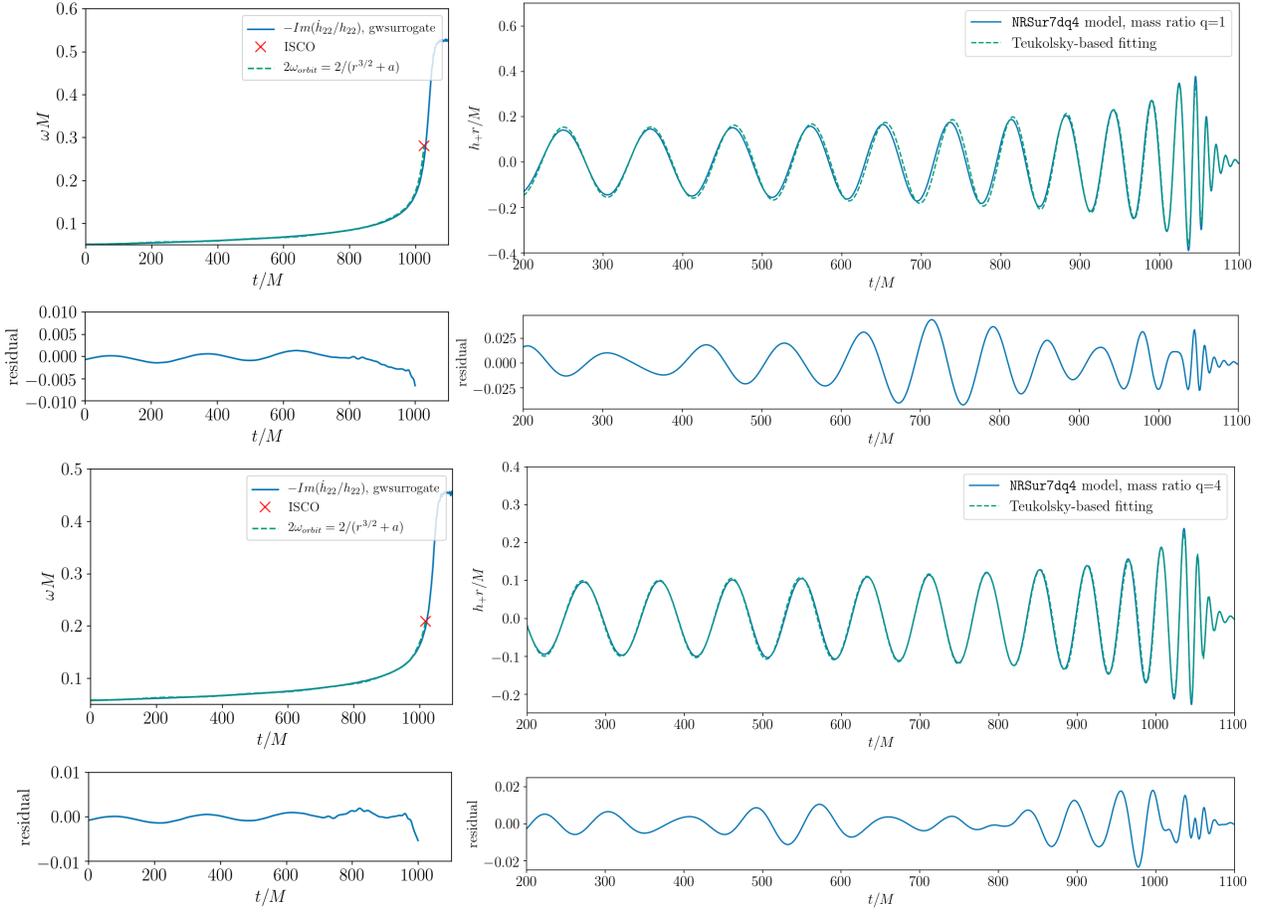

	\centering
	\includegraphics[height=4cm]{q1_omg.pdf}	
	\includegraphics[height=4cm]{q1_hp.pdf}	
	
	\hspace{-0.5cm}
	\includegraphics[height=2.05cm]{q1_residual.pdf}
	\hspace{-0.25cm}
	\includegraphics[height=1.93cm]{q1_hresidual.pdf}	
	\includegraphics[height=4cm]{q4_omg.pdf}	
	\includegraphics[height=4cm]{q4_hp.pdf}

	\hspace{-0.4cm}
	\includegraphics[height=2.05cm]{q4_residual.pdf}
	\hspace{-0.15cm}
	\includegraphics[height=1.91cm]{q4_hresidual.pdf}	
	\caption{ 
	Comparison between the NR surrogate model \texttt{NRSur7dq4} and waveforms generated from the Teukolsky code with phenomenological trajectory described by Eq.~\eqref{eq_radiationF} (see text for details), for $q=1$ (top two panels) and $q=4$ (bottom two panels). The time $t$ is shifted so that the waveform starts at a certain frequency during inspiral, which is consistent with the starting time of the phenomenological trajectory.
	}
	\label{fig_surrogate}
\end{figure*}


In order to obtain the trajectory that leads to a Teukolsky waveform matching the surrogate waveform, we take the $(2,2)$ component of the surrogate waveform, evaluate its instantaneous angular frequency $\omega_{(2,2)}$, as well as its rate of change $\dot\omega_{(2,2)}$, and make sure that, before reaching the Innermost Stable Circular Orbit (ISCO), the {\it orbital} frequency of the particle follows the a corresponding evolution with $\omega_{\rm orb} = 1/2\omega_{(2,2)}$ by adjusting the radiation reaction forces.  We turn off the radiation reaction forces after the particle reaches ISCO. The amplitude of the final waveform from this orbit is rescaled so that the normalization is the same as that of the surrogate model. Note that the time axis of the surrogate model waveform is also rescaled to units of remnant mass. 


To make comparisons, we take $(\ell,m)=(2,2)$, compute $Z_{22\omega}^{\infty\,{\rm BH}}$, obtain the time-domain waveform via an inverse Fourier transform [see ~\eqref{BHwaveform}], and then compare the waveform with the NR surrogate model. 
Fig.~\ref{fig_surrogate} shows the wavforms with mass ratio $q=1$ and $q=4$ obtained by ``NRSur7dq4'' model~\cite{PhysRevResearch.1.033015}, compared with our approximate Teukolsky-based waveforms.  For the portion of the waveform shown in the figure, using Advanced LIGO noise spectrum at the design sensitivity, 
the match~\cite{sathyaprakash2009physics}  between the Teukolsky and the NR surrogate waveforms are above 0.99 for all $M$ between $10\,M_\odot$ and $500\,M_\odot$. 

\section{Constructing echoes}
\label{p_waveforms}

In this paper, we assume that the ECO spacetime is identical to a Kerr spacetime except for $ r_* /M\ll -1$.  
Since GW echoes are mainly sourced by the plunge part of the trajectory, which is not significantly affected by the radiation reaction, we can neglect the modification to the trajectory due to the ECO surface.  For these reasons, as a particle falls towards the ECO, we keep the same Teukolsky equation, with the source term in Eq.~\eqref{Teukolsky} given by Eq.~\eqref{eq:T:detail}. 
However, we need to change the boundary condition for $\rho^{-4}\psi_4$ to a more general form
 \begin{align}
 \label{eq_asym}
  R^{{\rm ECO}}_{\ell m \omega}
  =\left\{
  \begin{array}{ll}
Z^{\infty \rm ECO}_{\ell m \omega}r^3 e^{i\omega
 	r_*} & r\rightarrow \infty \\
 	\\
 	Z^{{\rm in}}_{\ell m \omega}\Delta^2 e^{-ip
 	r_*} +Z^{{\rm out}}_{\ell m \omega} e^{ip
 	r_*}  & r\rightarrow r_+,
  \end{array}\right.
 \end{align}
where $Z^{\rm out}_{\ell m \omega}$ appears due to the ``reflection'' from the ECO surface, while $Z^{\rm in}_{\ell m \omega}$ is modified since additional waves propagate toward the ECO upon reflection from the inner side of the Kerr potential barrier near the light ring. 
 
In Sec.~\ref{subsec:bcsn}, we first prescribe reflectivity within the SN framework, following previous literature.  However, it turns out that for Kerr spacetime, it has a more direct physical meaning to impose boundary conditions in terms of $\psi_0$ and $\psi_4$, which are tied to curvature perturbations experienced by observers near the future and past horizons, respectively.  We obtain echo formulas from such boundary conditions in Sec.~\ref{subsec:teu}.
In Sec.~\ref{p_reflectivity}, we review the reflectivity models used in subsequent sections.  

\subsection{Boundary condition imposed on SN functions}
\label{subsec:bcsn}
 
\subsubsection{SN formalism and boundary condition}
The simplest way to describe the ECO's reflection of GWs is to use the SN formalism (see Appendix~\ref{p_perturbation}). We use fields $X_{\ell m \omega}$ that satisfies the SN equation
\begin{align}
\label{eq_SN1}
\frac{d^2
	X_{\ell m \omega}}{dr_*^2}-F(r)\frac{dX_{\ell m \omega}}{dr_*}-U(r)X_{\ell m \omega}=0.
\end{align}  
The transformation between the SN function and Teukolsky radial function is given by
\begin{align}
\label{eq_SN2T}
R^{}_{\ell m \omega}=\frac{1}{\eta}\left[\left(\alpha+\frac{\beta_{,r}}{\Delta}\right)\frac{\Delta
	X^{}_{\ell m \omega}}{\sqrt{r^2+a^2}}-\frac{\beta}{\Delta}\frac{d}{dr}\frac{\Delta
	X^{}_{\ell m \omega}}{\sqrt{r^2+a^2}}\right].
\end{align}  
The potentials $F(r)$ and $U(r)$ in Eq.~\eqref{eq_SN1} and functions $\alpha$, $\beta$, and $\eta$ in Eq.~\eqref{eq_SN2T} can be found in Eqs.~(3.4)--(3.9) of Ref.~\cite{mino1997black} (also see Appendix~\ref{p_perturbation}). 
To derive $R^{H}$ and $R^{\rm \infty}$, we use two homogeneous solutions of the SN equation, which have purely sinusoidal dependence on $r_*$ due to the short-ranged potential:
\begin{align}
\label{eq_XH}
X^{{\rm H}}_{\ell m \omega}(r)& =\left\{
\begin{array}{ll}
A^{{\rm hole}}_{\ell m \omega}e^{-ipr_*} &  r\rightarrow r_+ \\
\\
A^{\rm{out}}_{\ell m \omega}e^{i\omega
	r_*}+A^{\rm{in}}_{\ell m \omega}e^{-i\omega r_*} & r\rightarrow
\infty,
\end{array}\right. \\
\label{eq_Xinf}
X^{\infty}_{\ell m \omega}(r)
& =\left\{
\begin{array}{ll}
C^{\rm{out}}_{\ell m \omega}e^{ip
	r_*}+C^{\rm{in}}_{\ell m \omega}e^{-ip r_*} & r\rightarrow r_+ \\
	\\
C^{\infty}_{\ell m \omega}e^{i\omega r_*} & r\rightarrow
\infty.
\end{array} 
\right.
\end{align}
Because these $X$'s are directly used to compute the corresponding $R$'s, there are relations between the amplitudes $A$, $C$ here and $B$, $D$ in Eqs.~(\ref{eq_RH}) and (\ref{eq_Rinf}), given in Appendix \ref{p_perturbation}. For convenience, we can set $C^{\infty}_{\ell m \omega}=A^{{\rm hole}}_{\ell m \omega}=1$. Using linearity, we can also write

\begin{equation}
\label{bcSN}
X^{{\rm ECO}}_{\ell m \omega} = \xi_{\ell m \omega}^{\rm in} e^{-ipr_*} +\xi_{\ell m \omega}^{\rm out}    e^{ipr_* }\,, \quad r_* \rightarrow -\infty
\end{equation}
with 
\begin{equation}
\xi_{\ell m \omega}^{\rm out} = {\mathcal{R}}^{\rm ECO}_{\ell m \omega}   
\xi_{\ell m \omega}^{\rm in},
\end{equation}
where ${\mathcal{R}}^{\rm ECO}_{\ell m \omega}$ is the ECO reflectivity.
This is the baseline approach taken by most previous literature~\cite{PRL18,micchi2021loud,Maggio,wang2020echoes}, except for~\cite{tanaka}, where using energy reflectivity is proposed.
As discussed in more detail later in this paper, although $|\mathcal{R}_{\ell m \omega}^{\rm ECO}|^2$ corresponds to energy reflectivity in the Schwarzschild case, it is not generally true for Kerr BHs, or spinning ECOs.  We introduce more physically motivated reflectivites in Sec.~\ref{p_reflectivity} and comment on additional subtleties of the SN formalism.  At this stage, by comparing Eq.~\eqref{bcSN} with Eq.~\eqref{eq_asym}, as well as Eqs.~\eqref{eq_RH}, \eqref{eq_Rinf}, \eqref{eq_XH} and \eqref{eq_Xinf}, we write
\begin{equation}
    \frac{\xi^{\rm in}_{\ell m \omega}}{Z^{\rm in}_{\ell m \omega}} = \frac{C^{\rm in}_{\ell m \omega}}{D^{\rm in}_{\ell m \omega}}=\frac{A^{\rm hole}_{\ell m \omega}}{B^{\rm hole}_{\ell m \omega}}=\frac{1}{B^{\rm hole}_{\ell m \omega}}\,,\quad
    \frac{\xi^{\rm out}_{\ell m \omega}}{Z^{\rm out}_{\ell m \omega}}=\frac{C^{\rm out}_{\ell m \omega}}{D^{\rm out}_{\ell m \omega}}\,.
\end{equation}
Since the conversion relation~\eqref{eq_SN2T} between SN and Teukolsky functions involves derivatives, these formulas are only correct when the source terms vanish rapidly enough --- which is true for sources that are bounded outside the horizon, but not necessarily true for a particle that plunges into the horizon. See Sec.~\ref{sub} for more discussions. 



\subsubsection{Echoes in the SN Formalism}

In the BH case,  we obtain
 \begin{align}
 \label{eq_asym_BH}
 X^{{\rm BH}}_{\ell m \omega}(r) 
  =\left\{
  \begin{array}{ll}
\xi^{\infty\,{\rm BH}}_{\ell m \omega} e^{i\omega
 	r_*} & r\rightarrow \infty \\
 	\\
 	\xi^{\rm H\,BH}_{\ell m \omega}e^{-ip
 	r_*} & r\rightarrow r_+,
  \end{array}\right.
 \end{align}
in the SN formalism.
Then in the ECO case, we need to superimpose an additional homogeneous solution to form a new solution, 
\begin{equation}
     X^{{\rm ECO}}_{\ell m \omega}(r) = X^{{\rm BH}}_{\ell m \omega}(r)  + c X^{\infty}_{\ell m \omega}(r).
\end{equation}
We assume that the source term does not change for ECOs.  After imposing boundary condition~\eqref{bcSN} at the ECO surface, we obtain
\begin{equation}
    c=\frac{\mathcal{R}_{\ell m \omega}^{\rm ECO}}{C^{\rm out}_{\ell m \omega}-\mathcal{R}_{\ell m \omega}^{\rm ECO} C^{\rm in}_{\ell m \omega}}\xi_{\ell m \omega}^{\rm H\,BH} \equiv \mathcal{K}_{\ell m \omega} \xi_{\ell m \omega}^{\rm H\,BH}.
\end{equation}
Using the asymptotic form of $X_{\ell m \omega}^{\infty}$, we obtain
\begin{equation}
\label{eqechoSN}
	\xi^{\infty\,{\rm ECO}}_{\ell m \omega} = \mathcal{K}_{\ell m \omega}\xi^{\rm H\,BH}_{\ell m \omega} + \xi^{\infty\, \rm BH}_{\ell m \omega}.
\end{equation}
We can further write
\begin{equation}
\label{eq_K}
	\mathcal{K}_{\ell m \omega} = \frac{{\mathcal{R}}^{\rm ECO}_{\ell m \omega}   \mathcal{T}^{\rm BH}_{\ell m \omega} }{1- {\mathcal{R}}^{\rm ECO}_{\ell m \omega} \mathcal{R}^{{\rm BH}}_{\ell m \omega}} ,
\end{equation}
where $\mathcal{R}^{{\rm BH}}_{\ell m \omega}$ and $ \mathcal{T}^{\rm BH}_{\ell m \omega}$ are amplitude {\it reflecitity} and {\it transmissivity} of the BH potential barrier --- or the SN potential in Eq.~\eqref{eq_SN1}; in terms of asymptotic expressions of $X_{\ell m \omega}^{\rm \infty}$, they can be written as 
\begin{equation}
\mathcal{T}^{\rm BH}_{\ell m \omega}=\frac{1}{C^{\rm out}_{\ell m \omega}}, \,\quad \mathcal{R}^{\rm BH}_{\ell m \omega} = \frac{C^{\rm in}_{\ell m \omega}}{C^{\rm out}_{\ell m \omega}}.
\end{equation}
In Eq.~\eqref{eqechoSN}, the out-going wave at infinity, in the case of an ECO, is the out-going wave in the case of a BH plus echoes, which are determined by the horizon-going wave in the BH case.


\subsubsection{Echoes in Teukolsky functions}

\begin{figure*}
\includegraphics[width=1\textwidth]{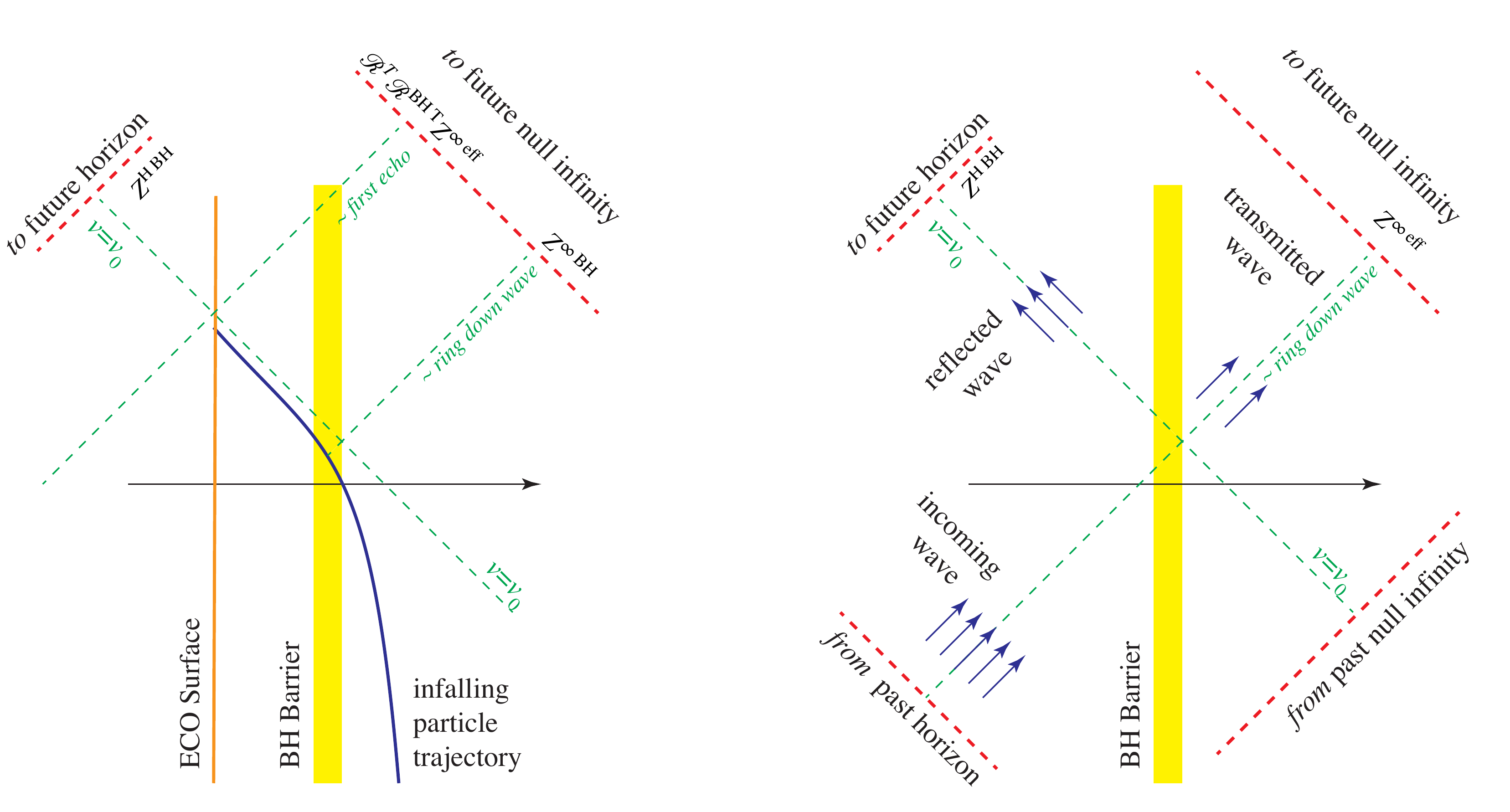}
\caption{Spacetime diagrams illustrating generation of echoes when a particle plunges into a massive ECO (left panel), and the construction of $Z^{\infty\,\rm eff}$ from $Z^{\rm H\,BH}$ (right panel), with $Z^{\infty\,\rm eff}$ being the transmitted wave at future null infinity. In the left panel, we label $v=v_0$ at which the particle plunges into the future horizon; at this point $Z^{\rm H\,BH}$ has a local feature. In the upper-right region of the panel, we also label the approximate locations of ringdown and the first echo in $Z^{\infty\,\rm ECO}$. In the right panel, we indicate that one needs to supply the incoming wave from the past horizon, in order for the BH barrier to reflect it and generate $Z^{\rm H\,BH}$. 
\label{stdiagram}}
\end{figure*}

Using linearity of the transformation between the SN and Teukolsky functions, we have
\begin{equation}
    \frac{\xi^{\infty}_{\ell m \omega}}{Z^{\infty}_{\ell m \omega}}= \frac{A^{\rm out}_{\ell m \omega}}{B^{\rm out}_{\ell m \omega}}= \frac{C^{\infty}_{\ell m \omega}}{D^{\infty}_{\ell m \omega}}=\frac{1}{D^{\infty}_{\ell m \omega}},
\end{equation}
and
\begin{equation}
    \frac{\xi^{\rm BH}_{\ell m \omega}}{Z^{\rm BH}_{\ell m \omega}}=\frac{C^{\rm in}_{\ell m \omega}}{D^{\rm in}_{\ell m \omega}}= \frac{A^{\rm hole}_{\ell m \omega}}{B^{\rm hole}_{\ell m \omega}}= \frac{1}{B^{\rm hole}_{\ell m \omega}}\,.
\end{equation}
Here we use the conventions $C^{\infty}_{\ell m \omega}=A^{\rm hole}_{\ell m \omega} =1$. 
We can rewrite Eq.~\eqref{eqechoSN} in terms of the Teukolsky functions
\begin{equation}
\label{eq_zeffrelation}
	Z^{\infty \, {\rm ECO}}_{\ell m \omega} =  Z^{\infty\, {\rm BH}}_{\ell m \omega}+\frac{\mathcal{R}_{\ell m \omega}^{\rm ECO}\mathcal{R}_{\ell m \omega}^{\rm BH}}{1-\mathcal{R}_{\ell m \omega}^{\rm ECO}\mathcal{R}_{\ell m \omega}^{\rm BH}}Z_{\ell m \omega}^{\infty\,{\rm eff}},
\end{equation}
with 
\begin{equation}
\label{eq:Zeff}
    Z_{\ell m \omega}^{\infty\,{\rm eff}} \equiv \frac{D^{\infty}_{\ell m \omega}}{D^{\rm in}_{\ell m \omega}}Z^{\rm H\,BH}_{\ell m \omega} \,.
\end{equation}
For very compact ECOs, substantial phase shift exists in $\mathcal{R}^{\rm ECO}_{\ell m \omega}$, causing substantial time delays between neighboring echoes, which allows 
\begin{equation}
\label{eq_ZeffExpand}
	Z^{\infty \, {\rm ECO}}_{\ell m \omega} =  Z^{\infty\, {\rm BH}}_{\ell m \omega} 
+\sum_{n=1}^{+\infty}\left(\mathcal{R}_{\ell m \omega}^{\rm ECO}\mathcal{R}_{\ell m \omega}^{\rm BH}\right)^n Z^{\infty\, {\rm eff}}_{\ell m \omega}.   	\end{equation}
Here the $n^{\rm th}$ item is effectively the $n^{\rm th}$ echo in the full waveform. The physical understanding of the expansion above is that the $n^{\rm th}$ echo is reflected $n$ times by the ECO surface and $n-1$ times by the potential barrier (of SN or Teukolsky equations), propagates for an additional $2n$ times the distance between the potential barrier and the ECO surface (in terms of $r_*$), and finally transmits through the potential barrier.  

We can obtain GW strain $h$ from $Z$ via Eq.~\eqref{BHwaveform}. Eqs.~\eqref{eq_zeffrelation} and \eqref{eq_ZeffExpand} reduce to those in Maggio \etal{}~\cite{Maggio} and Wang \etal{}~\cite{wang2020echoes} (in their ``inside'' prescription), if we make the substitution of
\begin{equation}
\label{eq:inside}
    \mbox{``inside'' prescription:} \quad Z_{\ell m \omega}^{\infty\,{\rm eff}} \leftarrow Z_{\ell m \omega}^{\infty\,{\rm BH\,{\rm RD}}}.
\end{equation}
Here $Z_{\ell m \omega}^{\infty\,{\rm BH\,{\rm RD}}}$ is the ringdown part of the waveform at infinity in the BH case. 

The quantity $Z_{\ell m \omega}^{\infty\,{\rm eff}}$, as defined in Eq.~\eqref{eq:Zeff}, has the following physical meaning.
If we replace the spacetime of a particle plunging into Kerr with a linear perturbation of Kerr, and replicate the waveform at infinity by sending in a wave from the past horizon, in such a way that the waveform toward the future horizon agrees with $Z^{\rm H\,BH}$, then the waveform at infinity is given by $Z_{\ell m \omega}^{\infty\,{\rm eff}}$.


The situation is illustrated in Fig.~\ref{stdiagram}.  As we can see from the figure, in the region ``above'' the trajectory of the particle, we have a vacuum spacetime, and $[Z^{\rm H\,BH}]_{v>v_0} =[(D^{\rm in}/D^{\infty}) Z^{\infty\,\rm BH}]_{v>v_0}$.   It is then plausible that $Z_{\ell m \omega}^{\infty\,{\rm eff}}$ can be approximated by the ringdown part of $Z_{\ell m \omega}^{\infty\,{\rm BH}}$, or $Z_{\ell m \omega}^{\infty\,{\rm BH\,RD}}$, as  Maggio \etal{}~\cite{Maggio} and Wang \etal{}~\cite{wang2020echoes} have done in the ``from inside'' prescription [see Eq.~\eqref{eq:inside}].

However, our results turn out to differ rather significantly from these prescriptions. More details are given in Sec.~\ref{featuresechoes}. Here we point out two features, namely $D^{\infty}_{\ell m \omega}/D^{\rm in}_{\ell m \omega} \sim 0$ when $\omega\sim m\Omega_+$, and $D^{\infty}_{\ell m \omega}/D^{\rm in}_{\ell m \omega}$ diverges quickly when $\omega \rightarrow +\infty$. This means that $Z_{\ell m \omega}^{\infty\,{\rm eff}}$ vanishes when $\omega\sim m\Omega_+$ (i.e., for radiations that are co-rotating with the horizon), and tends to infinity as $\omega\rightarrow +\infty$~\footnote{This does not lead to a diverging echo when $\omega\rightarrow +\infty$ since $\mathcal{R}^{\rm BH}_{\ell m \omega}$ converges to zero quickly as $\omega \rightarrow +\infty$.}. This is clearly different from $Z_{\ell m \omega}^{\infty \rm BH}$.

Therefore, we write
\begin{align}
\label{eq_zeffrelation2}
	Z^{\infty \, {\rm ECO}}_{\ell m \omega}& =  Z^{\infty\, {\rm BH}}_{\ell m \omega}+\frac{\mathcal{R}_{\ell m \omega}^{\rm ECO}\mathcal{T}_{\ell m \omega}^{\rm BH}}{1-\mathcal{R}_{\ell m \omega}^{\rm ECO}\mathcal{R}_{\ell m \omega}^{\rm BH}} \tilde Z^{\rm H\,BH}, \nonumber\\
\end{align}
where we define
\begin{equation}
\label{ztilde}
    \tilde Z^{\rm H\,BH} \equiv {\frac{D^{\infty}_{\ell m \omega}}{B^{\rm hole}_{\ell m \omega}}Z^{\rm H\,BH}_{\ell m \omega}},
\end{equation}
which remains finite as $\omega \rightarrow +\infty$.  By absorbing the ${D^{\infty}_{\ell m \omega}}/{B^{\rm hole}_{\ell m \omega}}$ factor, we have 
\begin{align}
\label{ztildeint}
    \tilde Z^{\rm H\,BH} &= \frac{1}{2i\omega B_{\rm \ell m\omega}^{\rm in}}\int_{r_+}^{+\infty} dr' \frac{R^{\infty}(r')\mathcal{T}_{\ell m \omega}(r')}{\Delta^2(r')} \nonumber\\
    &= \frac{1}{2i\omega B^{in}_{\ell m \omega}} \int_{-\infty}^{+\infty} dt  e^{i[\omega t - m\phi(t)]} \nonumber\\
 &\qquad\qquad\qquad \Bigg\{   R^\infty_{\ell m \omega}(r(t)) [A_{nn0} + A_{n\bar m0}+A_{\bar m \bar m 0}] \nonumber \\
&\qquad\qquad\qquad - \frac{d R^\infty_{\ell m \omega}}{dr}\bigg|_{r(t)} [A_{n\bar m 1} + A_{\bar m \bar m 1} ] \nonumber\\
&\qquad\qquad\qquad +\frac{d^2 R^\infty_{\ell m \omega}}{dr^2}\bigg|_{r(t)} A_{\bar m\bar m 2} \Bigg\}.
\end{align}
This is related to $\tilde Z^{\infty \,\rm BH}$ by replacing the Green function $R^H$ with $R^\infty$ [cf.~Eqs.~\eqref{eq_ZH}--\eqref{eq_Zinf}]. 

\begin{figure*}
    \centering
    \subfloat[$a/M = 0$ (non-spinning)]{\includegraphics[width=1.0\columnwidth]{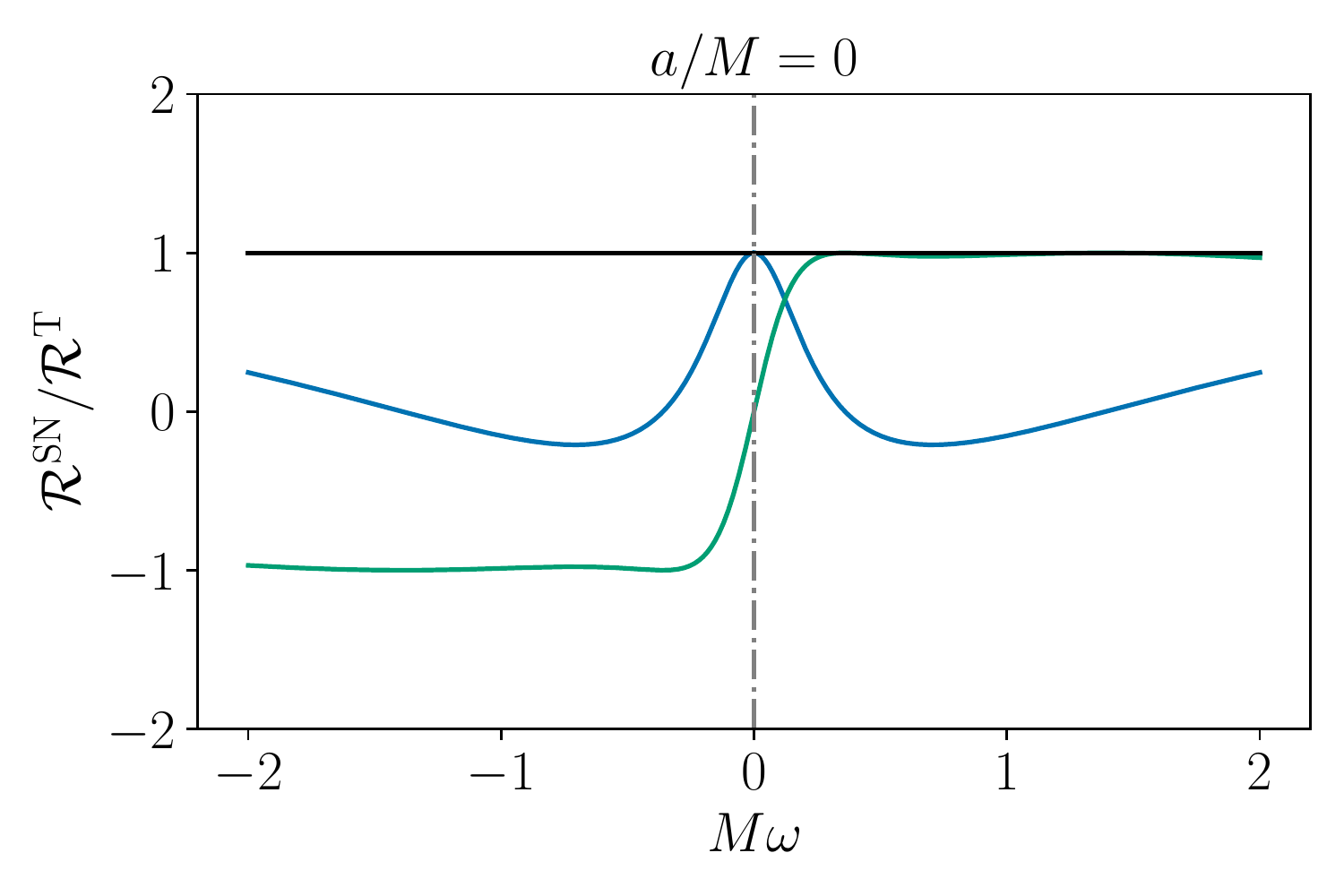}}
    \subfloat[$a/M = 0.7$ (spinning)]{\includegraphics[width=1.0\columnwidth]{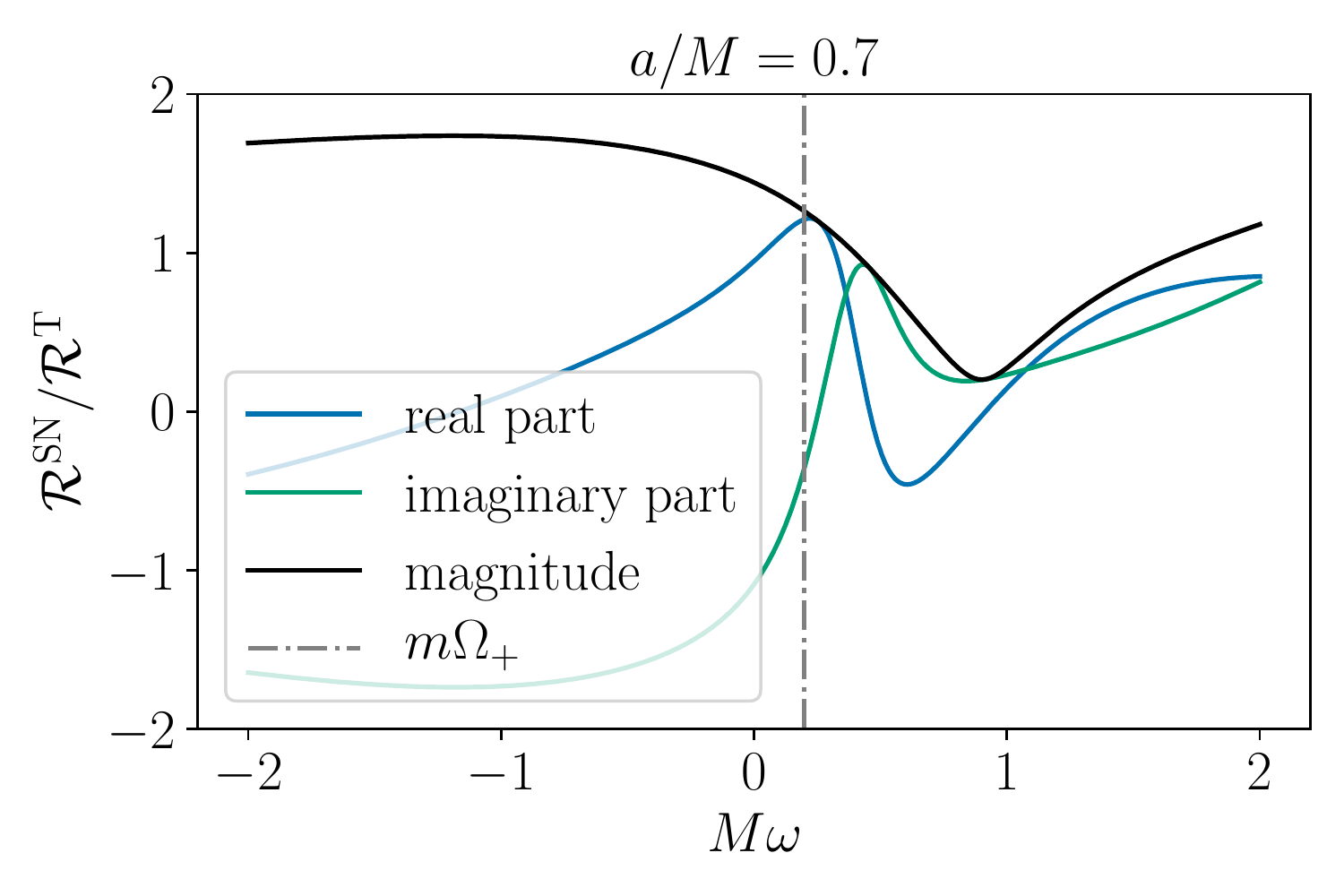}}
    \caption{The real (blue) and imaginary (green) parts and absolute values (black) of the reflectivity conversion factor [for $(\ell,m)=(2,2)$] as a function of $\omega$ for (a) $a/M=0$ and (b) $a/M=0.7$.
    The vertical line at non-zero $\omega$ in the right panel indicates $\omega = m \Omega_+$. For a non-spining ECO, the SN and Teukolsky reflectivities are related to each other with a pure phase shift; while for a spinning ECO, the factors have substantial frequency dependence, but remain at the order of unity at frequencies interested for compact binary mergers.}
    \label{fig:conversion}
\end{figure*}

\subsection{Boundary conditions directly imposed on Teukolsky functions}
\label{subsec:teu}

In this section, we discuss some subtleties in the SN formalism, and directly impose boundary conditions on Teukolsky functions.  The SN formalism remains as a computational tool in obtaining homogeneous solutions to the Teukolsky equation.

\subsubsection{Subtleties in the SN formalism}
\label{sub}
As argued by Nakano \etal{}, from the Wronskian of the radial equations, the energy reflectivity, in terms of the SN reflectivity $\mathcal{R}^{\rm ECO}$, is given by~\cite{tanaka}
{
\begin{equation}
\frac{\dot{E}_{\rm out}}{\dot{E}_{\rm in}} = \frac{|C_{\ell m \omega}d_{\ell m \omega}|^2  }{16 (2Mr_+)^5 (p^2+4\epsilon^2) (p^2+16\epsilon^2)  |  \eta^2(r_+)|}|{\mathcal{R}}^{\rm ECO}_{\ell m \omega} |^2,
\end{equation}
}
with
\begin{equation}
    \epsilon = \sqrt{M^2-a^2}/(4Mr_+) 
\end{equation}
equal to half the surface gravity. More specifically, this is the energy flux that emerges from the past horizon divided by the energy flux that goes into the future horizon~\cite{chen2020tidal}.  Note that the modulus of this conversion factor is not unity for $a/M\neq 0$ (see Fig.~\ref{fig:conversion}).

As we use the SN formalism to derive echoes, we need to note two important subtleties: (i) as we convert between $X$ and $R$, it is important to keep both the first two leading orders in $r-r_+$ near the horizon, and keep those in $1/r$ near infinity; (ii) the particle plunges toward the horizon, and therefore the source term for Teukolsky and SN equations does not vanish as \mbox{$r_*\rightarrow -\infty$}.  Such subtleties affect the evaluation of in-going energies down the horizon, as well as the boundary conditions near the ECO surface.  

A particular example of such subtleties is as follows. As one tries to evaluate the SN source terms, there are two degrees of freedom in the form of the integration constants. They lead to different leading-order field amplitudes at infinity (in terms of $1/r$) and at the horizon (in terms of $r-r_-$). Only after accounting for the correction terms up to the second order, one can obtain the correct Teukolsky amplitudes.  This presents an ambiguity for how to impose boundary conditions for $X$.  

\subsubsection{Physical boundary conditions from $\psi_0$ and $\psi_4$}

In a companion paper~\cite{chen2020tidal}, we have determined the relation between $Z^{\rm out}_{\ell m \omega}$ and $Z^{\rm in}_{\ell m \omega}$ by considering tidal tensor fields of fiducial observers near the horizon.  To summarize the results, we connect $\psi_0$ and $\psi_4$ to the tidal tensors of ficucial observers near the horizon:
\begin{equation}
    \mathcal{E} \sim -\frac{\Delta}{4\Sigma}\psi_0-\frac{\Sigma}{\Delta}\psi_4^*\,,\quad \Sigma = r^2+a^2\cos^2\theta\,.
\end{equation}
Note that it is the in-going piece of $\psi_0$ ($\sim \Delta^{-2} e^{-ipr_*}$) and the out-going piece of $\psi_4$ ($\sim  e^{+ipr_*}$) that dominate this expression, with both contributing to $\mathcal{E}$ at the order of $1/\Delta$ since the effect of GWs is heavily blue-shifted for near-horizon observers. Since the in-going piece of $\psi_0$ is externally applied to the ECO, while the out-going piece is generated by the ECO, the ratio of these two terms can then be viewed as a local {\it tidal Love number} of the ECO.

We can find the the in-going $\psi_0$ components via the Teukolsky-Starobinsky identity 
\begin{equation}
    Y^{\rm in}_{\ell m \omega}= {\sigma}_{\ell m \omega} Z^{\rm in}_{\ell m \omega},
\end{equation}
with
\begin{equation}
    \sigma_{\ell m \omega}= \frac{64(2Mr_+)^4 ip (p^2+4\epsilon^2)(-ip+4\epsilon)}{C_{\ell m \omega}}.
\end{equation}
Here $C_{\ell m \omega}$ is the Starobinsky constant, given by 
\begin{align}
    |C_{\ell m \omega}|^2 &= (Q^2+4 a\omega m - 4 a^2\omega^2)[(Q-2)^2 +36 a\omega m - 36 a^2\omega^2]\nonumber\\
    &+(2Q-1)(96a^2\omega^2-48 a\omega m)\nonumber\\
    &+144\omega^2(M^2-a^2),
    \nonumber\\
    \mathrm{Im} \; C_{\ell m \omega} &=12 M \omega, \nonumber\\
    \mathrm{Re} \; C_{\ell m \omega}& = +\sqrt{|C_{\ell m \omega}|^2-(\mathrm{Im}C_{\ell m \omega})^2},
\end{align}
with $Q= E_{\ell m} +a^2\omega^2-2a\omega m$, where $E_{\ell m}$ is the spheroidal eigenvalue [see discussions below Eq.~\eqref{eq:psi4}]. This has previously been applied to computing energy and angular momentum carried by GWs into the horizon.  However, we need to be careful here because the Starobinsky-Teukolsky identity may not work in the presence of source terms --- while here we do have a particle plunging into the horizon. This remains an unaddressed issue in this paper. 


By considering tidal distortions of zero-angular-momentum fiducial observers very close to the horizon, we obtain
\begin{equation}
    Z^{\rm out}_{\ell m \omega}=\frac{(-1)^{m+1}}{4} \mathcal{R}^{\rm ECO\,T}_{\ell m \omega}Y^{\rm in\,*}_{\ell-m-\omega}.
\end{equation}
Here the Teukolsky reflectivity $\mathcal{R}^{\rm ECO\,T}_{\ell m \omega}$ can be related to the response of the ECO to external driving. Its modulus, $|\mathcal{R}^{\rm ECO\,T}_{\ell m \omega}|$, corresponds to the energy reflectivity of the ECO surface.  Here we have ignored the mixing between different $\ell$-modes, which is a general feature due to the distortion of spacetime geometry by the spin of the ECO.  As it turns out, we also need to consider 
$(Z^{\rm out}_{\ell m \omega},Z^{\rm out\,*}_{\ell-m-\omega})$ and $(Z^{\rm in}_{\ell m \omega},Z^{\rm in\,*}_{\ell m-\omega})$, since generically an ECO couples between these modes.    
Nevertheless, in the most commonly considered situation of an equatorial, quasi-circular orbit, we have
 \begin{equation}
 \label{eq:zlm}
     Z_{\ell m \omega}=Z_{\ell-m-\omega}^*,
 \end{equation}
as described by Maggio \etal{}~\cite{Maggio}.  This indicates that only one reflectivity needs to be considered.   However, when the particle has an inclined orbit, relation~\eqref{eq:zlm} no longer holds, and thus the echoes have a more complex form. 

Assuming Eq.~\eqref{eq:zlm} to hold, we have
\begin{equation}
    Z_{\ell m \omega}^{\rm out} = \frac{(-1)^{m+1}}{4} \sigma_{\ell m \omega} \mathcal{R}_{\ell m \omega}^{\rm ECO\,T}    Z_{\ell m \omega}^{\rm in}.
\end{equation}
Assuming a linear relation between the SN and Teukolsky functions, we can write
\begin{equation}
    \mathcal{R}^{\rm ECO}_{\ell m \omega} =\frac{X^{\rm out}_{\ell m \omega}}{X^{\rm in}_{\ell m \omega}}
= \frac{C^{\rm out}_{\ell m \omega}}{D^{\rm out}_{\ell m \omega}} \frac{D^{\rm in}_{\ell m \omega}}{C^{\rm in}_{\ell m \omega}} 
\frac{Z^{\rm out}_{\ell m \omega}}{Z^{\rm in}_{\ell m \omega}}.
\end{equation}
This leads to 
\begin{align}
    \label{eq_RB2R}
  \mathcal{R}^{\rm ECO}_{\ell m \omega}  &=\frac{(-1)^{m+1}}{4} \sigma_{\ell m \omega} \frac{C^{\rm out}_{\ell m \omega}}{D^{\rm out}_{\ell m \omega}} \frac{D^{\rm in}_{\ell m \omega}}{C^{\rm in}_{\ell m \omega}}\mathcal{R}_{\ell m \omega}^{\rm ECO\,T} \nonumber \\
   &=
(-1)^m\frac{4(2Mr_+)^{5/2}\eta(r_+)(p-2i\epsilon)(p+4i\epsilon)}{C_{\ell m \omega}d_{\ell m \omega}}\mathcal{R}_{\ell m \omega}^{\rm ECO\,T}. 
\end{align}
In Fig.~\ref{fig:conversion}, we plot the real, imaginary parts and modulus of $\mathcal{R}^{\rm ECO}/\mathcal{R}^{\rm ECO\,T}$. 

At this stage, aside from subtleties of the Teukolsky-Starobinsky relation, Eq.~\eqref{eq_RB2R} provides the reflectivity of the ECO in the SN frame work, $\mathcal{R}^{\rm ECO}_{\ell m \omega}$, in terms of the physically defined ECO reflectivity, $\mathcal{R}^{\rm ECO\,T}_{\ell m \omega}$.  We can insert Eq.~\eqref{eq_RB2R} into Eqs.~\eqref{eq_zeffrelation} and \eqref{eq_zeffrelation2} to obtain echoes that arise from these physical boundary conditions.

\subsubsection{Echoes in terms of Teukolsky reflectivity}

We can now write echo waveforms in terms of the amplitudes of Teukolsky functions. In terms of reflectivity, we have
\begin{equation}
    \mathcal{R}^{\rm ECO}_{\ell m \omega}    \mathcal{R}^{\rm BH}_{\ell m \omega}=    \mathcal{R}^{\rm ECO\,T}_{\ell m \omega}    \mathcal{R}^{\rm BH\,T}_{\ell m \omega},  
\end{equation}
where
\begin{equation}
    \mathcal{R}^{\rm BH\,T}_{\ell m \omega}=\frac{(-1)^{m+1}}{4}\sigma_{\ell m \omega}
    \frac{D^{\rm in}_{\ell m \omega}}{D^{\rm out}_{\ell m \omega}} .
\end{equation}
We can also write 
\begin{equation}
	Z^{\infty \, {\rm ECO}}_{\ell m \omega} =  Z^{\infty\, {\rm BH}}_{\ell m \omega} 
+\sum_{n=1}^{+\infty}\left(\mathcal{R}_{\ell m \omega}^{\rm ECO\, T}\mathcal{R}_{\ell m \omega}^{\rm BH\,T}\right)^n Z^{\infty\, {\rm eff}}_{\ell m \omega}.   	\end{equation}
Here  $   | \mathcal{R}^{\rm BH\,T}_{\ell m \omega}|^2$ directly gives the energy reflectivity of the BH potential barrier, including superradiance at frequencies $\omega<m\Omega_+$. 
The condition $ | \mathcal{R}^{\rm BH\,T}_{\ell m \omega}  \mathcal{R}_{\ell m \omega}^{\rm ECO\,T} | < 1$ needs to be satisfied such that the instability does not happen in the ECO.

We can also express the echo waveform in terms of the in-going $\psi_0$ component toward the horizon in the BH case, $Y_{\ell m \omega}^{\rm H\,BH}$, as
\begin{equation}
\label{zechopsi0}
       Z^{\infty \, {\rm ECO}}_{\ell m \omega} =
       Z^{\infty\, {\rm BH}}_{\ell m \omega}+\frac{\mathcal{R}_{\ell m \omega}^{\rm ECO\, T}  \mathcal{J}_{\ell m \omega}}{1-\mathcal{R}_{\ell m \omega}^{\rm ECO\, T}\mathcal{R}_{\ell m \omega}^{\rm BH\, T}} Y^{\rm H\,BH}_{\ell m \omega},
\end{equation}
where we define
\begin{equation}
    \mathcal{J}_{\ell m \omega}=\frac{(-1)^{m+1}}{4}
    \frac{D_{\ell m \omega}^{\infty}}{D_{\ell m \omega}^{\rm out}} .
\end{equation}
Here, the in-going $\psi_0$ takes the form of curvature perturbations of fiducial observers near the horizon. Response of structures in the observers' frame gives rise to a local reflectivity $\mathcal{R}_{\ell m \omega}^{\rm ECO}$, leading to the out-going $\psi_4$ with a conversion factor of $(-1)^{m+1} \mathcal{R}_{\ell m \omega}^{\rm ECO}/4$.  This $\psi_4$ is then transmitted to infinity with a factor of $D^{\infty}_{\ell m \omega}/D^{\rm out}_{\ell m \omega}$ applied. Since Eq.~\eqref{zechopsi0} does not use the Teukolsky-Starobinsky transformation (which is only valid for homogeneous solutions), it provides a more straightforward way to compute echoes.

\subsection{Models for ECO reflectivity}
\label{p_reflectivity}

We consider two types of reflectivity, namely a parameterized Lorentzian reflectivity and a Boltzman-type reflectivity~\cite{wang2020echoes}.  
 
\subsubsection{Lorentzian reflectivity} 
 
In the Lorentzian case, we assume the reflection takes place at a fixed position of $r=b$, or $r_*=b_*$ [Eq.~\eqref{rs}]. At position $r=b$, the proper distance $\delta$ along the radial direction toward the horizon is given by 
\begin{equation}
    \delta = \int_{r_+}^b \sqrt{g_{rr}} dr\approx \sqrt{\frac{r_+^2 +a^2\cos\theta^2}{Mr_+\kappa}} \sqrt{b-r_+} ,
\end{equation}
for $\delta \ll M$. For BHs with $a/M$ not too close to unity, this leads to
\begin{align}
    b_* &\approx r_+ + \frac{1}{2\kappa}\log\frac{b-r_+}{2M}-\frac{r_-}{2\kappa r_+}\log\frac{r_+-r_-}{2M} \nonumber\\
    & \approx \frac{1}{\kappa}\log\frac{\delta }{\sqrt{r_+^2+a^2\cos^2\theta}} .
\end{align}
Another way of measuring the closeness to the horizon is via the red-shift of zero-angular-momentum observers at a constant $r=b$, with
\begin{equation}
    \alpha = \sqrt{\frac{4Mr_+ \kappa}{r_+^2 +a^2\cos^2\theta}}\sqrt{b-r_+}  .
\end{equation}
For $a\neq 0$, both $\delta$ and $\alpha$ depend on $\theta$.  This can be understood as the deformation of spherical symmetry due to the spin. In Ref.~\cite{chen2020tidal}, we choose to set the reflection surface at a constant red shift $\alpha$ (i.e., the reflectivity has the same phase for all values of $\theta$ when $\alpha$ is a constant), which leads to mixing between the modes with different $\ell$.  In this paper, for simplicity, we assume that the Lorentzian reflectivity is a constant at $r=b$ (or $r_*=b_*$) and can be written as
\begin{equation}
     \mathcal{R}^{L}_{\ell m \omega} =\varepsilon \left(\frac{ i \Gamma}{p +i\Gamma}\right) e^{-2i  b_* p}.
 \end{equation}
Here the quantity $\varepsilon \in (0, 1)$ parametrizes the amplitude reflectivity of the ECO surface.  Note that $\mathcal{R}$ depends on $\omega$ only via $p=\omega-m\Omega_+$, the frequency of oscillations measured by observers co-rotating with the horizon of the Kerr spacetime (even though it is covered by the ECO surface at $r=b$).  The quantity $\Gamma$ characterizes a relaxation rate of the ECO surface, 
which corresponds to an impulse response function $\sim e^{-\Gamma t}$ in the time domain and imposes a low-pass filtering of waves upon reflection in the frequency domain.  For distant observers, GWs with frequencies $|\omega-m\Omega_+| \lesssim \Gamma$ have the highest reflectivity.  Note in particular, that peak reflectivity takes place at $\omega\sim 2\Omega_+$ for $m=2$ and $\omega\sim -2\Omega_+$ for $m=-2$.  As argued by Refs.~\cite{maggio2017exotic,wang2020echoes}, as long as $\varepsilon$ is not too close to unity, the ECO is stable under the Lorentzian reflecitivity. 


The phase factor $e^{-2i b_*p}$ in $\mathcal{R}_{\ell m \omega}^{\rm L}$  corresponds to a time delay of $-2b_*$.  If we consider that the ringdown wave is generated roughly at $r_*\approx 0$, the term $-2b_*$ provides an  estimate of the time delay between the main wave and the first echo, as well as time delays between neighbouring echoes.

\subsubsection{Boltzmann reflectivity}
 
 
Considering wave reflection by a thermal atmosphere, Wang \etal{}~\cite{wang2020echoes} and Oshita \etal{}~\cite{Oshita_2020} proposed the following Boltzmann reflectivity, given by
\begin{equation}
\label{eq:RB}
    \mathcal{R}_{\ell m \omega}^{B} =\exp\left(-\frac{|p|}{2 T_H}\right)
    \exp\left[
    -i\frac{p}{\pi T_H}\log(\gamma|p|)\right],
\end{equation}
with Hawking temperature
\begin{equation}
    T_H =  \frac{\kappa}{2\pi} = \frac{r_+-r_-}{4\pi(r_+^2+a^2)} =\frac{\sqrt{1-a^2}}{4\pi M ( 1+\sqrt{1-a^2})}.
\end{equation}
We may replace $T_{\rm H}$ with a free parameter $T_{\rm QH}$ to generalize the Boltzmann reflectivity,
\begin{equation}
    \mathcal{R}_{\ell m \omega}^{B} =\exp\left(-\frac{|p|}{2T_{\rm QH}}\right)
    \exp\left[
    -i\frac{p}{\pi T_{\rm QH}}\log(\gamma|p|)\right].
\end{equation}
We fix the temperature to the Hawking temperature, i.e., $T_{QH}=T_H$, until in Sec.~\ref{p_det}, where we relax this condition to explore the detectability of echoes that arise from a broader class of reflectivity models.
Similar to the Lorentzian reflectivity, $\mathcal{R}^{\rm B}_{\ell m \omega}$ depends on $\omega$ via $p=\omega-m\Omega_+$, leading to the peak reflectivity (equal to unity) for modes with zero frequency viewed by observers co-rotating with the horizon, or $\omega \sim m\Omega_+$, and vanishing reflectivity for $|\omega-m\Omega_+| \gg  T_H$.


Boltzmann reflection does not take place at a fixed point, but for waves oscillating near the quasi-normal mode (QNM) frequency, $\omega \approx \Re[ \omega_{\rm QNM} ]$. The effective distance traveled by waves at this frequency, in terms of $r_*$, due to the phase factor in Eq.~\eqref{eq:RB}, is given by
 \begin{equation}
 \label{eq:rstarB}
 2 r_*^{\rm eff\,B}= \log(\gamma|\Re[ \omega_{\rm QNM} ]-m\Omega_+|)/\pi T_H.
\end{equation}
Similar to the Lorentzian case, here $-2r_*^{\rm eff\,B}$ corresponds to the time lag between echoes.  

As shown in Ref.~\cite{wang2020echoes}, the Boltzmann reflectivity leads to a stable ECO. Basically, the BH potential barrier has a reflectivity higher than unity for $\omega <  m\Omega_+$, and the ECO simply needs to have a reflectivity that decreases fast enough as $p$ increases from zero.


\section{Horizon waveforms and echoes}
\label{Section:Results}

In this section, we discuss numerical features of GWs that go down the horizon and echoes generated using several models of the ECO reflectivity. 

\subsection{Prescriptions for computing echoes}

 \begin{figure*}[ht]
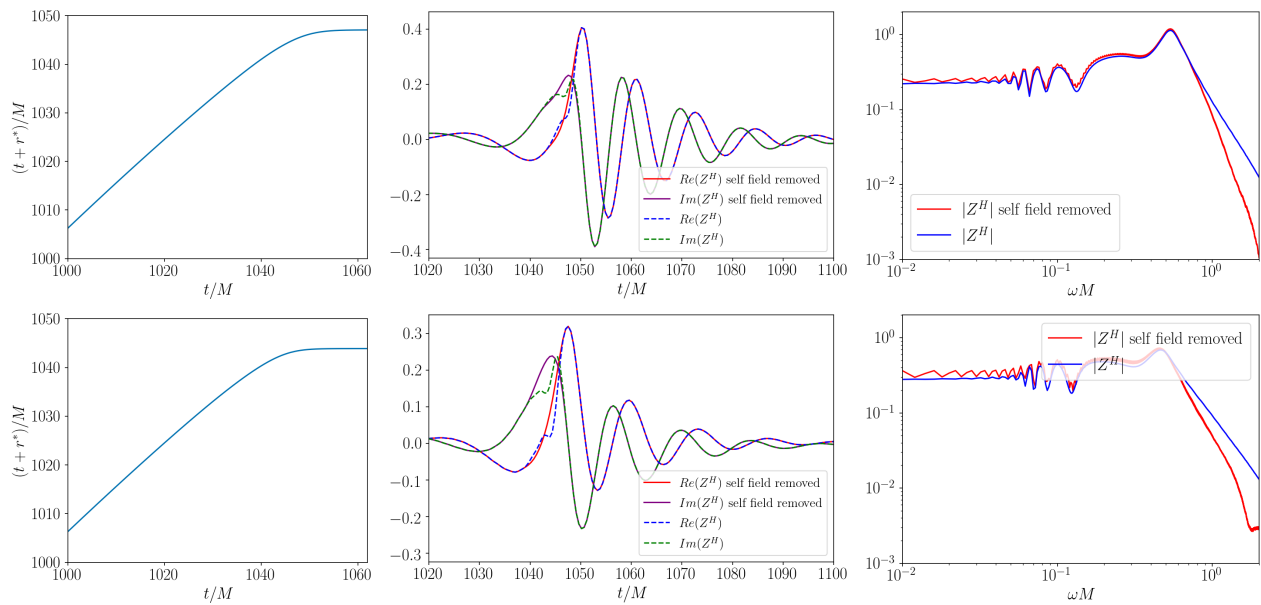

	\centering
	\includegraphics[height=4cm]{advtime.pdf}	
	\includegraphics[height=4cm]{ZH_q1_pr.pdf}
	\includegraphics[height=4cm]{ZHfreq_q1_pr.pdf}
	\includegraphics[height=4cm]{advtime_q4.pdf}	
	\includegraphics[height=4cm]{ZH_q4_pr.pdf}		
	\includegraphics[height=4cm]{ZHfreq_q4_pr.pdf}	
	\caption{Advanced-time trajectory and horizon waveform $Z^{\rm H\,BH}$, with and without self field, corresponding to a trajectory fitted to the NR waveform shown in Fig.~\ref{fig_surrogate} (Top row: $q=1$, bottom row: $q=4$). The left panels show the advanced time verses coordinate time. In the late stage of the particle plunge, advanced time converges to a constant. All the late-time pieces in the integration for $Z^{\rm H\,BH}$ accumulate near this epoch, resulting in a dip near $t/M=1047$, as shown in the middle panels. The right panels show the frequency spectrum $|Z^{\rm H\,BH}(\omega)|$, which peaks at the QNM frequency of the final BH.}
	\label{fig_peak}
\end{figure*}

We first review existing prescriptions and show how our prescription is connected to and differs from these previous studies. In existing literature, in particular Wang \etal{}~\cite{wang2020echoes} and Maggio \etal{}~\cite{Maggio}, echoes are obtained from the out-going GWs at infinity and the SN transmissivities and reflectivities.  In the ``inside prescription'' of \cite{wang2020echoes} and \cite{Maggio}, one has
\begin{equation}
\label{hinside}
    h^{\infty \rm ECO} =h^{\infty  \rm BH}+ \frac{
    \mathcal{R}^{\rm ECO}
    \mathcal{R}^{\rm BH}
    }{1-\mathcal{R}^{\rm ECO}\mathcal{R}^{\rm BH}}[h^{\infty  \rm BH}]_{\rm RD},
\end{equation}
where ``RD'' represents the ringdown part of the binary black hole coalescence waveform $h^{\infty  \rm BH}$. 

In this work, following Ref.~\cite{chen2020tidal}, we propose that it is in fact $\mathcal{R}^{\rm ECO\,T}$ that follows the reflectivity models described in Sec.~\ref{p_reflectivity}, since $\mathcal{R}^{\rm ECO\,T}$ is directly connected to the tidal fields measured by fiducial observers near the horizon.  One must take a reflectivity from Sec.~\ref{p_reflectivity}, and {\it convert} it into a SN reflectivity $\mathcal{R}^{\rm ECO}$ using  Eq.~\eqref{eq_RB2R}.

The second difference between this study and previous work is that we obtain the in-going $\psi_4$ wave toward the horizon directly from Teukolsky formulation.  This is equivalent to obtaining it directly from the SN formalism for $\psi_4$, e.g., done in Ref.~\cite{sago2020gravitational} for radially in-falling particles. We insert the in-going Teukolsky amplitude $Z^{\rm H\,BH}$, obtained from Eq.~\eqref{source_term_inf}, into Eq.~\eqref{eq_zeffrelation} to compute echoes. This is equivalent to using Eqs.~\eqref{eq_zeffrelation2}--\eqref{ztildeint}. 

We would like to mention that instead of computing in-going wave via $\psi_4$ (or the SN formalism for $\psi_4$), one can also compute the in-going $\psi_0$ directly, and then use Eq.~\eqref{zechopsi0} to compute echoes.  As discussed, since the reflection on the ECO surface is really a relation between the in-going $\psi_0$ and the out-going $\psi_4$, this approach is more direct and not subject to uncertainties of whether the in-going $\psi_4$ can be converted into the in-going $\psi_0$ reliably using the Teukolsky-Starobinsky relation when the point particle plunges into the horizon.  We leave this for future work. 

\begin{figure}[t]
    \centering
    \includegraphics[height=5.0cm]{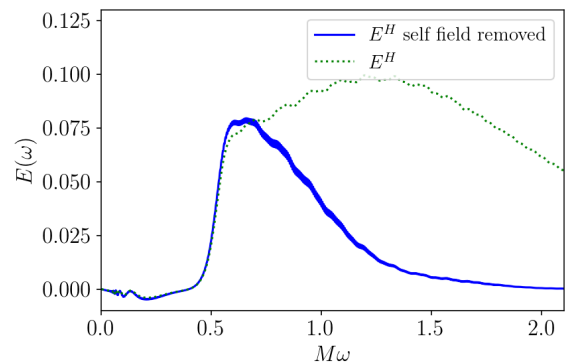}	
    \caption{The in-going energy spectrum of the merger waveform with $q=1$. The dashed curve shows the energy spectrum directly computed. The solid curve shows the energy spectrum of the interpolated waveform after removing the self-field part.}  
    \label{fig_energycontent}
\end{figure}

  \begin{figure*}[tbh!]
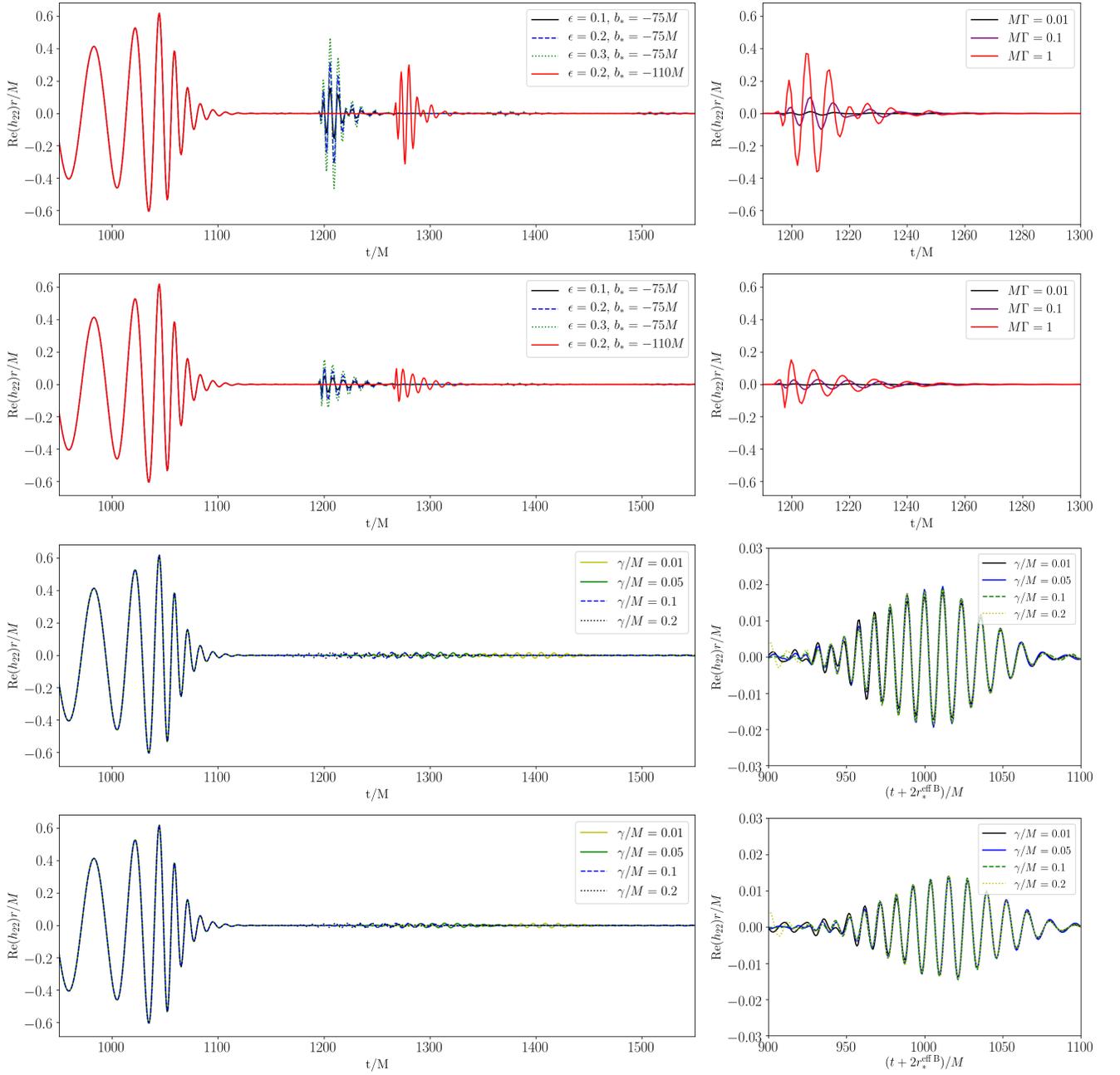

    \centering
    \includegraphics[height=4.3cm]{L_eb_q1.pdf}	
	\includegraphics[height=4.3cm]{L_G_q1.pdf}
	\includegraphics[height=4.3cm]{L_eb_q1_Tref.pdf}	
	\includegraphics[height=4.3cm]{L_G_q1_Tref.pdf}

    \includegraphics[height=4.3cm]{B_full_q1.pdf}	
	\includegraphics[height=4.3cm]{B_1echo_q1_phaseshifted.pdf}
     \includegraphics[height=4.3cm]{B_full_q1_Tref.pdf}	
	\includegraphics[height=4.3cm]{B_1echo_q1_Tref_phaseshifted.pdf}    
    \caption{
    Echoes for an equal-mass binary merger ($q=1$) with Lorentzian (top two rows) and Boltzmann (bottom two rows) reflectivities imposed on SN or Teukolsky radial functions. Top row: Impose Lorentzian reflectivity on SN functions. The left panel shows how $\epsilon$ and $b_*/M$ in Lorentzian reflectivity change the magnitude and separation of echoes ($M\Gamma = 0.5$). The right panel shows how $\Gamma$ impacts the shape of the first echo (with $\varepsilon=0.2$, $b_*/M=-75$). Second row: (similar to the top row) Impose Lorentzian reflectivity on Teukolsky radial functions.
    Third row: Impose Boltzmann reflectivity on SN functions. The left panel shows how $\gamma$ in the Boltzmann reflectivity changes the echoes. The right panel shows how $\gamma$ impacts the shape of the first echo [time is shifted by $2 r_*^{\rm eff \,B}$, as defined in Eq.~\eqref{eq:rstarB}, to align with the first echo]. Bottom row: (similar to the third row) Impose Boltzmann reflectivity on Teukolsky radial functions.  
    Imposing reflectivities on Teukolsky functions (the more physical approach) tends to generate echoes with lower amplitudes than imposing the same reflectivities on the SN functions; this can be understood from the frequency content of $Z^{\rm H\,BH}$ and the conversion factors shown in Fig.~\ref{fig:conversion}.}
    \label{fig_LBonSN}
\end{figure*} 

\subsection{Features of horizon waveforms}

In Fig.~\ref{fig_peak}, we plot the horizon waveform $Z^{\rm H\,BH}_{22}$ for $q=1$ and $q=4$. 
There is a feature in the time-domain horizon waveform, right at the advanced time when the particle plunges into the horizon.  This differs from the discussion for scalar fields by Mark \etal{}~\cite{echoSchw}, and has to do with curvature perturbations due to a point particle.  
As shown in the figure, the feature occurs at a rather early time in the horizon waveform. Therefore most of the echo does arise from GWs that hit the ECO after the point particle plunges into the horizon. 

To compute the energy of the GWs going down the horizon, we need to convert $\psi_4$ to $\psi_0$, using the Teukolsky-Starobinsy identity and then the Hartle formula of BH area increase, which leads to Eq.~(4.44) in Ref.~\cite{teukolsky_energy}:\begin{equation}
    \frac{dE^{\rm hole}}{d\omega} =  
 \sum_{\ell m} \frac{64\omega p (p^2+4\epsilon^2)(p^2+16\epsilon^2) (2Mr_+)^5 }{\pi | C_{\ell m \omega}|^2} \vert Z^{\rm H\,BH}_{\ell m \omega} \vert^2\,.
\end{equation} 
This leads to diverging energy going down the horizon, even for each individual $\ell$, as indicated in Fig.~\ref{fig_energycontent}. 
However, at least in the Schwarzschild case, this energy should not diverge for each $\ell$, as shown in Ref.~\cite{davis1972pulses}.   We suspect that this is due to the fact that the Teukolsky-Starobinsky identity does not apply to our case, where the particle plunges into the horizon, but leave the detailed study for future work.  In our current formulation, the divergence of the energy flux is a direct consequence of the ``self-field'' feature of $Z^{\rm H\,BH}$ near the location where the particle plunges.  Smoothing out this feature in the time domain (see middle panels of Fig.~\ref{fig_peak}) can fix the energy divergence (blue curve in Fig.~\ref{fig_energycontent}) without significantly affecting echo waveforms, as discussed later in Sec.~\ref{subsec:remove_selffield}.

\begin{figure*}[tbh!]
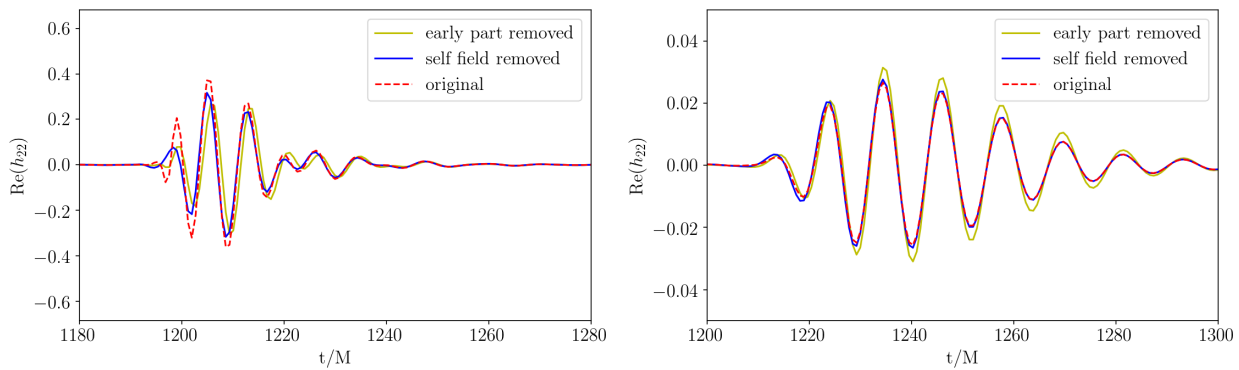

    \centering
	\includegraphics[height=5cm]{q1_1echo_ew.pdf}
	\includegraphics[height=5cm]{q1_1echo_Boltzman.pdf}
    \caption{Impact of the particle self-field on echoes using the Lorentzian reflectivity for SN functions with $\epsilon=0.2,\, M\Gamma=1, \, b_*=-75M$ (left) and the Boltzman reflectivity for SN functions with $\gamma/M=0.2$ (right).
    Both panels display the first echo directly computed using the particle plunging orbit (red dashed curve), after removing self-field region (blue solid curve), and after removing all early-time waves (yellow solid curve). } 
    \label{fig_peakremove}
\end{figure*}

\subsection{Features of echoes}
\label{featuresechoes}

In Fig.~\ref{fig_LBonSN}, we plot the GR and echo waveforms for binaries with $q=1$, using Lorentzian (the first and second rows) and Boltzman reflectivities (the third and fourth rows), either imposing them on the SN functions (the first and third rows), or on Teukolsky functions (the second and fourth rows) using the prescription described in Ref.~\cite{chen2020tidal}.  We zoom in and show the details of the first echo in each row on the right panels.

\subsubsection{Lorentzian reflectivity}

For Lorentzian reflectivity $\mathcal{R}^{L}$, $\varepsilon$ simply scales the magnitude of the $n$-th echo by $\varepsilon^n$, while $|2b_*|$ shifts the time-domain separation between echoes (effects of $\varepsilon$ and $b$ on the first echo are shown in the left panels of the first two rows in Fig.~\ref{fig_LBonSN}).  The bandwidth $\Gamma$ of the reflectivity acts as a low-pass filter in the reference frame of the ECO surface, therefore it filters out frequency components with $|\omega- m\Omega_+| \lesssim \Gamma$.  By comparing the first and second rows of Fig.~\ref{fig_LBonSN}, we find that the new boundary condition, imposed on curvature perturbations (the second row), gives rise to slightly weaker echoes than those with the condition imposed on the SN functions (the first row), but does not modify the qualitative features of the echoes.  This is consistent with the conversion factor in Eq.~\eqref{eq_RB2R} with an absolute value smaller than unity in the frequency band of the echoes, i.e., around $\omega\sim m\Omega_+$ and higher toward $\omega_{\rm QNM}$ (as shown in Fig.~\ref{fig:conversion}).

\subsubsection{Boltzmann reflectivity}

The Boltzmann reflectivity $\mathcal{R}^B$ only has one free parameter $\gamma$, which simply shifts the separation between echoes (as well as between the first echo and the GR wave) in the time domain by $2r_*^{\rm eff\,B}$, as given by Eq.~\eqref{eq:rstarB}.  Similar to the Lorentzian case, the new boundary condition on curvature perturbations leads to slightly weaker echoes, for the same reason discussed above. 

\subsubsection{Removal of ``self field''}
\label{subsec:remove_selffield}
In Fig.~\ref{fig_peakremove}, we investigate the impact of removing the feature of $Z^{\rm H\,BH}$ near the location where the particle plunges into the future horizon.  We compare the original first echo, the one resulted from a smoothed version of $Z^{\rm H\,BH}$, as well as the one obtained by completely removing the part before the plunge from $Z^{\rm H\,BH}$. The differences caused by the removal of these features are small. 

\subsubsection{Polarization of the echoes}

\begin{figure*}[tbh!]
    \centering
    \subfloat[]{\includegraphics[width=\columnwidth]{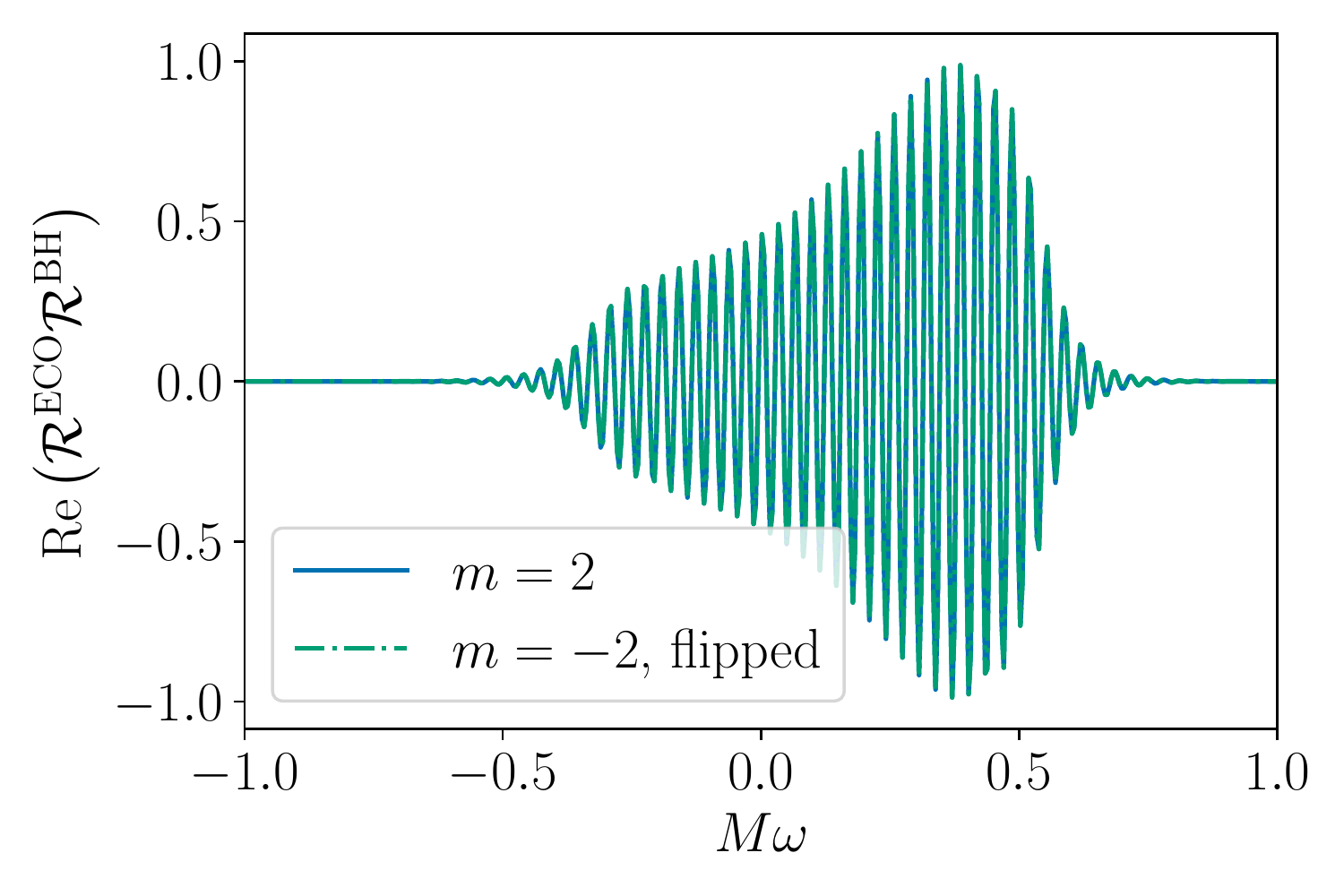}}
    \subfloat[]{\includegraphics[width=\columnwidth]{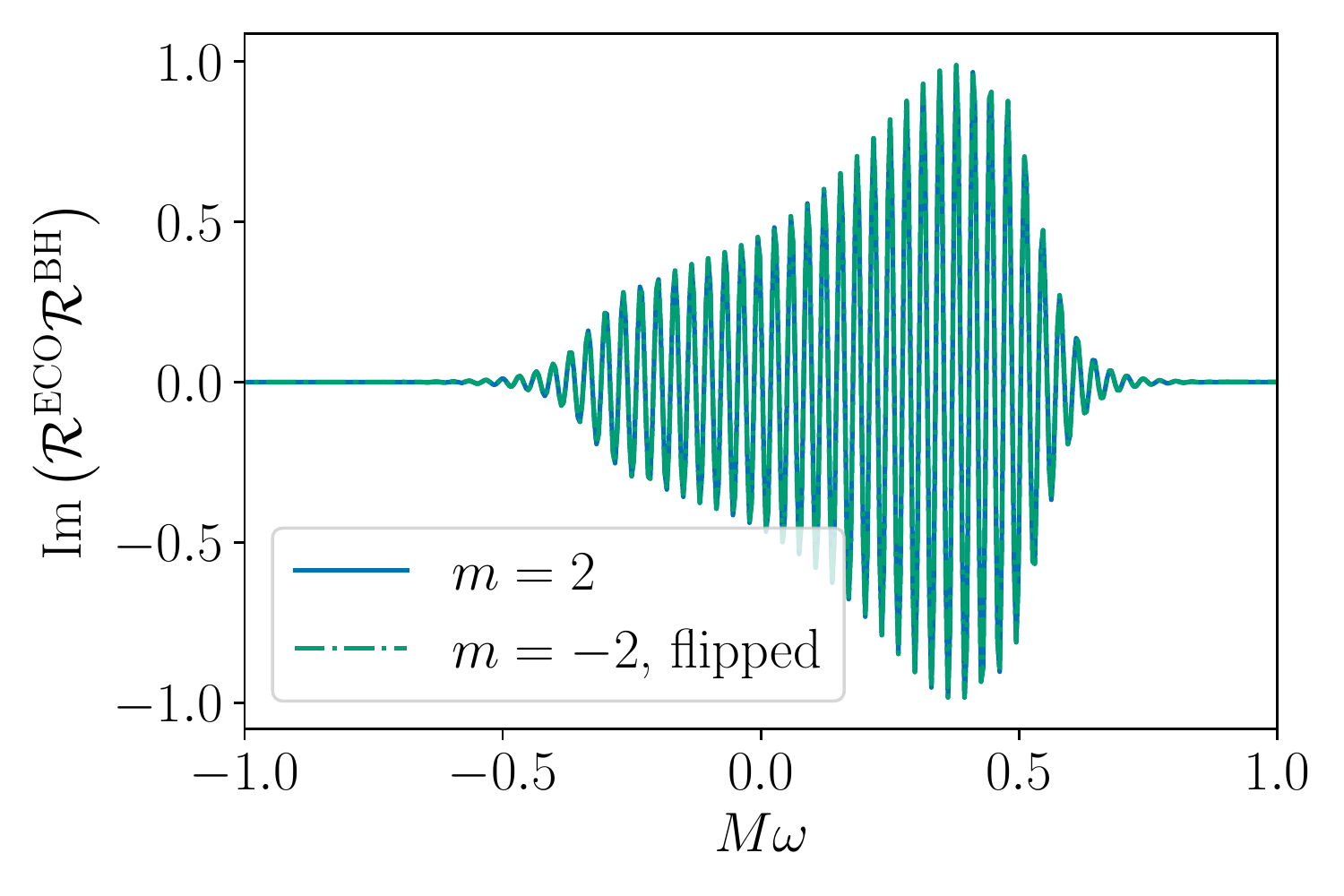}}
    \caption{The real part (left panel) and the imaginary part (right panel) of $\mathcal{R}_{\ell m\omega}^{\rm ECO}\mathcal{R}_{\ell m\omega}^{\rm BH}$ when $\ell = |m| = 2$ using the Lorentzian reflectivity with $\epsilon = 1$ and $M\Gamma=0.21$. For $m=-2$, the real part is flipped such that $\rm{Re} \left[\mathcal{R}^{\rm ECO}(-\omega)\mathcal{R}^{\rm BH}(-\omega)\right]$ is shown instead, and the imaginary part is also flipped such that $-\rm{Im} \left[\mathcal{R}^{\rm ECO}(-\omega)\mathcal{R}^{\rm BH}(-\omega)\right]$ is plotted. This shows explicitly that the equatorial symmetry is preserved such that $\mathcal{R}_{\ell m\omega}^{\rm ECO}\mathcal{R}_{\ell m\omega}^{\rm BH} = \left( \mathcal{R}_{\ell -m-\omega}^{\rm ECO}\mathcal{R}_{\ell -m-\omega}^{\rm BH} \right)^{*}$.}
    \label{Fig:RECO_RBH}
\end{figure*}

For the waveforms and reflectivity models considered here, we have $\mathcal{R}_{\ell -m -\omega}=\mathcal{R}^*_{\ell m \omega}$ for both ECO and BH reflecitivites,
and $h_{\ell -m}(-\omega)= h_{\ell m}^*(\omega)$ for all waveforms, including the main wave and the echoes. For example, the fact that $\mathcal{R}_{\ell m\omega}^{\rm ECO}\mathcal{R}_{\ell m\omega}^{\rm BH} = \left( \mathcal{R}_{\ell -m-\omega}^{\rm ECO}\mathcal{R}_{\ell -m-\omega}^{\rm BH} \right)^{*}$ is shown explicitly in Fig.~\ref{Fig:RECO_RBH}.  This means that, for the models considered here, we do not see the polarization mixing pointed out by Maggio \etal{}~\cite{Maggio}. Note that the effect of polarization mixing is absent as long as we consider pairs of $\pm m$ simultaneously, while for an individual $m$, such polarization mixing can still exist. 

\begin{figure*}[tbh!]
    \centering
    \subfloat[]{\includegraphics[width=\columnwidth]{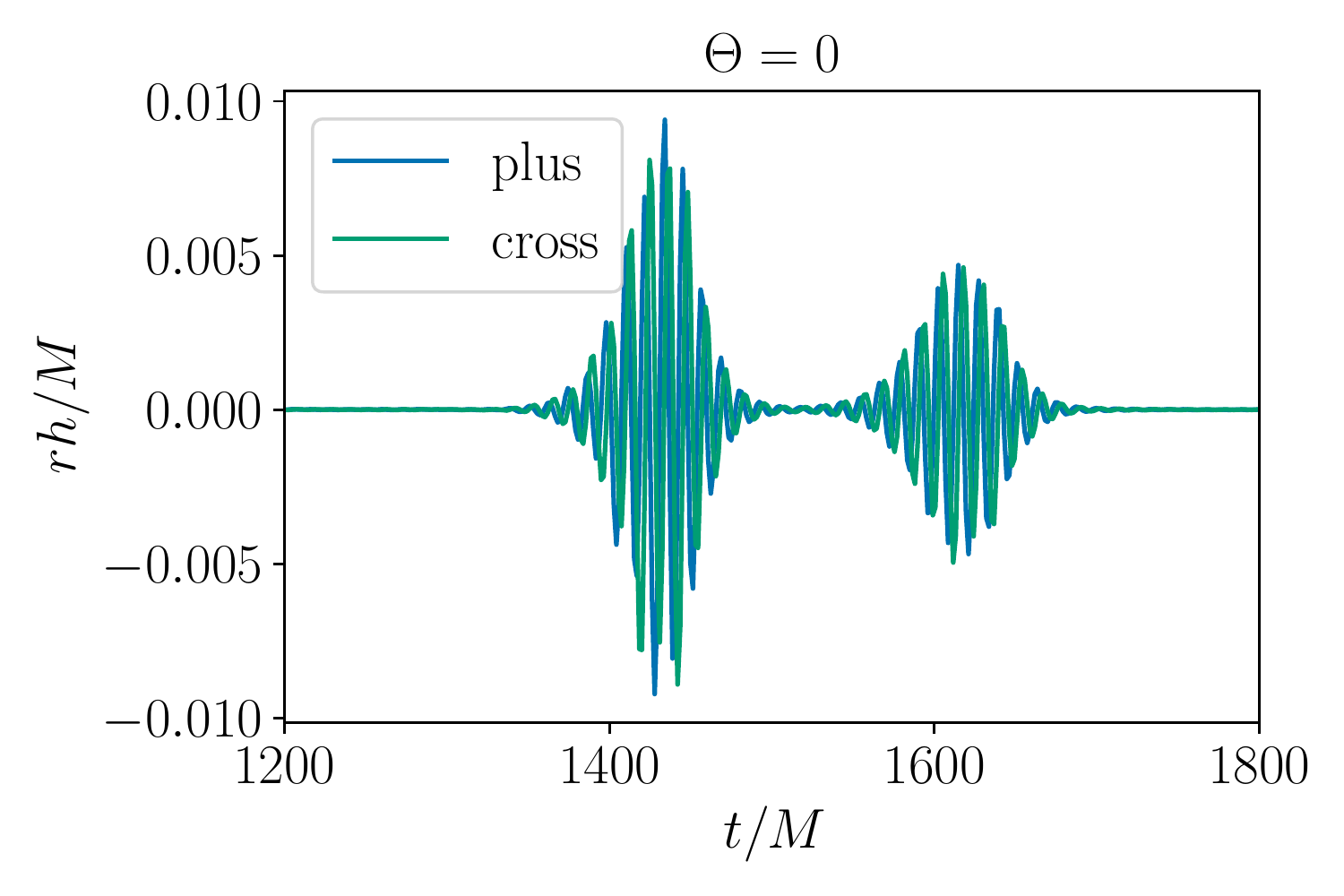}}
    \subfloat[]{\includegraphics[width=\columnwidth]{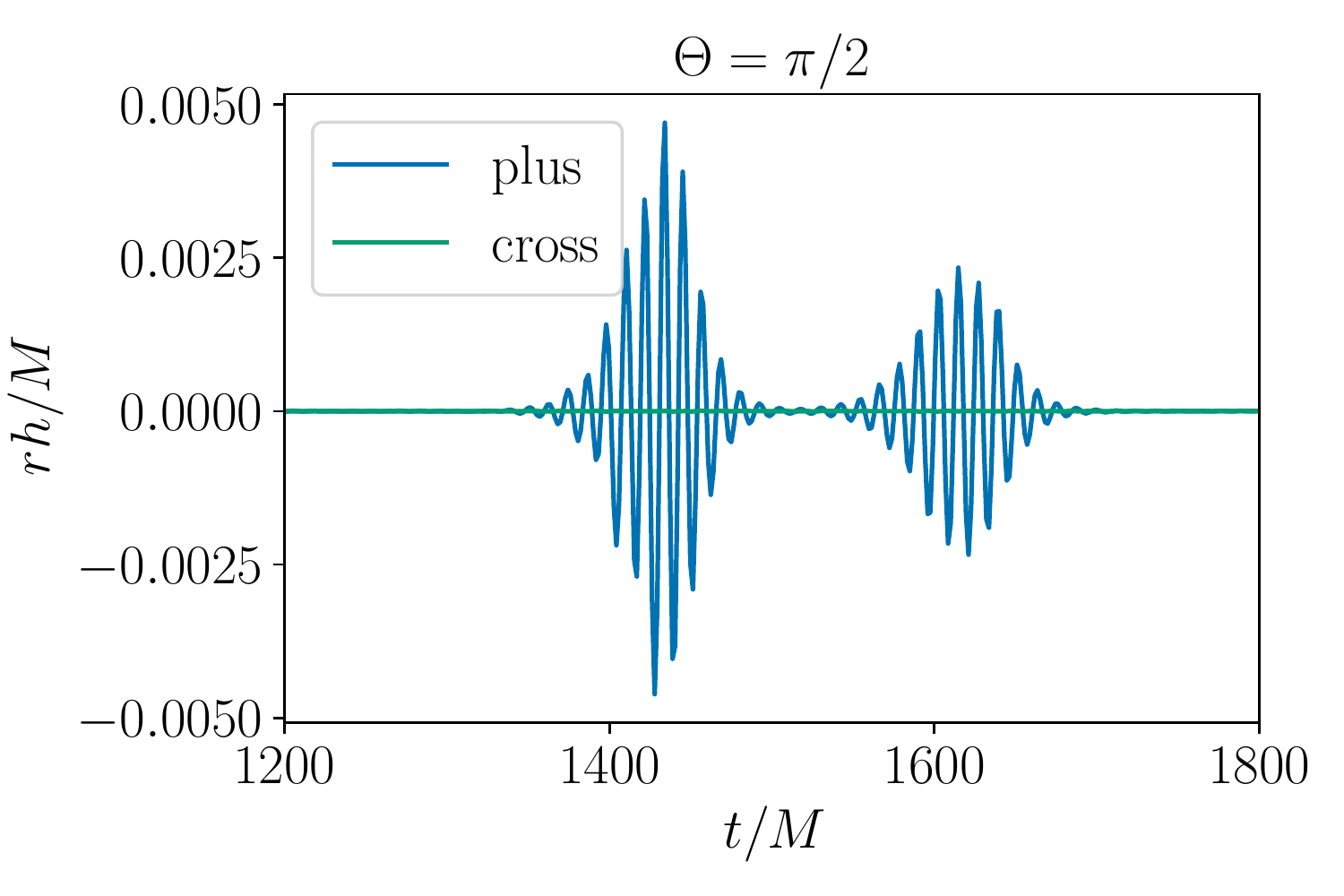}}
    \caption{The plus and cross polarizations of echoes when viewing the ECO at two different angles, $\Theta = 0$ (left panel) where the echoes are circularly polarized and $\Theta = \pi/2$ (right panel) where the echoes are linearly polarized (no cross polarization) using our prescription and the Lorentzian reflectivity model with $\epsilon=1$ and $M\Gamma=0.21$. }
    \label{Fig:Polarization Mixing}
\end{figure*}

In Fig.~\ref{Fig:Polarization Mixing}, we plot both the $+$ and $\times$ polarizations, for the GR wave and echoes, and for inclination angle $\Theta=0$ (face on) and $\Theta=\pi/2$ (edge on) --- indeed, the polarization state of the echos traces that of the main wave.  The echoes are approximately circularly polarized for the face-on case, and linearly polarized for the edge-on case.   

We also see that even by accounting for the $\ell$-$\ell'$ mixing effects due to the ECO rotation, as considered in Ref.~\cite{chen2020tidal}, polarization mixing is still absent. This is because the equatorial symmetry is not broken by the deformation of the spacetime geometry due to the ECO rotation.  To observe polarization mixing proposed by Maggio \etal{}~\cite{Maggio}, one can consider precessing binaries, whose source modes can be explicitly decomposed into those with $h_{\ell -m}(-\omega)= h_{\ell m}^*(\omega)$ and $h_{\ell -m}(-\omega)=- h_{\ell m}^*(\omega)$~\cite{chen2020tidal}.

\subsection{Comparison between prescriptions}
\label{comparison}
\begin{figure*}[t]
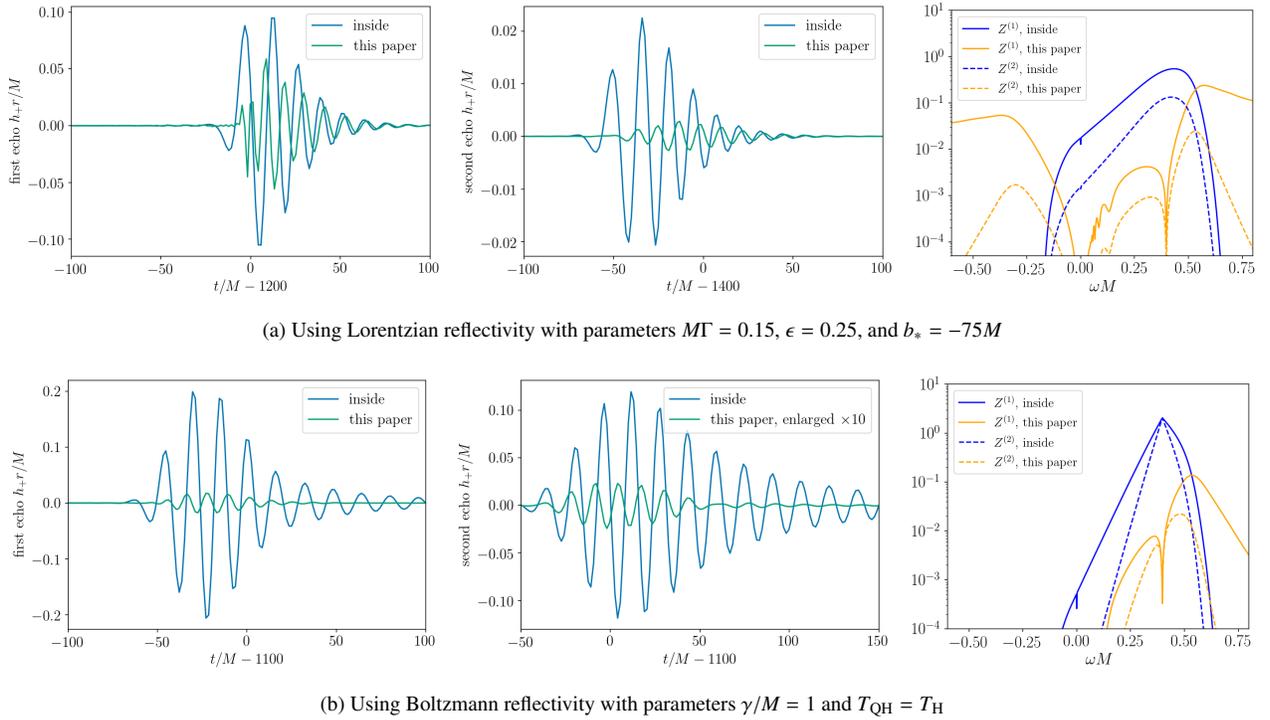

    \centering
    \subfloat[Using Lorentzian reflectivity with parameters $M\Gamma=0.15$, $\epsilon=0.25$, and $b_*=-75M$]{
        \begin{tabular}{ccc}
         \includegraphics[height=4cm]{comp3_echo1.pdf} & \includegraphics[height=4cm]{comp3_echo2.pdf}  & \includegraphics[height=4cm]{inout_freq_L.pdf}
        \end{tabular}
    }\\
    \subfloat[Using Boltzmann reflectivity with parameters $\gamma/M=1$ and $T_{\rm QH}=T_{\rm H}$]{
        \begin{tabular}{ccc}
         \includegraphics[height=4cm]{compB_echo1.pdf} & \includegraphics[height=4cm]{compB_echo2.pdf} & \includegraphics[height=4cm]{inout_freq.pdf}
        \end{tabular}
    }
    \caption{Comparison between the echoes generated using the ``inside" prescription and the method in this paper. Left panel: the first echoes in time domain; middle panel: the second echoes in time domain; right panel: the first and second echoes in frequency domain. Note that Refs.~\cite{Maggio} and \cite{micchi2021loud} both correspond to the ``inside'' prescription.  The echoes in this paper are weaker than those obtained using the ``inside" formulation.
    \label{fig_comp1}}
\end{figure*}

We finally compare our numerical waveforms with the results from the ``inside'' formulation \cite{Maggio,wang2020echoes}, as outlined in Eq.~\eqref{hinside}.
The comparison is shown in Fig.~\ref{fig_comp1} using (a) a Lorentzian reflectivity and (b) a Boltzmann reflectivity.  We can see that the magnitude of echoes are substantially different. The differences between these models in frequency domain are shown on the right panels.
%
%
%
The echoes obtained using our model are weaker than those from the ``inside" prescription. This is because $Z^{\rm H\,BH}$ does not peak at $\omega\sim m\Omega_+$ in our method (it peaks at $\omega_{\rm QNM}$, as indicated in Fig.~\ref{fig_peak}); while for the ``inside" model, it naturally peaks near $\omega\sim m\Omega_+$ by using $(D^{\rm in}/D^{\rm \infty})Z^{\infty\,\rm BH}$ (or $Z^{\infty\,\rm BH}$) to estimate the in-going $\psi_4$ toward the horizon.  Thus, the echoes from our method are suppressed by $\omega-m\Omega_+$ when $\omega\sim m\Omega_+$, compared to the inside model. 

It seems peculiar that the results from our model and the inside model have a large discrepancy. As clarified above, one can see that the spacetime region {\it above} the plunge trajectory in the spacetime diagram (left panel of Fig.~\ref{stdiagram}) is free of sources, and therefore $\psi_4$ in this region should be characterized by a homogeneous solution with no incoming wave from the past null infinity, or $R^{\infty}$ in Eq.~\eqref{eq_Rinf}.  
However, since the conversion from $\psi_4$ at infinity to $\psi_4$ on the future horizon is done in the Fourier domain, and the homogeneous solution is only valid for part of the future horizon, additional transient waves may need to be added to this conversion, e.g., corresponding to the poles of the frequency-domain conversion factor. Thus, the $Z^{\rm H\,BH}$ computed using our method is a more faithful representation of $\psi_4$ going down the horizon.  


However, a more faithful horizon-going $\psi_4$ does not guarantee a better approximation for curvature perturbations for the fiducial observers, and hence a better approximation for the echoes.  This is because $\psi_0$ associated with the in-going wave is directly responsible for tidal perturbations for the fiducial observers, and we obtain in-going $\psi_0$ by applying the Teukolsky-Starobinsky identity to the in-going $\psi_4$. However, strictly speaking, the  Teukolsky-Starobinsky identity only applies to homogeneous solutions. One hint that this approach may not be sound is the fact that the amount of energy going down the horizon computed this way diverges for the $(2,2)$ mode alone; the same behavior is also seen in Ref.~\cite{sago2020gravitational} for radially plunging particles.  
As discussed before~\cite{davis1972pulses}, for a plunging point particle, even though the total in-going GW energy summed over all $\ell$'s are divergent, the energy corresponding to each individual $\ell$ does not diverge.  We believe it is necessary to directly compute the in-going $\psi_0$, in order to completely resolve the above issue.

\section{Detectability}
\label{p_det}
In this section, we discuss the detectability of echoes with current and future detectors. To quantify the detectability, one can compute the optimal signal-to-noise ratio (SNR) $\rho_{\rm opt}$, which is defined as \cite{Flanagan:1997sx}
\begin{equation}
\rho_{\rm opt}^2 = 4 \int_{0}^{\infty} df \; \frac{|\tilde{h}(f)|^2}{S_{\rm n}(f)},
\end{equation}
where $S_{\rm n}(f)$ is the one-sided noise power spectral density of a detector, and $\tilde{h}(f)$ is the strain measured by the detector which is given by
\begin{equation}
    \tilde{h}(f) = F_{+}\tilde{h}_{+}(f) + F_{\times}\tilde{h}_{\times}(f),
\end{equation}
with $F_{+,\times}$ being the detector response to the plus and the cross polarization respectively.

Following Ref.~\cite{Flanagan:1997sx}, we define a new quantity $H_{+,\times}(f)$ that factors out the $1/d_{\rm L}$ dependence on the luminosity distance $d_{\rm L}$ \footnote{To account for the expansion of the Universe, one can simply replace the the coordinate distance $r$ with the luminosity distance $d_{\rm L}$, and replace the total mass $M$ with the redshifted total mass $M(1+z)$ where $z = z(d_{\rm L})$ is the redshift of the source.} for each polarization, where
\begin{equation}
    \tilde{h}_{+,\times}(f) = \frac{1}{d_{\rm L}} H_{+,\times}(f).
\end{equation}
The direction and orientation averaged optimal SNR $\langle \rho^2 \rangle$ is then given by \cite{Flanagan:1997sx}
\begin{equation}
    \langle \rho^2 \rangle = \frac{4}{5} \frac{1}{d_{\rm L}^2} \int \frac{d\Omega}{4\pi} \int_{0}^{\infty} df \; \frac{|H_{+}(\Theta, \Phi, f)|^2 + |H_{\times}(\Theta, \Phi, f)|^2}{S_{\rm n}(f)},
\end{equation}
where the angle bracket $\langle ... \rangle$ denotes average over the sky location angles of the source with respect to the detector, the polarization angle and the polar angles of the detector with respect to the source (with $d\Omega = \sin \Theta d\Theta d\Phi$). If we only consider the $\ell = |m| = 2$ modes, the averaging over the orientation can also be done analytically. In fact, it is given by \cite{Finn:1992xs}
\begin{equation}
    \langle \rho^2 \rangle = \frac{16}{25} \frac{1}{d_{\rm L}^2} \int_{0}^{\infty} df \; \frac{|H_{+}(\Theta=0, \Phi, f)|^2}{S_{\rm n}(f)}.
\end{equation}
Similarly, we can also compute the maximal $\rho_{\rm opt}$ by setting the source to be face-on ($\Theta=0$) and directly above a detector ($F_{+,\times}=1$), i.e. both optimally oriented and optimally located.

We compute both the direction-and-orientation averaged, as well as the maximal optimal SNR of the first \emph{five} echoes in Advanced LIGO at the design sensitivity \cite{aLIGODesignNoiseCurve} and Cosmic Explorer \cite{Evans:2016mbw} with both the Lorentzian and Boltzmann reflectivities, using the inside prescription and the prescription in this paper. However, we only use the Teukolsky formulation for reflectivity; the corresponding transformed SN reflectivity becomes \emph{weaker} (see Fig.~\ref{fig:conversion} for $\ell = m =2$ where most of the wave contents are concentrated in the positive frequencies). These differ from the treatment in Ref.~\cite{micchi2021loud}.

\begin{figure*}[t]
    \centering
    \subfloat[Inside prescription]{\includegraphics[width=\columnwidth]{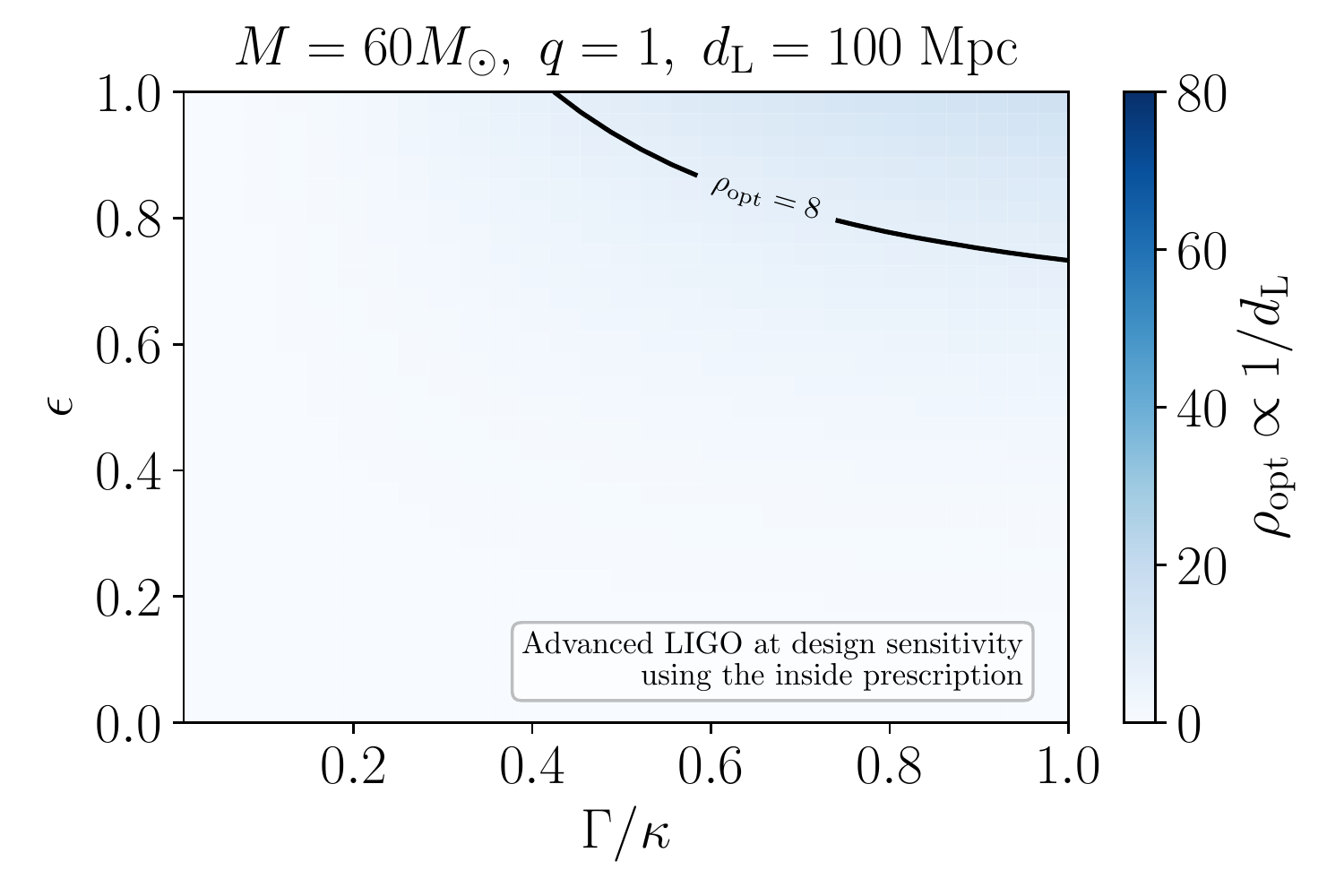}}
    \subfloat[Our prescription]{\includegraphics[width=\columnwidth]{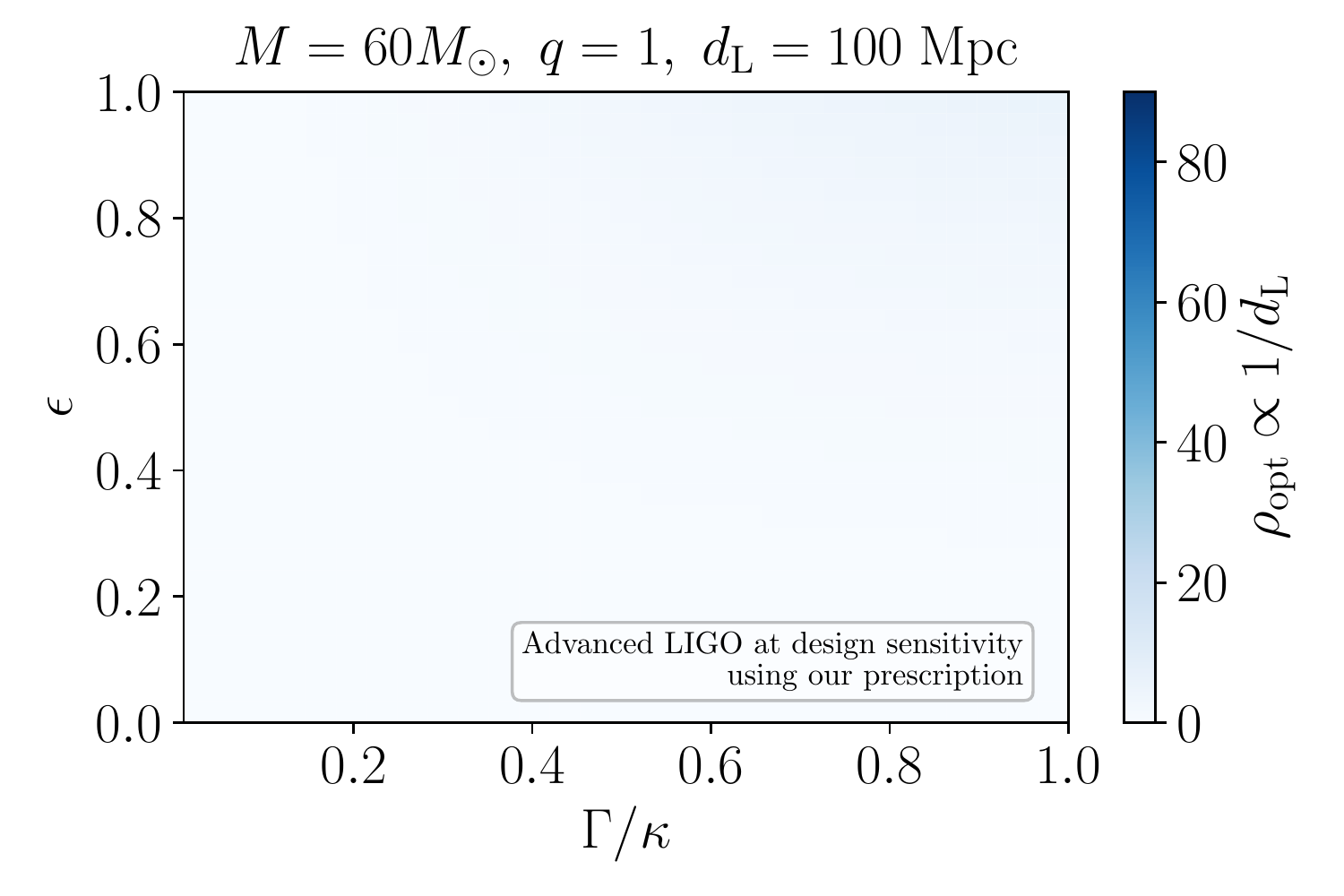}}
    \caption{SNR and detectability of echoes in the $\epsilon$--$\Gamma$ parameter space for the Lorentzian reflectivity model assuming the Advanced LIGO design sensitivity~\cite{aLIGODesignNoiseCurve} with $N_{\rm echo} = 5$ at a luminosity distance of $d_{\rm L} = 100\;{\rm Mpc}$. The solid contour corresponds to the \emph{maximal SNR} $\rho_{\rm opt} = 8$ as the detection threshold. Here we set $b_* = - 100M$. This choice of $b_*$ is not expected to affect the detectability as it mostly affects the time delay between echoes. We see that echo signals from the inside prescription are only detectable in a small part of the parameter space, while the echoes obtained using the method in this paper are generally too weak to be detected.}
    \label{fig:detectability_Lorentizan_aLIGOdesign}
\end{figure*}

\begin{figure*}[ht]
    \centering
    \subfloat[Inside prescription]{\includegraphics[width=\columnwidth]{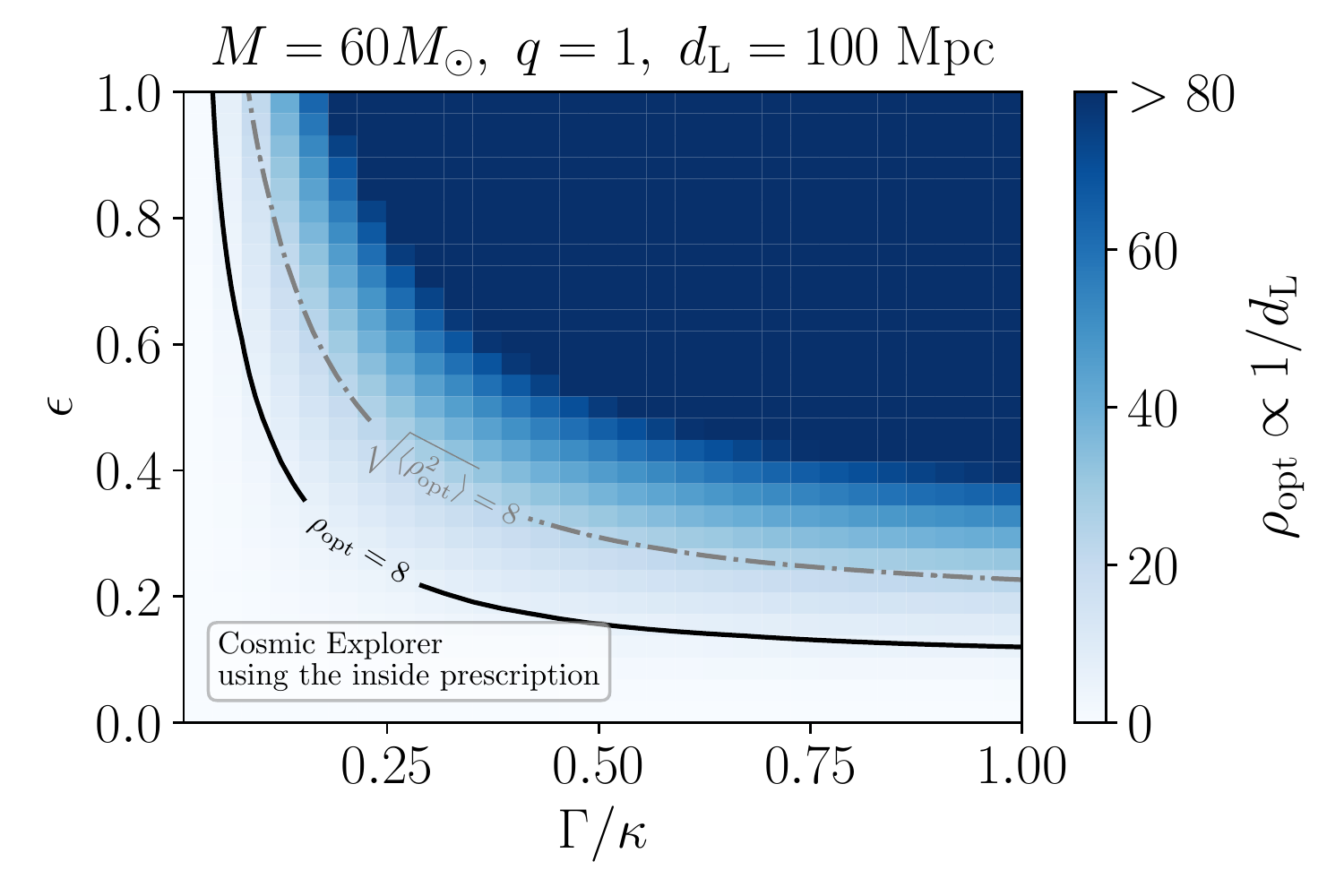}}
    \subfloat[Our prescription]{\includegraphics[width=\columnwidth]{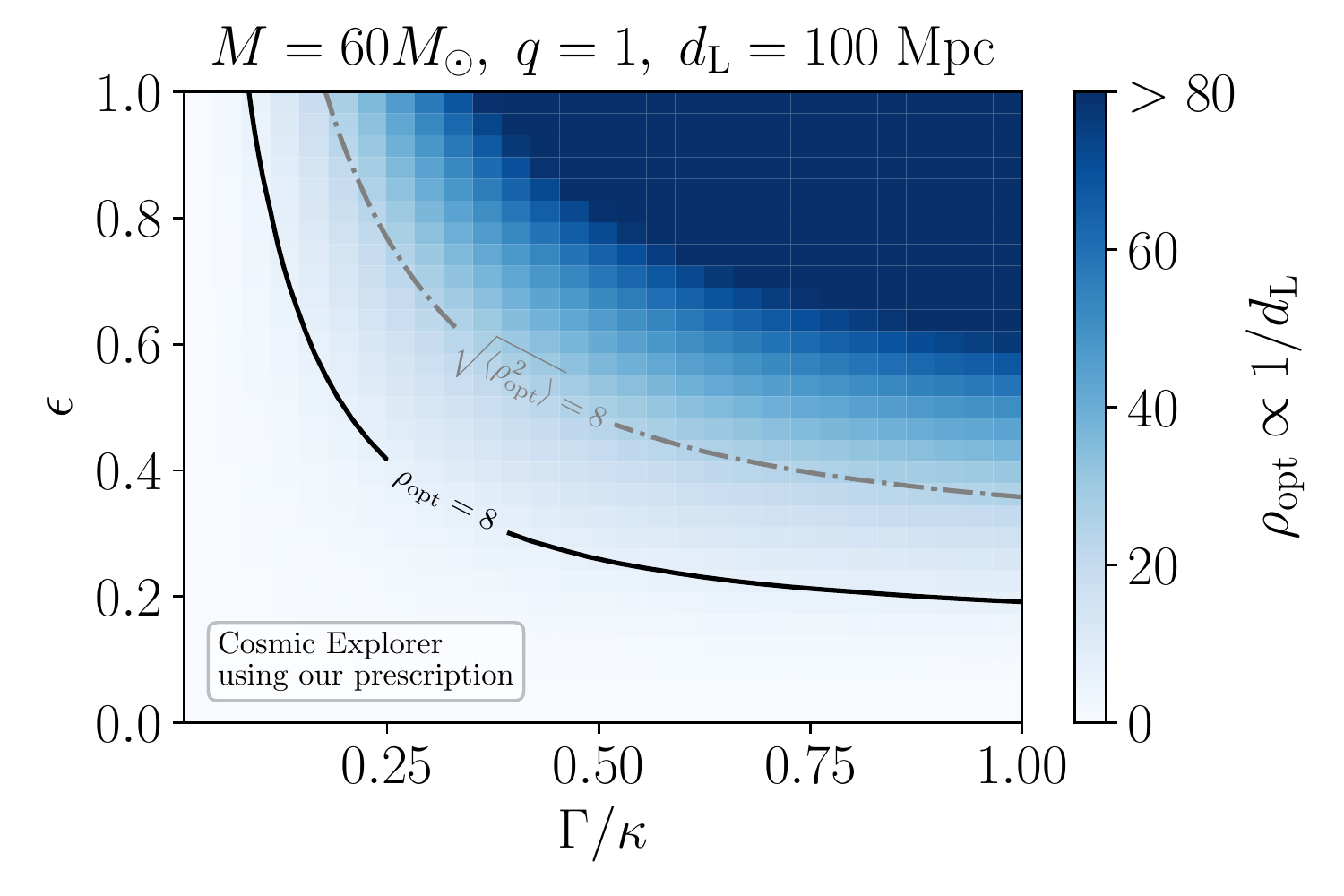}}
    \caption{Same as Fig.~\ref{fig:detectability_Lorentizan_aLIGOdesign} but with Cosmic Explorer at its design sensitivity \cite{Evans:2016mbw}. The solid contours correspond to the \emph{maximal SNR} $\rho_{\rm opt} = 8$ as the detection threshold, while the dash-dotted contours correspond to the location-and-orientation averaged SNR of $8$. We see that the detectable $\epsilon$--$\Gamma$ parameter space using the inside prescription is much larger, with the threshold of $\epsilon$ being detectable reaching as low as $\approx 0.2$. With Cosmic Explorer, the echoes computed using our prescription are now strong enough to be detected, but with a smaller detectable parameter space compared with that using the inside prescription.}
    \label{fig:detectability_Lorentizan_CE}
\end{figure*}

Figure \ref{fig:detectability_Lorentizan_aLIGOdesign} and \ref{fig:detectability_Lorentizan_CE} show the SNR and detectability of echoes in the $\epsilon$--$\Gamma$ parameter space for the Lorentzian reflectivity model assuming the Advanced LIGO and Cosmic Explorer at their design sensitivities, respectively. For Advanced LIGO, we see that the echoes computed using the inside prescription are only detectable in a small part of the parameter space, while the echoes obtained using the prescription in this paper are too weak to be detected in the parameter space that we explore here ($0 \leq \epsilon \leq 1$, $0 \leq \Gamma/\kappa \leq 1$). This implies that if our prescription is correct, we would \emph{not} be able to detect echoes with second-generation terrestrial detectors, and would require next-generation detectors in order to test the existance of ECOs via GW echoes. Indeed, Fig.~\ref{fig:detectability_Lorentizan_CE} indicates that with Cosmic Explorer, a much larger fraction of the $\epsilon$--$\Gamma$ parameter space allows detection for echoes from both the inside model and our prescription.

\begin{figure*}[ht]
    \centering
    \subfloat[Inside prescription]{\includegraphics[width=\columnwidth]{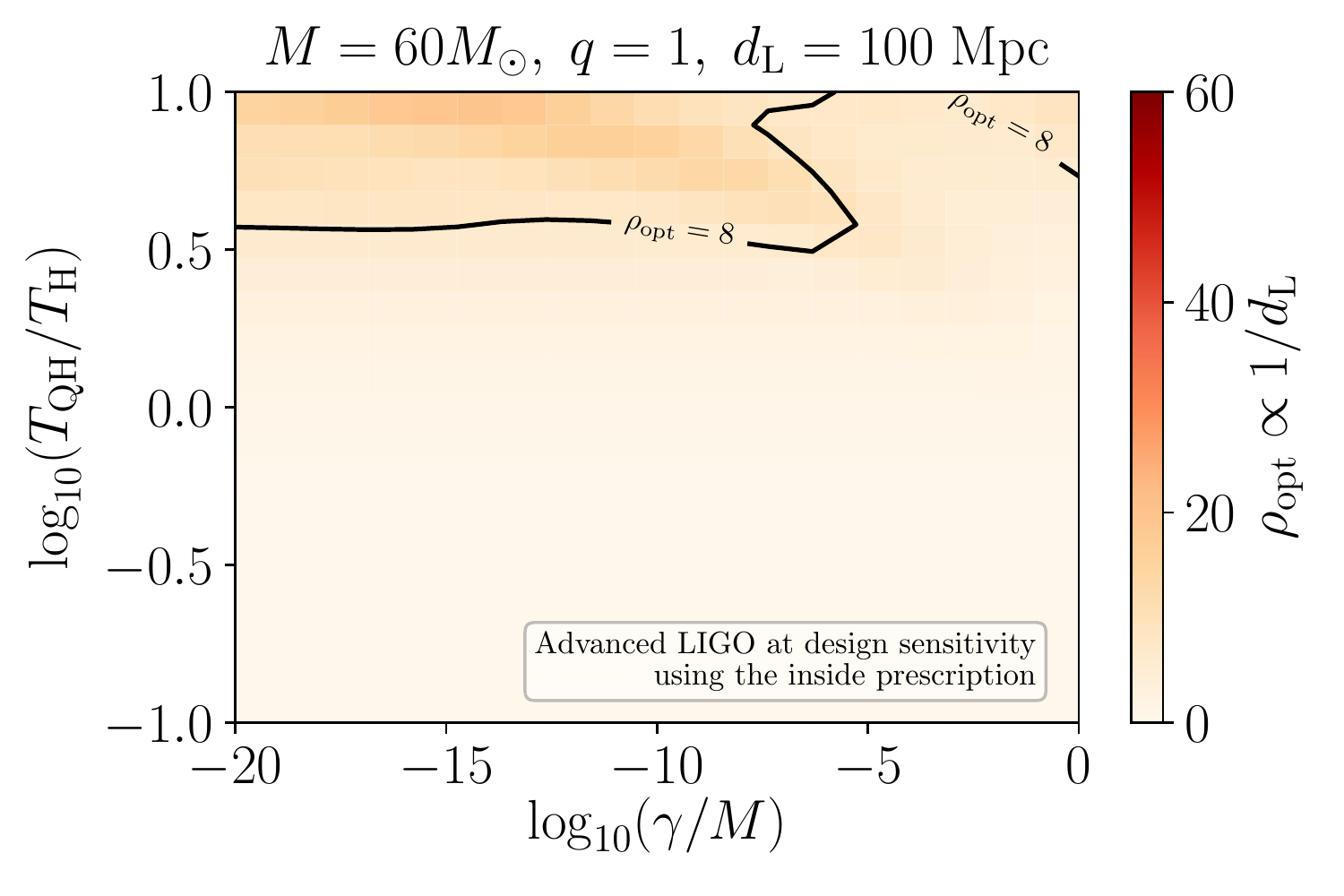}}
    \subfloat[Our prescription]{\includegraphics[width=\columnwidth]{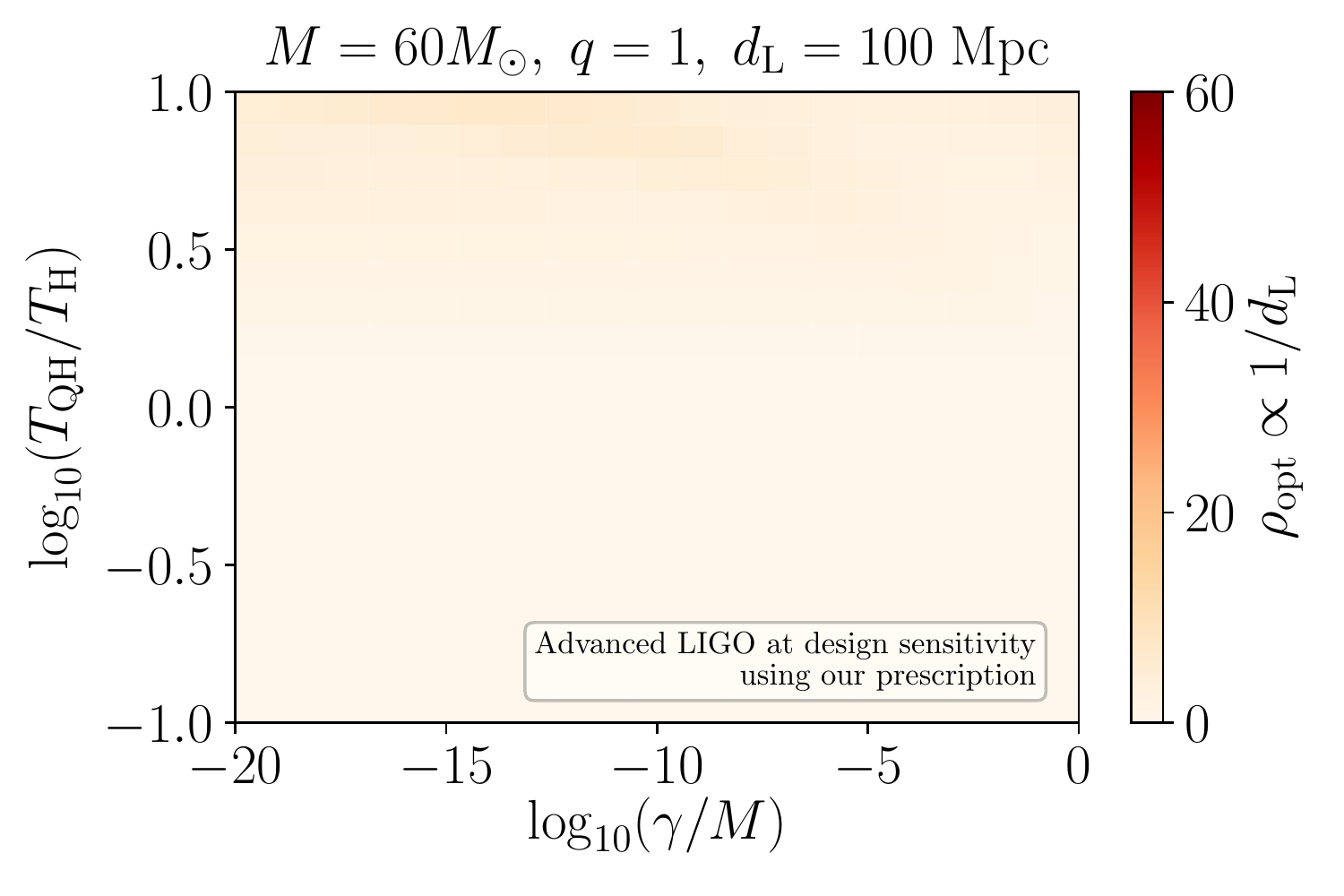}}
    \caption{SNR and detectability of echoes in the $T_{\rm QH}$--$\gamma$ parameter space for the Boltzmann reflectivity model assuming the Advanced LIGO design sensitivity \cite{aLIGODesignNoiseCurve} with $N_{\rm echo} = 5$ at a luminosity distance of $d_{\rm L} = 100\;{\rm Mpc}$. The solid contour corresponds to the \emph{maximal optimal SNR} $\rho_{\rm opt} = 8$ as the detection threshold. We see that echoes obtained using the inside prescription are louder and detectable in part of the $T_{\rm QH}$--$\gamma$ parameter space, while echoes obtained using our method are too weak to be detected.}
    \label{fig:detectability_Boltzmann_aLIGOdesign}
\end{figure*}

\begin{figure*}[ht]
    \centering
    \subfloat[Inside prescription]{\includegraphics[width=\columnwidth]{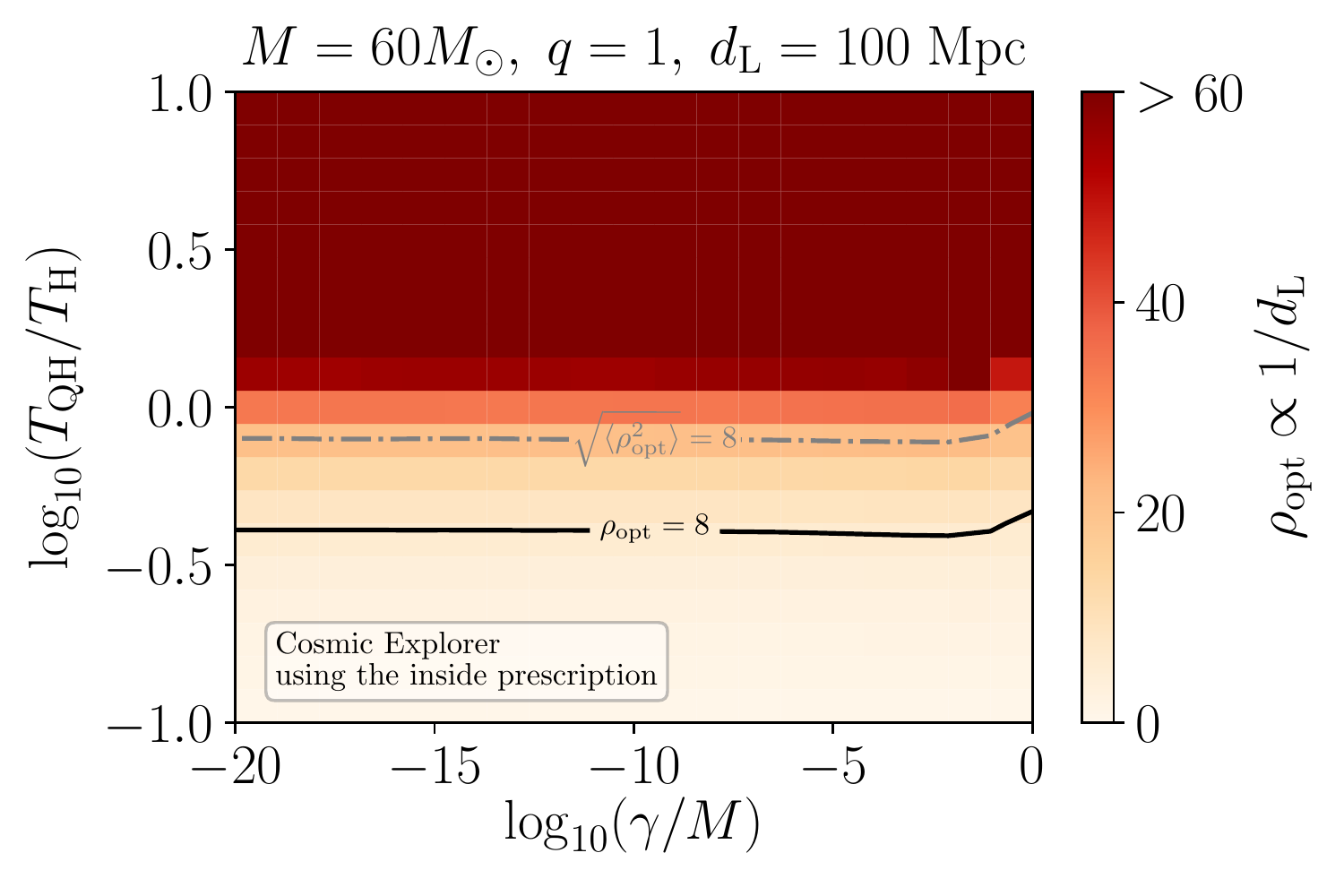}}
    \subfloat[Our prescription]{\includegraphics[width=\columnwidth]{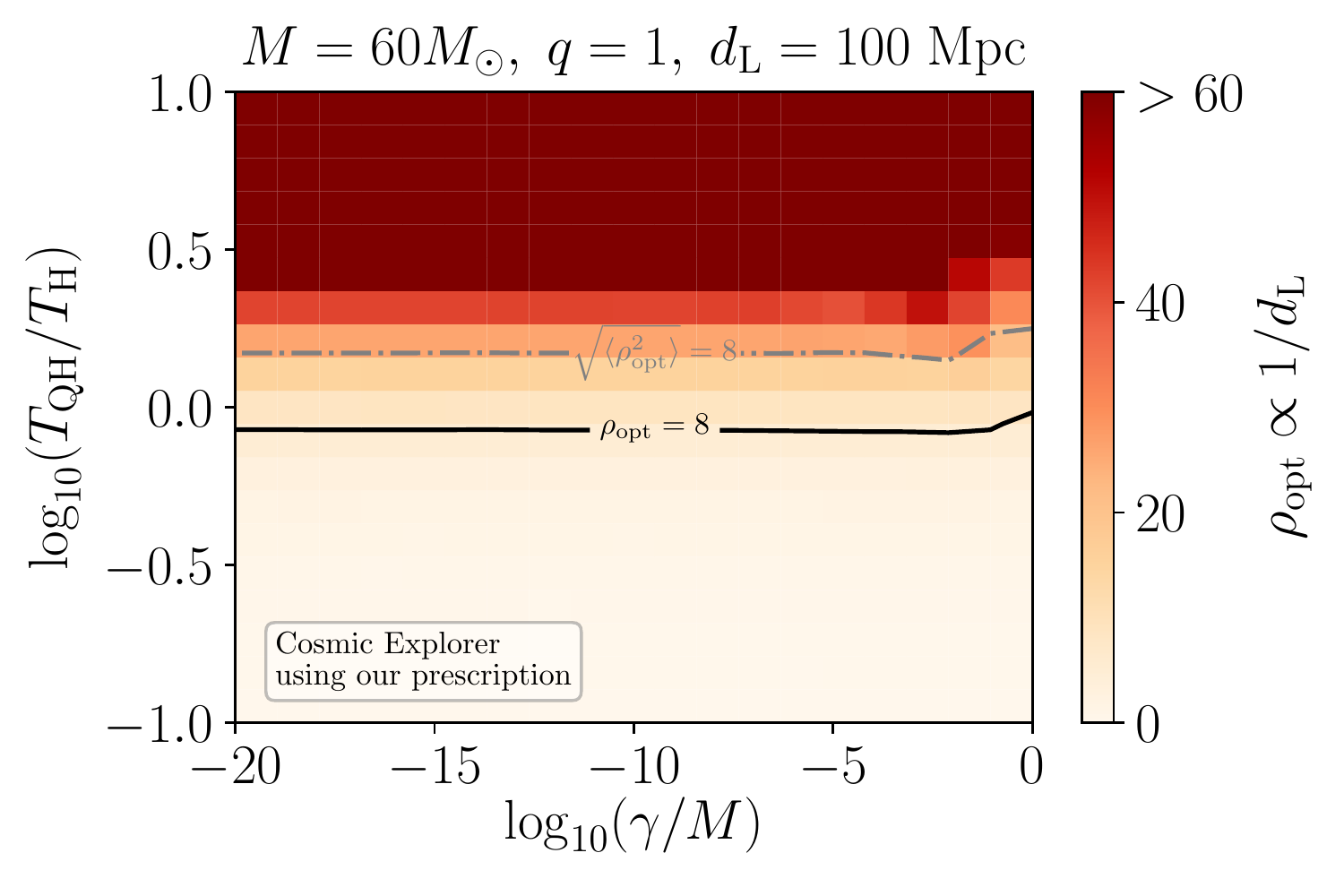}}
    \caption{Same as Fig.~\ref{fig:detectability_Boltzmann_aLIGOdesign} but with Cosmic Explorer at its design sensitivity \cite{Evans:2016mbw}. The solid contours correspond to the \emph{maximal optimal SNR} $\rho_{\rm opt} = 8$ as the detection threshold, while the dash-dotted contours correspond to the location-and-orientation averaged SNR of $8$. Again, we see that the detectable $T_{\rm QH}$--$\gamma$ parameter space with the inside prescription is much larger for Cosmic Explorer, compared with Fig.~\ref{fig:detectability_Boltzmann_aLIGOdesign} for Advanced LIGO. Echoes computed using our prescription are also loud enough to be detected, but again the detectable parameter space is smaller compared with the inside prescription. The plots also indicate that the detectability of echoes is generally \emph{independent} of the value of $\gamma$.}
    \label{fig:detectability_Boltzmann_CE}
\end{figure*}

Figure \ref{fig:detectability_Boltzmann_aLIGOdesign} and \ref{fig:detectability_Boltzmann_CE} show the SNR and detectability of echoes in the $T_{\rm QH}$--$\gamma$ parameter space for the Boltzmann reflectivity model assuming the Advanced LIGO and Cosmic Explorer at their design sensitivities, respectively. Similar to the Lorentzian reflectivity model, echoes obtained from the inside prescription are only detectable in a small part of the parameter space explored ($-1 \leq \log_{10} \left( T_{\rm QH}/T_{\rm H} \right) \leq 1$, $-20 \leq \log_{10} \gamma \leq 0$) with Advanced LIGO, while we would not see any echoes from our model with second-generation detectors. Detecting echoes would become more promising with next-generation detectors.
Interesting, from the plots we see that the detectability of echoes is generally \emph{independent} of $\gamma$.

\section{Conclusions}
\label{sec:conclusions}


In this paper, we compute GW echoes from merging ECOs that arise from the waves reflected by the surface of spinning ECOs.  The exterior spacetime of a spinning ECO is modeled as Kerr spacetime except in a small region above the horizon where a reflecting boundary exists. We obtain the echo waveforms by first computing the $\psi_4$ of the GWs that travel toward the horizon of the final BH in the case of BBH mergers, and then computing the subsequent reflection from the ECO surface and the Kerr potential barrier, in the case of ECO. More specifically, we solve the Teukolsky equation for $\psi_4$ sourced by an inspiraling particle that eventually plunges into the horizon of a Kerr BH. 
In order to model binaries with comparable masses, which is the most interesting case with existing GW events, we have adopted an EOB-like approach: by modifying the trajectory of the infalling particle and calibrating the GWs at infinity to match NR surrogate waveforms, we can obtain the horizon-going GWs that approximate those of comparable-mass mergers. 
For comparable-mass binaries, our approach is only designed for echoes reflected by the surface of the final ECO.  Nevertheless, from Fig.~\ref{fig_peakremove} (and the associated discussions), we show that most of the echoes arise from the reflection of waves that propagate toward the final horizon after the particle plunges. This justifies our approximation. 

In the comparison with previous work, we use prescriptions of the ECO reflectivity that are better connected to spacetime geometry near the ECO surface (obtained from a companion paper~\cite{chen2020tidal}).  More specifically, the reflectivity $\mathcal{R}^{\rm ECO\,T}$ in our method is directly related to the tidal response of the ECO surface to the external curvature perturbations due to incoming GWs. This reflectivity can also be converted into an effective reflectivity for SN functions using Eq.~\eqref{eq_RB2R}. Numerically, this conversion factor leads to discrepancies in the resulting waveforms, although it does not qualitatively modify the main features of the echo waveform. 

The echo waveforms we obtain here are significantly weaker than those obtained by Maggio \etal{} and from the ``inside" formulation of Wang \etal{}, because the $\psi_4$ we obtain by directly solving the Teukolsky equation turns out to be much smaller in magnitude than those obtained in previous studies (see Sec.~\ref{comparison} for details). As shown in Sec.~\ref{p_det}, the echoes obtained in this paper are not expected to be detectable using the second-generation detectors, and we would need the next-generation detectors to test ECOs via GW echoes.





One subtlety in our calculation is about obtaining the reflectivity for $\psi_4$ on the ECO surface. When applying the $\mathcal{R}^{\rm ECO\,T}$ between the in-going and out-going $\psi_4$, the Starobinsky-Teukolsky relation between the in-going $\psi_0$ and the in-going $\psi_4$ has been assumed. Strictly speaking, this relation only applies to homogeneous solutions of the Teukolsky equation and may not apply to the situation here.  A more direct approach, to be studied in future work, is to compute the in-going $\psi_0$, and then obtain the $\psi_4$ of the echoes by applying reflectivities to the ECO surface.     




\begin{acknowledgements}
Research of B. C. and Y. C. are supported by the Simons Foundation  (Award Number 568762), the Brinson Foundation, and the National Science Foundation (Grants 
PHY--2011968, PHY--2011961 and PHY--1836809).  Research of S. X. and Q. W. was done in part during their visits to the California Institute of Technology, which was supported by funds from the Simons Foundation. 
R. K. L. L. and L. S. acknowledge support of the National Science Foundation and the LIGO Laboratory.
LIGO was constructed by the California Institute of Technology and Massachusetts Institute of Technology with funding from the United States National Science Foundation, and operates under cooperative agreement PHY--1764464. Advanced LIGO was built under award PHY--0823459. 
R. K. L. L. would also like to acknowledge support from the Croucher Foundation. 
L. S. also acknowledges the support of the Australian Research Council Centre of Excellence for Gravitational Wave Discovery (OzGrav), project number CE170100004.
{W.H.\ is supported by by NSFC (National Natural Science Foundation of China) no.\ 11773059}
The computations presented here were conducted on the Caltech High Performance Cluster partially supported by a grant from the Gordon and Betty Moore Foundation.
\end{acknowledgements} 
\appendix

\section{Source term in Teukolsky equation}
\label{app_source}
Consider a massive point particle with the trajectory $x^\mu = (t,r(t),\theta(t),\phi(t)) $(expressed in Boyer-Lindquist coordinates) in Kerr spacetime with instantaneous energy $E$ and angular momentum in $z$ direction $L_z$. Following \cite{SasakiReview}, the source term $\mathcal{T}_{\ell m \omega}$ induced by this trajectory is (see Eq. (2.24) of \cite{SasakiReview}):

\begin{align}
	\mathscr{T}_{\ell m \omega} (r') 
	= \int_{-\infty}^\infty dt & e^{i[\omega t - m \phi(t)]}  \Delta^2(r') \nonumber\\
	& 
	\Big\{ [A_{nn0} + A_{n\bar m 0} + A_{\bar{m}\bar m 0} ] \delta(r'-r(t)) \nonumber \\
	&+\partial_{r'} ([A_{n\bar m 1} + A_{\bar m \bar m 1} ]\delta (r'-r(t)) ) \nonumber\\
	&+\partial_{r'}^2 [ A_{\bar m \bar m 2} \delta(r'-r(t))] \Big\},
\end{align}
where the coefficients are given by 
\begin{equation}
\begin{aligned}
A_{nn0} =& \frac{-2}{\Delta^2} C_{nn} \rho^{-2}\bar\rho^{-1} L_1^\dagger\{\rho^{-4} L_2^\dagger (\rho^3 S)\},\\
A_{\bar m n 0} =& \frac{2\sqrt{2}}{\Delta} C_{\bar m n} \rho^{-3} \left[\left( \frac{iK}{\Delta} + \rho +\bar \rho \right) (L_2^\dagger S) - a\sin\theta S\frac{K}{\Delta} (\bar \rho - \rho)  \right],\\
A_{\bar m\bar m 0} =& -\rho^{-3} \bar \rho C_{\bar m\bar m} S \left[ - i \partial_r\left( \frac{K}{\Delta} \right) - \frac{K^2}{\Delta^2} + 2i\rho \frac{K}{\Delta} \right],\\
A_{\bar m n 1} =& \frac{2\sqrt{2}}{\Delta} \rho^{-3} C_{\bar m n} [ L_2^\dagger S  + ia\sin\theta (\bar\rho - \rho) S],\\
A_{\bar m\bar m 1} =& -2 \rho^{-3} \bar \rho C_{\bar m\bar m} S(i(\frac{K}{\Delta} + \rho),\\
A_{\bar m\bar m 2} =& - \rho^{-3} \bar \rho C_{\bar m\bar m} S,
\end{aligned}
\end{equation}
with
\begin{equation}
\begin{aligned}
    C_{nn} &= \frac{1}{4\Sigma^3 dt/d\tau} \left[ E(r^2+a^2) - aL_z + \Sigma \frac{dr}{d\tau} \right]^2, \\
    C_{\bar m n} &= \frac{-\rho}{2\sqrt{2}\Sigma^2 dt/d\tau} \left[E(r^2+a^2) - aL_z + \Sigma \frac{dr}{d\tau} \right]\times\\
    &\quad\quad\left[ i\sin\theta \left( aE-\frac{L_z}{\sin^2\theta} \right) \right],\\
    C_{\bar m\bar m} &= \frac{\rho^2}{2\Sigma dt/d\tau} \left[ i\sin\theta \left( aE-\frac{L_z}{\sin^2\theta} \right) \right]^2.
\end{aligned}
\end{equation}
The operators appearing in the coefficients are defined by
\begin{equation}
    L_s^\dagger = \partial_\theta - \frac{m}{\sin\theta} + a\omega \sin\theta + s{\rm cot} \theta.
\end{equation}
The coefficients are different from \cite{SasakiReview} by a factor of $1/\sqrt{2\pi}$ due to normalization. The spin-weighted spheroidal harmonics functions we use are normalized as (while in \cite{SasakiReview} this expression is normalized to 1):
\begin{equation}
    \int_0^{\pi} [_{-2}S^{a\omega}_{\ell m \omega} (\theta)]^2 \sin\theta d\theta = \frac{1}{2\pi}.
\end{equation}

\section{Sasaki-Nakamura formalism}
\label{p_perturbation}

Transformation between Teukolsky function $R_{\ell m \omega}$ and SN function $X^{}_{\ell m \omega}$ is given by\cite{Hughes}:
\begin{align}
\label{SN2T}
R^{}_{\ell m \omega}=\frac{1}{\eta}\left[\left(\alpha+\frac{\beta_{,r}}{\Delta}\right)\frac{\Delta
	X^{}_{\ell m \omega}}{\sqrt{r^2+a^2}}-\frac{\beta}{\Delta}\frac{d}{dr}\frac{\Delta
	X^{}_{\ell m \omega}}{\sqrt{r^2+a^2}}\right].
\end{align}
Taking Eq. (\ref{SN2T}) into Teukolsky equation, one can find the equation for SN function $X^{}_{\ell m \omega}$:
\begin{align}
\label{eq_SN}
\frac{d^2
	X_{\ell m \omega}}{(dr^{*})^2}-F(r)\frac{dX_{\ell m \omega}}{dr^{*}}-U(r)X_{\ell m \omega}=0.
\end{align}
A detailed discussion of the SN formalism can be found in Ref.~\cite{Hughes}. Expressions for functions $\alpha, \beta, \eta$ and the potentials $F(r), U(r)$ can be found in Eqs.~(3.4)--(3.9) of Ref.~\cite{mino1997black}. For completeness, we list $\eta$ here:
\begin{equation}
 \eta=c_0+\frac{c_1}{r}+\frac{c_2}{r^2}+\frac{c_3}{r^3}+\frac{c_4}{r^4}\,,
 \end{equation}
 with
 \begin{align}
     c_0&=-12 i\omega M +\lambda(\lambda+2)-12 a\omega(a\omega-m), \\
     c_1&=8ia[3a\omega-\lambda(a\omega-m)], \\
     c_2&=-24ia  M(a\omega-m)+12a^2[1-2(a\omega-m)^2], \\
     c_3&=24ia^3(a\omega-m)-24Ma^2, \\
     c_4&=12a^4.
 \end{align}
The SN equation admits two homogeneous solutions having purely sinusoidal asymptotic behavior since the potential $U(r)$ is short-ranged:
\begin{equation}
\begin{aligned}
\label{eq_XH_app}
X^{{\rm H}}_{\ell m \omega}&=A^{{\rm hole}}_{\ell m \omega}e^{-ipr*},\quad r\rightarrow r_+,\\
X^{{\rm H}}_{\ell m \omega}&=A^{\rm{out}}_{\ell m \omega}e^{i\omega
	r*}+A^{\rm{in}}_{\ell m \omega}e^{-i\omega r*},\quad r\rightarrow
\infty,
\end{aligned}
\end{equation}
and
\begin{equation}
\begin{aligned}
\label{eq_Xinf_app}
X^{\infty}_{\ell m \omega}&=C^{\rm{out}}_{\ell m \omega}e^{ip
	r*}+C^{\rm{in}}_{\ell m \omega}e^{-ip r*},\quad r\rightarrow r_+,\\
X^{\infty}_{\ell m \omega}&=C^{\infty}_{\ell m \omega}e^{i\omega r*},\quad r\rightarrow
\infty.
\end{aligned}
\end{equation}
The homogeneous equations $X^{H,\infty}$ are related to $R^{H,\infty}$ by Eq. (\ref{SN2T}) and thus the asymptotic amplitudes $A^{{\rm hole}}, A^{{\rm in}}, A^{{\rm out}}$, $C^{{\rm in}}, C^{{\rm out}}, C^{\infty} $, $B^{{\rm hole}}, B^{{\rm in}}, B^{{\rm out}}$, and $D^{{\rm in}}, D^{{\rm out}}, D^{\infty} $ have following relations~\cite{SasakiReview}:
	\begin{equation}
	\label{trans_AB}
	\begin{aligned}
	B^{{\rm in}}_{\ell m \omega} =& -\frac{1}{4\omega^2} A^{{\rm in}}_{\ell m \omega},\\
	B^{{\rm out}}_{\ell m \omega} =& -\frac{4\omega^2}{-12i\omega M + \lambda(\lambda+2)-12a\omega(a\omega-m)} A^{{\rm out}}_{\ell m \omega},\\
	B^{{\rm hole}}_{\ell m \omega} =& d_{\ell m \omega}^{-1} A^{{\rm hole}}_{\ell m \omega},
	\end{aligned}
	\end{equation}
	\begin{equation}
	\begin{aligned}
	D^{{\rm in}}_{\ell m \omega} =& d_{\ell m \omega}^{-1}C^{{\rm in}}_{\ell m \omega},\\
	D^{{\rm out}}_{\ell m \omega} =& -\frac{4p\sqrt{2Mr_+} (2Mr_+ p+i\sqrt{M^2-a^2})}{\eta(r_+)}  C^{{\rm out}}_{\ell m \omega},\\
	D^{\infty}_{\ell m \omega} =& -\frac{4\omega^2}{-12i\omega M + \lambda(\lambda+2)-12a\omega(a\omega-m)} C^{\infty}_{\ell m \omega}.\\
	\end{aligned}\label{SN_D}
	\end{equation}
where
	\begin{align}
    d_{\ell m \omega}= &	 \sqrt{2Mr_+} \Big[ (8-24iM\omega - 16 M^2\omega^2)r^2_+ \nonumber\\
	&\qquad\quad + (12iam-16M+16amM\omega +24iM^2\omega)r_+ \nonumber\\ 
	&\qquad\quad  - 4a^2m^2 - 12iamM +8M^2 \Big].
	\end{align}
We note the subtlety that since the transformation between $X$ and $R$ contains derivatives, in regions where the source does not vanish, one has to expand the coefficients of $X$ and $R$ up to the second order in $1/r$ at infinity and $\Delta^2$ near horizon in order to obtain the correct transformations.

\bibliography{cit}

\begin{thebibliography}{80}%
\makeatletter
\providecommand \@ifxundefined [1]{%
 \@ifx{#1\undefined}
}%
\providecommand \@ifnum [1]{%
 \ifnum #1\expandafter \@firstoftwo
 \else \expandafter \@secondoftwo
 \fi
}%
\providecommand \@ifx [1]{%
 \ifx #1\expandafter \@firstoftwo
 \else \expandafter \@secondoftwo
 \fi
}%
\providecommand \natexlab [1]{#1}%
\providecommand \enquote  [1]{``#1''}%
\providecommand \bibnamefont  [1]{#1}%
\providecommand \bibfnamefont [1]{#1}%
\providecommand \citenamefont [1]{#1}%
\providecommand \href@noop [0]{\@secondoftwo}%
\providecommand \href [0]{\begingroup \@sanitize@url \@href}%
\providecommand \@href[1]{\@@startlink{#1}\@@href}%
\providecommand \@@href[1]{\endgroup#1\@@endlink}%
\providecommand \@sanitize@url [0]{\catcode `\\12\catcode `\$12\catcode
  `\&12\catcode `\#12\catcode `\^12\catcode `\_12\catcode `\%12\relax}%
\providecommand \@@startlink[1]{}%
\providecommand \@@endlink[0]{}%
\providecommand \url  [0]{\begingroup\@sanitize@url \@url }%
\providecommand \@url [1]{\endgroup\@href {#1}{\urlprefix }}%
\providecommand \urlprefix  [0]{URL }%
\providecommand \Eprint [0]{\href }%
\providecommand \doibase [0]{http://dx.doi.org/}%
\providecommand \selectlanguage [0]{\@gobble}%
\providecommand \bibinfo  [0]{\@secondoftwo}%
\providecommand \bibfield  [0]{\@secondoftwo}%
\providecommand \translation [1]{[#1]}%
\providecommand \BibitemOpen [0]{}%
\providecommand \bibitemStop [0]{}%
\providecommand \bibitemNoStop [0]{.\EOS\space}%
\providecommand \EOS [0]{\spacefactor3000\relax}%
\providecommand \BibitemShut  [1]{\csname bibitem#1\endcsname}%
\let\auto@bib@innerbib\@empty
\bibitem [{\citenamefont {Abbott}\ \emph {et~al.}(2016)\citenamefont {Abbott},
  \citenamefont {Abbott}, \citenamefont {Abbott}, \citenamefont {Abernathy},
  \citenamefont {Acernese}, \citenamefont {Ackley}, \citenamefont {Adams},
  \citenamefont {Adams}, \citenamefont {Addesso}, \citenamefont {Adhikari}
  \emph {et~al.}}]{GW150914}%
  \BibitemOpen
  \bibfield  {author} {\bibinfo {author} {\bibfnamefont {B.~P.}\ \bibnamefont
  {Abbott}}, \bibinfo {author} {\bibfnamefont {R.}~\bibnamefont {Abbott}},
  \bibinfo {author} {\bibfnamefont {T.}~\bibnamefont {Abbott}}, \bibinfo
  {author} {\bibfnamefont {M.}~\bibnamefont {Abernathy}}, \bibinfo {author}
  {\bibfnamefont {F.}~\bibnamefont {Acernese}}, \bibinfo {author}
  {\bibfnamefont {K.}~\bibnamefont {Ackley}}, \bibinfo {author} {\bibfnamefont
  {C.}~\bibnamefont {Adams}}, \bibinfo {author} {\bibfnamefont
  {T.}~\bibnamefont {Adams}}, \bibinfo {author} {\bibfnamefont
  {P.}~\bibnamefont {Addesso}}, \bibinfo {author} {\bibfnamefont
  {R.}~\bibnamefont {Adhikari}},  \emph {et~al.},\ }\href@noop {} {\bibfield
  {journal} {\bibinfo  {journal} {Physical review letters}\ }\textbf {\bibinfo
  {volume} {116}},\ \bibinfo {pages} {061102} (\bibinfo {year}
  {2016})}\BibitemShut {NoStop}%
\bibitem [{\citenamefont {Abbott}\ \emph
  {et~al.}(2017{\natexlab{a}})\citenamefont {Abbott}, \citenamefont {Abbott},
  \citenamefont {Abbott}, \citenamefont {Acernese}, \citenamefont {Ackley},
  \citenamefont {Adams}, \citenamefont {Adams}, \citenamefont {Addesso},
  \citenamefont {Adhikari}, \citenamefont {Adya} \emph {et~al.}}]{GW170817}%
  \BibitemOpen
  \bibfield  {author} {\bibinfo {author} {\bibfnamefont {B.~P.}\ \bibnamefont
  {Abbott}}, \bibinfo {author} {\bibfnamefont {R.}~\bibnamefont {Abbott}},
  \bibinfo {author} {\bibfnamefont {T.}~\bibnamefont {Abbott}}, \bibinfo
  {author} {\bibfnamefont {F.}~\bibnamefont {Acernese}}, \bibinfo {author}
  {\bibfnamefont {K.}~\bibnamefont {Ackley}}, \bibinfo {author} {\bibfnamefont
  {C.}~\bibnamefont {Adams}}, \bibinfo {author} {\bibfnamefont
  {T.}~\bibnamefont {Adams}}, \bibinfo {author} {\bibfnamefont
  {P.}~\bibnamefont {Addesso}}, \bibinfo {author} {\bibfnamefont
  {R.}~\bibnamefont {Adhikari}}, \bibinfo {author} {\bibfnamefont
  {V.}~\bibnamefont {Adya}},  \emph {et~al.},\ }\href@noop {} {\bibfield
  {journal} {\bibinfo  {journal} {Physical Review Letters}\ }\textbf {\bibinfo
  {volume} {119}},\ \bibinfo {pages} {161101} (\bibinfo {year}
  {2017}{\natexlab{a}})}\BibitemShut {NoStop}%
\bibitem [{\citenamefont {Aasi}\ \emph {et~al.}(2015)\citenamefont {Aasi} \emph
  {et~al.}}]{LIGO2014}%
  \BibitemOpen
  \bibfield  {author} {\bibinfo {author} {\bibfnamefont {J.}~\bibnamefont
  {Aasi}} \emph {et~al.} (\bibinfo {collaboration} {LSC}),\ }\href {\doibase
  10.1088/0264-9381/32/7/074001} {\bibfield  {journal} {\bibinfo  {journal}
  {Classical and Quantum Gravity}\ }\textbf {\bibinfo {volume} {32}},\ \bibinfo
  {pages} {074001} (\bibinfo {year} {2015})}\BibitemShut {NoStop}%
\bibitem [{\citenamefont {Acernese}\ \emph {et~al.}(2015)\citenamefont
  {Acernese} \emph {et~al.}}]{Virgo2014}%
  \BibitemOpen
  \bibfield  {author} {\bibinfo {author} {\bibfnamefont {F.}~\bibnamefont
  {Acernese}} \emph {et~al.} (\bibinfo {collaboration} {Virgo}),\ }\href
  {\doibase 10.1088/0264-9381/32/2/024001} {\bibfield  {journal} {\bibinfo
  {journal} {Classical and Quantum Gravity}\ }\textbf {\bibinfo {volume}
  {32}},\ \bibinfo {pages} {024001} (\bibinfo {year} {2015})}\BibitemShut
  {NoStop}%
\bibitem [{\citenamefont {Collaboration}\ \emph {et~al.}(2018)\citenamefont
  {Collaboration}, \citenamefont {Collaboration} \emph {et~al.}}]{catalog}%
  \BibitemOpen
  \bibfield  {author} {\bibinfo {author} {\bibfnamefont {L.~S.}\ \bibnamefont
  {Collaboration}}, \bibinfo {author} {\bibfnamefont {V.}~\bibnamefont
  {Collaboration}},  \emph {et~al.},\ }\href@noop {} {\bibfield  {journal}
  {\bibinfo  {journal} {arXiv preprint arXiv:1811.12907}\ } (\bibinfo {year}
  {2018})}\BibitemShut {NoStop}%
\bibitem [{\citenamefont {Collaboration}\ \emph {et~al.}(2020)\citenamefont
  {Collaboration}, \citenamefont {Collaboration} \emph {et~al.}}]{catalog2}%
  \BibitemOpen
  \bibfield  {author} {\bibinfo {author} {\bibfnamefont {L.~S.}\ \bibnamefont
  {Collaboration}}, \bibinfo {author} {\bibfnamefont {V.}~\bibnamefont
  {Collaboration}},  \emph {et~al.},\ }\href@noop {} {\bibfield  {journal}
  {\bibinfo  {journal} {arXiv preprint arXiv:2010.14527}\ } (\bibinfo {year}
  {2020})}\BibitemShut {NoStop}%
\bibitem [{\citenamefont {Abbott}\ \emph {et~al.}(2019)\citenamefont {Abbott},
  \citenamefont {Abbott}, \citenamefont {Abbott}, \citenamefont {Abraham},
  \citenamefont {Acernese}, \citenamefont {Ackley}, \citenamefont {Adams},
  \citenamefont {Adhikari}, \citenamefont {Adya}, \citenamefont {Affeldt} \emph
  {et~al.}}]{abbott2019tests}%
  \BibitemOpen
  \bibfield  {author} {\bibinfo {author} {\bibfnamefont {B.}~\bibnamefont
  {Abbott}}, \bibinfo {author} {\bibfnamefont {R.}~\bibnamefont {Abbott}},
  \bibinfo {author} {\bibfnamefont {T.}~\bibnamefont {Abbott}}, \bibinfo
  {author} {\bibfnamefont {S.}~\bibnamefont {Abraham}}, \bibinfo {author}
  {\bibfnamefont {F.}~\bibnamefont {Acernese}}, \bibinfo {author}
  {\bibfnamefont {K.}~\bibnamefont {Ackley}}, \bibinfo {author} {\bibfnamefont
  {C.}~\bibnamefont {Adams}}, \bibinfo {author} {\bibfnamefont {R.~X.}\
  \bibnamefont {Adhikari}}, \bibinfo {author} {\bibfnamefont {V.}~\bibnamefont
  {Adya}}, \bibinfo {author} {\bibfnamefont {C.}~\bibnamefont {Affeldt}},
  \emph {et~al.},\ }\href@noop {} {\bibfield  {journal} {\bibinfo  {journal}
  {Physical Review D}\ }\textbf {\bibinfo {volume} {100}},\ \bibinfo {pages}
  {104036} (\bibinfo {year} {2019})}\BibitemShut {NoStop}%
\bibitem [{\citenamefont {Abbott}\ \emph
  {et~al.}(2020{\natexlab{a}})\citenamefont {Abbott}, \citenamefont {Abbott},
  \citenamefont {Abraham}, \citenamefont {Acernese}, \citenamefont {Ackley},
  \citenamefont {Adams}, \citenamefont {Adams}, \citenamefont {Adhikari},
  \citenamefont {Adya}, \citenamefont {Affeldt} \emph
  {et~al.}}]{abbott2020tests}%
  \BibitemOpen
  \bibfield  {author} {\bibinfo {author} {\bibfnamefont {R.}~\bibnamefont
  {Abbott}}, \bibinfo {author} {\bibfnamefont {T.}~\bibnamefont {Abbott}},
  \bibinfo {author} {\bibfnamefont {S.}~\bibnamefont {Abraham}}, \bibinfo
  {author} {\bibfnamefont {F.}~\bibnamefont {Acernese}}, \bibinfo {author}
  {\bibfnamefont {K.}~\bibnamefont {Ackley}}, \bibinfo {author} {\bibfnamefont
  {A.}~\bibnamefont {Adams}}, \bibinfo {author} {\bibfnamefont
  {C.}~\bibnamefont {Adams}}, \bibinfo {author} {\bibfnamefont
  {R.}~\bibnamefont {Adhikari}}, \bibinfo {author} {\bibfnamefont
  {V.}~\bibnamefont {Adya}}, \bibinfo {author} {\bibfnamefont {C.}~\bibnamefont
  {Affeldt}},  \emph {et~al.},\ }\href@noop {} {\bibfield  {journal} {\bibinfo
  {journal} {arXiv preprint arXiv:2010.14529}\ } (\bibinfo {year}
  {2020}{\natexlab{a}})}\BibitemShut {NoStop}%
\bibitem [{\citenamefont {Berti}\ \emph
  {et~al.}(2018{\natexlab{a}})\citenamefont {Berti}, \citenamefont {Yagi},\
  and\ \citenamefont {Yunes}}]{berti2018extremea}%
  \BibitemOpen
  \bibfield  {author} {\bibinfo {author} {\bibfnamefont {E.}~\bibnamefont
  {Berti}}, \bibinfo {author} {\bibfnamefont {K.}~\bibnamefont {Yagi}}, \ and\
  \bibinfo {author} {\bibfnamefont {N.}~\bibnamefont {Yunes}},\ }\href@noop {}
  {\bibfield  {journal} {\bibinfo  {journal} {General Relativity and
  Gravitation}\ }\textbf {\bibinfo {volume} {50}},\ \bibinfo {pages} {1}
  (\bibinfo {year} {2018}{\natexlab{a}})}\BibitemShut {NoStop}%
\bibitem [{\citenamefont {Berti}\ \emph
  {et~al.}(2018{\natexlab{b}})\citenamefont {Berti}, \citenamefont {Yagi},
  \citenamefont {Yang},\ and\ \citenamefont {Yunes}}]{berti2018extremeb}%
  \BibitemOpen
  \bibfield  {author} {\bibinfo {author} {\bibfnamefont {E.}~\bibnamefont
  {Berti}}, \bibinfo {author} {\bibfnamefont {K.}~\bibnamefont {Yagi}},
  \bibinfo {author} {\bibfnamefont {H.}~\bibnamefont {Yang}}, \ and\ \bibinfo
  {author} {\bibfnamefont {N.}~\bibnamefont {Yunes}},\ }\href@noop {}
  {\bibfield  {journal} {\bibinfo  {journal} {General Relativity and
  Gravitation}\ }\textbf {\bibinfo {volume} {50}},\ \bibinfo {pages} {1}
  (\bibinfo {year} {2018}{\natexlab{b}})}\BibitemShut {NoStop}%
\bibitem [{\citenamefont {Palenzuela}\ \emph {et~al.}(2017)\citenamefont
  {Palenzuela}, \citenamefont {Pani}, \citenamefont {Bezares}, \citenamefont
  {Cardoso}, \citenamefont {Lehner},\ and\ \citenamefont
  {Liebling}}]{palenzuela2017gravitational}%
  \BibitemOpen
  \bibfield  {author} {\bibinfo {author} {\bibfnamefont {C.}~\bibnamefont
  {Palenzuela}}, \bibinfo {author} {\bibfnamefont {P.}~\bibnamefont {Pani}},
  \bibinfo {author} {\bibfnamefont {M.}~\bibnamefont {Bezares}}, \bibinfo
  {author} {\bibfnamefont {V.}~\bibnamefont {Cardoso}}, \bibinfo {author}
  {\bibfnamefont {L.}~\bibnamefont {Lehner}}, \ and\ \bibinfo {author}
  {\bibfnamefont {S.}~\bibnamefont {Liebling}},\ }\href@noop {} {\bibfield
  {journal} {\bibinfo  {journal} {Physical Review D}\ }\textbf {\bibinfo
  {volume} {96}},\ \bibinfo {pages} {104058} (\bibinfo {year}
  {2017})}\BibitemShut {NoStop}%
\bibitem [{\citenamefont {Mazur}\ and\ \citenamefont
  {Mottola}(2004)}]{mazur2004gravitational}%
  \BibitemOpen
  \bibfield  {author} {\bibinfo {author} {\bibfnamefont {P.~O.}\ \bibnamefont
  {Mazur}}\ and\ \bibinfo {author} {\bibfnamefont {E.}~\bibnamefont
  {Mottola}},\ }\href@noop {} {\bibfield  {journal} {\bibinfo  {journal}
  {Proceedings of the National Academy of Sciences}\ }\textbf {\bibinfo
  {volume} {101}},\ \bibinfo {pages} {9545} (\bibinfo {year}
  {2004})}\BibitemShut {NoStop}%
\bibitem [{\citenamefont {Pani}\ \emph {et~al.}(2009)\citenamefont {Pani},
  \citenamefont {Berti}, \citenamefont {Cardoso}, \citenamefont {Chen},\ and\
  \citenamefont {Norte}}]{pani2009gravitational}%
  \BibitemOpen
  \bibfield  {author} {\bibinfo {author} {\bibfnamefont {P.}~\bibnamefont
  {Pani}}, \bibinfo {author} {\bibfnamefont {E.}~\bibnamefont {Berti}},
  \bibinfo {author} {\bibfnamefont {V.}~\bibnamefont {Cardoso}}, \bibinfo
  {author} {\bibfnamefont {Y.}~\bibnamefont {Chen}}, \ and\ \bibinfo {author}
  {\bibfnamefont {R.}~\bibnamefont {Norte}},\ }\href@noop {} {\bibfield
  {journal} {\bibinfo  {journal} {Physical Review D}\ }\textbf {\bibinfo
  {volume} {80}},\ \bibinfo {pages} {124047} (\bibinfo {year}
  {2009})}\BibitemShut {NoStop}%
\bibitem [{\citenamefont {Giddings}\ and\ \citenamefont
  {Psaltis}(2018)}]{giddings2018event}%
  \BibitemOpen
  \bibfield  {author} {\bibinfo {author} {\bibfnamefont {S.~B.}\ \bibnamefont
  {Giddings}}\ and\ \bibinfo {author} {\bibfnamefont {D.}~\bibnamefont
  {Psaltis}},\ }\href@noop {} {\bibfield  {journal} {\bibinfo  {journal}
  {Physical Review D}\ }\textbf {\bibinfo {volume} {97}},\ \bibinfo {pages}
  {084035} (\bibinfo {year} {2018})}\BibitemShut {NoStop}%
\bibitem [{\citenamefont {Bianchi}\ \emph {et~al.}(2020)\citenamefont
  {Bianchi}, \citenamefont {Consoli}, \citenamefont {Grillo}, \citenamefont
  {Morales}, \citenamefont {Pani},\ and\ \citenamefont
  {Raposo}}]{bianchi2020distinguishing}%
  \BibitemOpen
  \bibfield  {author} {\bibinfo {author} {\bibfnamefont {M.}~\bibnamefont
  {Bianchi}}, \bibinfo {author} {\bibfnamefont {D.}~\bibnamefont {Consoli}},
  \bibinfo {author} {\bibfnamefont {A.}~\bibnamefont {Grillo}}, \bibinfo
  {author} {\bibfnamefont {J.~F.}\ \bibnamefont {Morales}}, \bibinfo {author}
  {\bibfnamefont {P.}~\bibnamefont {Pani}}, \ and\ \bibinfo {author}
  {\bibfnamefont {G.}~\bibnamefont {Raposo}},\ }\href@noop {} {\bibfield
  {journal} {\bibinfo  {journal} {Physical Review Letters}\ }\textbf {\bibinfo
  {volume} {125}},\ \bibinfo {pages} {221601} (\bibinfo {year}
  {2020})}\BibitemShut {NoStop}%
\bibitem [{\citenamefont {Bena}\ and\ \citenamefont
  {Mayerson}(2020{\natexlab{a}})}]{bena2020multipole}%
  \BibitemOpen
  \bibfield  {author} {\bibinfo {author} {\bibfnamefont {I.}~\bibnamefont
  {Bena}}\ and\ \bibinfo {author} {\bibfnamefont {D.~R.}\ \bibnamefont
  {Mayerson}},\ }\href@noop {} {\bibfield  {journal} {\bibinfo  {journal}
  {Physical Review Letters}\ }\textbf {\bibinfo {volume} {125}},\ \bibinfo
  {pages} {221602} (\bibinfo {year} {2020}{\natexlab{a}})}\BibitemShut
  {NoStop}%
\bibitem [{\citenamefont {Mukherjee}\ and\ \citenamefont
  {Chakraborty}(2020)}]{mukherjee2020multipole}%
  \BibitemOpen
  \bibfield  {author} {\bibinfo {author} {\bibfnamefont {S.}~\bibnamefont
  {Mukherjee}}\ and\ \bibinfo {author} {\bibfnamefont {S.}~\bibnamefont
  {Chakraborty}},\ }\href@noop {} {\bibfield  {journal} {\bibinfo  {journal}
  {Physical Review D}\ }\textbf {\bibinfo {volume} {102}},\ \bibinfo {pages}
  {124058} (\bibinfo {year} {2020})}\BibitemShut {NoStop}%
\bibitem [{\citenamefont {Wang}\ and\ \citenamefont
  {Afshordi}(2018)}]{ecology}%
  \BibitemOpen
  \bibfield  {author} {\bibinfo {author} {\bibfnamefont {Q.}~\bibnamefont
  {Wang}}\ and\ \bibinfo {author} {\bibfnamefont {N.}~\bibnamefont
  {Afshordi}},\ }\href {\doibase 10.1103/PhysRevD.97.124044} {\bibfield
  {journal} {\bibinfo  {journal} {Phys. Rev. D}\ }\textbf {\bibinfo {volume}
  {97}},\ \bibinfo {pages} {124044} (\bibinfo {year} {2018})}\BibitemShut
  {NoStop}%
\bibitem [{\citenamefont {Saravani}\ \emph {et~al.}(2014)\citenamefont
  {Saravani}, \citenamefont {Afshordi},\ and\ \citenamefont {Mann}}]{qc}%
  \BibitemOpen
  \bibfield  {author} {\bibinfo {author} {\bibfnamefont {M.}~\bibnamefont
  {Saravani}}, \bibinfo {author} {\bibfnamefont {N.}~\bibnamefont {Afshordi}},
  \ and\ \bibinfo {author} {\bibfnamefont {R.~B.}\ \bibnamefont {Mann}},\
  }\href {\doibase 10.1142/S021827181443007X} {\bibfield  {journal} {\bibinfo
  {journal} {International Journal of Modern Physics D}\ }\textbf {\bibinfo
  {volume} {23}},\ \bibinfo {pages} {1443007} (\bibinfo {year} {2014})},\
  \Eprint {http://arxiv.org/abs/https://doi.org/10.1142/S021827181443007X}
  {https://doi.org/10.1142/S021827181443007X} \BibitemShut {NoStop}%
\bibitem [{\citenamefont {Conklin}\ \emph {et~al.}(2018)\citenamefont
  {Conklin}, \citenamefont {Holdom},\ and\ \citenamefont
  {Ren}}]{spinechoSearch1}%
  \BibitemOpen
  \bibfield  {author} {\bibinfo {author} {\bibfnamefont {R.~S.}\ \bibnamefont
  {Conklin}}, \bibinfo {author} {\bibfnamefont {B.}~\bibnamefont {Holdom}}, \
  and\ \bibinfo {author} {\bibfnamefont {J.}~\bibnamefont {Ren}},\ }\href
  {\doibase 10.1103/PhysRevD.98.044021} {\bibfield  {journal} {\bibinfo
  {journal} {Phys. Rev. D}\ }\textbf {\bibinfo {volume} {98}},\ \bibinfo
  {pages} {044021} (\bibinfo {year} {2018})}\BibitemShut {NoStop}%
\bibitem [{\citenamefont {Li}\ and\ \citenamefont
  {Lovelace}(2008)}]{li2008generalization}%
  \BibitemOpen
  \bibfield  {author} {\bibinfo {author} {\bibfnamefont {C.}~\bibnamefont
  {Li}}\ and\ \bibinfo {author} {\bibfnamefont {G.}~\bibnamefont {Lovelace}},\
  }\href@noop {} {\bibfield  {journal} {\bibinfo  {journal} {Physical Review
  D}\ }\textbf {\bibinfo {volume} {77}},\ \bibinfo {pages} {064022} (\bibinfo
  {year} {2008})}\BibitemShut {NoStop}%
\bibitem [{\citenamefont {Fang}\ and\ \citenamefont
  {Lovelace}(2005)}]{PhysRevD.72.124016}%
  \BibitemOpen
  \bibfield  {author} {\bibinfo {author} {\bibfnamefont {H.}~\bibnamefont
  {Fang}}\ and\ \bibinfo {author} {\bibfnamefont {G.}~\bibnamefont
  {Lovelace}},\ }\href {\doibase 10.1103/PhysRevD.72.124016} {\bibfield
  {journal} {\bibinfo  {journal} {Phys. Rev. D}\ }\textbf {\bibinfo {volume}
  {72}},\ \bibinfo {pages} {124016} (\bibinfo {year} {2005})}\BibitemShut
  {NoStop}%
\bibitem [{\citenamefont {Datta}(2020)}]{Datta:2020rvo}%
  \BibitemOpen
  \bibfield  {author} {\bibinfo {author} {\bibfnamefont {S.}~\bibnamefont
  {Datta}},\ }\href {\doibase 10.1103/PhysRevD.102.064040} {\bibfield
  {journal} {\bibinfo  {journal} {Phys. Rev. D}\ }\textbf {\bibinfo {volume}
  {102}},\ \bibinfo {pages} {064040} (\bibinfo {year} {2020})},\ \Eprint
  {http://arxiv.org/abs/2002.04480} {arXiv:2002.04480 [gr-qc]} \BibitemShut
  {NoStop}%
\bibitem [{\citenamefont {Datta}\ \emph {et~al.}(2020)\citenamefont {Datta},
  \citenamefont {Brito}, \citenamefont {Bose}, \citenamefont {Pani},\ and\
  \citenamefont {Hughes}}]{datta2020tidal}%
  \BibitemOpen
  \bibfield  {author} {\bibinfo {author} {\bibfnamefont {S.}~\bibnamefont
  {Datta}}, \bibinfo {author} {\bibfnamefont {R.}~\bibnamefont {Brito}},
  \bibinfo {author} {\bibfnamefont {S.}~\bibnamefont {Bose}}, \bibinfo {author}
  {\bibfnamefont {P.}~\bibnamefont {Pani}}, \ and\ \bibinfo {author}
  {\bibfnamefont {S.~A.}\ \bibnamefont {Hughes}},\ }\href@noop {} {\bibfield
  {journal} {\bibinfo  {journal} {Physical Review D}\ }\textbf {\bibinfo
  {volume} {101}},\ \bibinfo {pages} {044004} (\bibinfo {year}
  {2020})}\BibitemShut {NoStop}%
\bibitem [{\citenamefont {Cardoso}\ and\ \citenamefont
  {Pani}(2017)}]{nature17}%
  \BibitemOpen
  \bibfield  {author} {\bibinfo {author} {\bibfnamefont {V.}~\bibnamefont
  {Cardoso}}\ and\ \bibinfo {author} {\bibfnamefont {P.}~\bibnamefont {Pani}},\
  }\href@noop {} {\bibfield  {journal} {\bibinfo  {journal} {Nature Astronomy}\
  }\textbf {\bibinfo {volume} {1}},\ \bibinfo {pages} {586} (\bibinfo {year}
  {2017})}\BibitemShut {NoStop}%
\bibitem [{\citenamefont {Abedi}\ \emph
  {et~al.}(2017{\natexlab{a}})\citenamefont {Abedi}, \citenamefont {Dykaar},\
  and\ \citenamefont {Afshordi}}]{abedi2017echoes}%
  \BibitemOpen
  \bibfield  {author} {\bibinfo {author} {\bibfnamefont {J.}~\bibnamefont
  {Abedi}}, \bibinfo {author} {\bibfnamefont {H.}~\bibnamefont {Dykaar}}, \
  and\ \bibinfo {author} {\bibfnamefont {N.}~\bibnamefont {Afshordi}},\
  }\href@noop {} {\bibfield  {journal} {\bibinfo  {journal} {Physical Review
  D}\ }\textbf {\bibinfo {volume} {96}},\ \bibinfo {pages} {082004} (\bibinfo
  {year} {2017}{\natexlab{a}})}\BibitemShut {NoStop}%
\bibitem [{\citenamefont {Ashton}\ \emph {et~al.}(2016)\citenamefont {Ashton},
  \citenamefont {Birnholtz}, \citenamefont {Cabero}, \citenamefont {Capano},
  \citenamefont {Dent}, \citenamefont {Krishnan}, \citenamefont {Meadors},
  \citenamefont {Nielsen}, \citenamefont {Nitz},\ and\ \citenamefont
  {Westerweck}}]{ashton2016comments}%
  \BibitemOpen
  \bibfield  {author} {\bibinfo {author} {\bibfnamefont {G.}~\bibnamefont
  {Ashton}}, \bibinfo {author} {\bibfnamefont {O.}~\bibnamefont {Birnholtz}},
  \bibinfo {author} {\bibfnamefont {M.}~\bibnamefont {Cabero}}, \bibinfo
  {author} {\bibfnamefont {C.}~\bibnamefont {Capano}}, \bibinfo {author}
  {\bibfnamefont {T.}~\bibnamefont {Dent}}, \bibinfo {author} {\bibfnamefont
  {B.}~\bibnamefont {Krishnan}}, \bibinfo {author} {\bibfnamefont {G.~D.}\
  \bibnamefont {Meadors}}, \bibinfo {author} {\bibfnamefont {A.~B.}\
  \bibnamefont {Nielsen}}, \bibinfo {author} {\bibfnamefont {A.}~\bibnamefont
  {Nitz}}, \ and\ \bibinfo {author} {\bibfnamefont {J.}~\bibnamefont
  {Westerweck}},\ }\href@noop {} {\bibfield  {journal} {\bibinfo  {journal}
  {arXiv preprint arXiv:1612.05625}\ } (\bibinfo {year} {2016})}\BibitemShut
  {NoStop}%
\bibitem [{\citenamefont {Abedi}\ \emph
  {et~al.}(2017{\natexlab{b}})\citenamefont {Abedi}, \citenamefont {Dykaar},\
  and\ \citenamefont {Afshordi}}]{echoADA}%
  \BibitemOpen
  \bibfield  {author} {\bibinfo {author} {\bibfnamefont {J.}~\bibnamefont
  {Abedi}}, \bibinfo {author} {\bibfnamefont {H.}~\bibnamefont {Dykaar}}, \
  and\ \bibinfo {author} {\bibfnamefont {N.}~\bibnamefont {Afshordi}},\ }\href
  {\doibase 10.1103/PhysRevD.96.082004} {\bibfield  {journal} {\bibinfo
  {journal} {Phys. Rev. D}\ }\textbf {\bibinfo {volume} {96}},\ \bibinfo
  {pages} {082004} (\bibinfo {year} {2017}{\natexlab{b}})}\BibitemShut
  {NoStop}%
\bibitem [{\citenamefont {Westerweck}\ \emph {et~al.}(2018)\citenamefont
  {Westerweck}, \citenamefont {Nielsen}, \citenamefont {Fischer-Birnholtz},
  \citenamefont {Cabero}, \citenamefont {Capano}, \citenamefont {Dent},
  \citenamefont {Krishnan}, \citenamefont {Meadors},\ and\ \citenamefont
  {Nitz}}]{echoLowSig}%
  \BibitemOpen
  \bibfield  {author} {\bibinfo {author} {\bibfnamefont {J.}~\bibnamefont
  {Westerweck}}, \bibinfo {author} {\bibfnamefont {A.~B.}\ \bibnamefont
  {Nielsen}}, \bibinfo {author} {\bibfnamefont {O.}~\bibnamefont
  {Fischer-Birnholtz}}, \bibinfo {author} {\bibfnamefont {M.}~\bibnamefont
  {Cabero}}, \bibinfo {author} {\bibfnamefont {C.}~\bibnamefont {Capano}},
  \bibinfo {author} {\bibfnamefont {T.}~\bibnamefont {Dent}}, \bibinfo {author}
  {\bibfnamefont {B.}~\bibnamefont {Krishnan}}, \bibinfo {author}
  {\bibfnamefont {G.}~\bibnamefont {Meadors}}, \ and\ \bibinfo {author}
  {\bibfnamefont {A.~H.}\ \bibnamefont {Nitz}},\ }\href {\doibase
  10.1103/PhysRevD.97.124037} {\bibfield  {journal} {\bibinfo  {journal} {Phys.
  Rev. D}\ }\textbf {\bibinfo {volume} {97}},\ \bibinfo {pages} {124037}
  (\bibinfo {year} {2018})}\BibitemShut {NoStop}%
\bibitem [{\citenamefont {Nielsen}\ \emph {et~al.}(2019)\citenamefont
  {Nielsen}, \citenamefont {Capano}, \citenamefont {Birnholtz},\ and\
  \citenamefont {Westerweck}}]{PhysRevD.99.104012}%
  \BibitemOpen
  \bibfield  {author} {\bibinfo {author} {\bibfnamefont {A.~B.}\ \bibnamefont
  {Nielsen}}, \bibinfo {author} {\bibfnamefont {C.~D.}\ \bibnamefont {Capano}},
  \bibinfo {author} {\bibfnamefont {O.}~\bibnamefont {Birnholtz}}, \ and\
  \bibinfo {author} {\bibfnamefont {J.}~\bibnamefont {Westerweck}},\ }\href
  {\doibase 10.1103/PhysRevD.99.104012} {\bibfield  {journal} {\bibinfo
  {journal} {Phys. Rev. D}\ }\textbf {\bibinfo {volume} {99}},\ \bibinfo
  {pages} {104012} (\bibinfo {year} {2019})}\BibitemShut {NoStop}%
\bibitem [{\citenamefont {Maselli}\ \emph {et~al.}(2017)\citenamefont
  {Maselli}, \citenamefont {V\"olkel},\ and\ \citenamefont
  {Kokkotas}}]{echoSearch1}%
  \BibitemOpen
  \bibfield  {author} {\bibinfo {author} {\bibfnamefont {A.}~\bibnamefont
  {Maselli}}, \bibinfo {author} {\bibfnamefont {S.~H.}\ \bibnamefont
  {V\"olkel}}, \ and\ \bibinfo {author} {\bibfnamefont {K.~D.}\ \bibnamefont
  {Kokkotas}},\ }\href {\doibase 10.1103/PhysRevD.96.064045} {\bibfield
  {journal} {\bibinfo  {journal} {Phys. Rev. D}\ }\textbf {\bibinfo {volume}
  {96}},\ \bibinfo {pages} {064045} (\bibinfo {year} {2017})}\BibitemShut
  {NoStop}%
\bibitem [{\citenamefont {Tsang}\ \emph {et~al.}(2018)\citenamefont {Tsang},
  \citenamefont {Rollier}, \citenamefont {Ghosh}, \citenamefont {Samajdar},
  \citenamefont {Agathos}, \citenamefont {Chatziioannou}, \citenamefont
  {Cardoso}, \citenamefont {Khanna},\ and\ \citenamefont {Van
  Den~Broeck}}]{echoSearch2}%
  \BibitemOpen
  \bibfield  {author} {\bibinfo {author} {\bibfnamefont {K.~W.}\ \bibnamefont
  {Tsang}}, \bibinfo {author} {\bibfnamefont {M.}~\bibnamefont {Rollier}},
  \bibinfo {author} {\bibfnamefont {A.}~\bibnamefont {Ghosh}}, \bibinfo
  {author} {\bibfnamefont {A.}~\bibnamefont {Samajdar}}, \bibinfo {author}
  {\bibfnamefont {M.}~\bibnamefont {Agathos}}, \bibinfo {author} {\bibfnamefont
  {K.}~\bibnamefont {Chatziioannou}}, \bibinfo {author} {\bibfnamefont
  {V.}~\bibnamefont {Cardoso}}, \bibinfo {author} {\bibfnamefont
  {G.}~\bibnamefont {Khanna}}, \ and\ \bibinfo {author} {\bibfnamefont
  {C.}~\bibnamefont {Van Den~Broeck}},\ }\href {\doibase
  10.1103/PhysRevD.98.024023} {\bibfield  {journal} {\bibinfo  {journal} {Phys.
  Rev. D}\ }\textbf {\bibinfo {volume} {98}},\ \bibinfo {pages} {024023}
  (\bibinfo {year} {2018})}\BibitemShut {NoStop}%
\bibitem [{\citenamefont {Nielsen}\ \emph {et~al.}(2018)\citenamefont
  {Nielsen}, \citenamefont {Capano},\ and\ \citenamefont
  {Westerweck}}]{echoSearch3}%
  \BibitemOpen
  \bibfield  {author} {\bibinfo {author} {\bibfnamefont {A.~B.}\ \bibnamefont
  {Nielsen}}, \bibinfo {author} {\bibfnamefont {C.~D.}\ \bibnamefont {Capano}},
  \ and\ \bibinfo {author} {\bibfnamefont {J.}~\bibnamefont {Westerweck}},\
  }\href@noop {} {\bibfield  {journal} {\bibinfo  {journal} {arXiv preprint
  arXiv:1811.04904}\ } (\bibinfo {year} {2018})}\BibitemShut {NoStop}%
\bibitem [{\citenamefont {Lo}\ \emph {et~al.}(2018)\citenamefont {Lo},
  \citenamefont {Li},\ and\ \citenamefont {Weinstein}}]{echoSearch4}%
  \BibitemOpen
  \bibfield  {author} {\bibinfo {author} {\bibfnamefont {R.~K.~L.}\
  \bibnamefont {Lo}}, \bibinfo {author} {\bibfnamefont {T.~G.}\ \bibnamefont
  {Li}}, \ and\ \bibinfo {author} {\bibfnamefont {A.~J.}\ \bibnamefont
  {Weinstein}},\ }\href@noop {} {\bibfield  {journal} {\bibinfo  {journal}
  {arXiv preprint arXiv:1811.07431}\ } (\bibinfo {year} {2018})}\BibitemShut
  {NoStop}%
\bibitem [{\citenamefont {Tsang}\ \emph {et~al.}(2020)\citenamefont {Tsang},
  \citenamefont {Ghosh}, \citenamefont {Samajdar}, \citenamefont
  {Chatziioannou}, \citenamefont {Mastrogiovanni}, \citenamefont {Agathos},\
  and\ \citenamefont {Van Den~Broeck}}]{Tsang:2019zra}%
  \BibitemOpen
  \bibfield  {author} {\bibinfo {author} {\bibfnamefont {K.~W.}\ \bibnamefont
  {Tsang}}, \bibinfo {author} {\bibfnamefont {A.}~\bibnamefont {Ghosh}},
  \bibinfo {author} {\bibfnamefont {A.}~\bibnamefont {Samajdar}}, \bibinfo
  {author} {\bibfnamefont {K.}~\bibnamefont {Chatziioannou}}, \bibinfo {author}
  {\bibfnamefont {S.}~\bibnamefont {Mastrogiovanni}}, \bibinfo {author}
  {\bibfnamefont {M.}~\bibnamefont {Agathos}}, \ and\ \bibinfo {author}
  {\bibfnamefont {C.}~\bibnamefont {Van Den~Broeck}},\ }\href {\doibase
  10.1103/PhysRevD.101.064012} {\bibfield  {journal} {\bibinfo  {journal}
  {Phys. Rev. D}\ }\textbf {\bibinfo {volume} {101}},\ \bibinfo {pages}
  {064012} (\bibinfo {year} {2020})},\ \Eprint
  {http://arxiv.org/abs/1906.11168} {arXiv:1906.11168 [gr-qc]} \BibitemShut
  {NoStop}%
\bibitem [{\citenamefont {Uchikata}\ \emph {et~al.}(2019)\citenamefont
  {Uchikata}, \citenamefont {Nakano}, \citenamefont {Narikawa}, \citenamefont
  {Sago}, \citenamefont {Tagoshi},\ and\ \citenamefont
  {Tanaka}}]{Uchikata:2019frs}%
  \BibitemOpen
  \bibfield  {author} {\bibinfo {author} {\bibfnamefont {N.}~\bibnamefont
  {Uchikata}}, \bibinfo {author} {\bibfnamefont {H.}~\bibnamefont {Nakano}},
  \bibinfo {author} {\bibfnamefont {T.}~\bibnamefont {Narikawa}}, \bibinfo
  {author} {\bibfnamefont {N.}~\bibnamefont {Sago}}, \bibinfo {author}
  {\bibfnamefont {H.}~\bibnamefont {Tagoshi}}, \ and\ \bibinfo {author}
  {\bibfnamefont {T.}~\bibnamefont {Tanaka}},\ }\href {\doibase
  10.1103/PhysRevD.100.062006} {\bibfield  {journal} {\bibinfo  {journal}
  {Phys. Rev. D}\ }\textbf {\bibinfo {volume} {100}},\ \bibinfo {pages}
  {062006} (\bibinfo {year} {2019})},\ \Eprint
  {http://arxiv.org/abs/1906.00838} {arXiv:1906.00838 [gr-qc]} \BibitemShut
  {NoStop}%
\bibitem [{\citenamefont {Abbott}\ \emph
  {et~al.}(2020{\natexlab{b}})\citenamefont {Abbott} \emph
  {et~al.}}]{Abbott:2020jks}%
  \BibitemOpen
  \bibfield  {author} {\bibinfo {author} {\bibfnamefont {R.}~\bibnamefont
  {Abbott}} \emph {et~al.} (\bibinfo {collaboration} {LIGO Scientific,
  Virgo}),\ }\href@noop {} {\  (\bibinfo {year} {2020}{\natexlab{b}})},\
  \Eprint {http://arxiv.org/abs/2010.14529} {arXiv:2010.14529 [gr-qc]}
  \BibitemShut {NoStop}%
\bibitem [{\citenamefont {Mark}\ \emph {et~al.}(2017)\citenamefont {Mark},
  \citenamefont {Zimmerman}, \citenamefont {Du},\ and\ \citenamefont
  {Chen}}]{echoSchw}%
  \BibitemOpen
  \bibfield  {author} {\bibinfo {author} {\bibfnamefont {Z.}~\bibnamefont
  {Mark}}, \bibinfo {author} {\bibfnamefont {A.}~\bibnamefont {Zimmerman}},
  \bibinfo {author} {\bibfnamefont {S.~M.}\ \bibnamefont {Du}}, \ and\ \bibinfo
  {author} {\bibfnamefont {Y.}~\bibnamefont {Chen}},\ }\href {\doibase
  10.1103/PhysRevD.96.084002} {\bibfield  {journal} {\bibinfo  {journal} {Phys.
  Rev. D}\ }\textbf {\bibinfo {volume} {96}},\ \bibinfo {pages} {084002}
  (\bibinfo {year} {2017})}\BibitemShut {NoStop}%
\bibitem [{\citenamefont {Du}\ and\ \citenamefont {Chen}(2018)}]{PRL18}%
  \BibitemOpen
  \bibfield  {author} {\bibinfo {author} {\bibfnamefont {S.~M.}\ \bibnamefont
  {Du}}\ and\ \bibinfo {author} {\bibfnamefont {Y.}~\bibnamefont {Chen}},\
  }\href {\doibase 10.1103/PhysRevLett.121.051105} {\bibfield  {journal}
  {\bibinfo  {journal} {Phys. Rev. Lett.}\ }\textbf {\bibinfo {volume} {121}},\
  \bibinfo {pages} {051105} (\bibinfo {year} {2018})}\BibitemShut {NoStop}%
\bibitem [{\citenamefont {Huang}\ \emph {et~al.}(2019)\citenamefont {Huang},
  \citenamefont {Xu},\ and\ \citenamefont {Zhou}}]{huang2019fredhom}%
  \BibitemOpen
  \bibfield  {author} {\bibinfo {author} {\bibfnamefont {Y.-X.}\ \bibnamefont
  {Huang}}, \bibinfo {author} {\bibfnamefont {J.-C.}\ \bibnamefont {Xu}}, \
  and\ \bibinfo {author} {\bibfnamefont {S.-Y.}\ \bibnamefont {Zhou}},\
  }\href@noop {} {\bibfield  {journal} {\bibinfo  {journal} {arXiv preprint
  arXiv:1908.00189}\ } (\bibinfo {year} {2019})}\BibitemShut {NoStop}%
\bibitem [{\citenamefont {Maggio}\ \emph {et~al.}(2020)\citenamefont {Maggio},
  \citenamefont {Buoninfante}, \citenamefont {Mazumdar},\ and\ \citenamefont
  {Pani}}]{maggio2020does}%
  \BibitemOpen
  \bibfield  {author} {\bibinfo {author} {\bibfnamefont {E.}~\bibnamefont
  {Maggio}}, \bibinfo {author} {\bibfnamefont {L.}~\bibnamefont {Buoninfante}},
  \bibinfo {author} {\bibfnamefont {A.}~\bibnamefont {Mazumdar}}, \ and\
  \bibinfo {author} {\bibfnamefont {P.}~\bibnamefont {Pani}},\ }\href@noop {}
  {\bibfield  {journal} {\bibinfo  {journal} {arXiv preprint arXiv:2006.14628}\
  } (\bibinfo {year} {2020})}\BibitemShut {NoStop}%
\bibitem [{\citenamefont {Farr}\ \emph {et~al.}(2018)\citenamefont {Farr},
  \citenamefont {Holz},\ and\ \citenamefont {Farr}}]{LIGOspin}%
  \BibitemOpen
  \bibfield  {author} {\bibinfo {author} {\bibfnamefont {B.}~\bibnamefont
  {Farr}}, \bibinfo {author} {\bibfnamefont {D.~E.}\ \bibnamefont {Holz}}, \
  and\ \bibinfo {author} {\bibfnamefont {W.~M.}\ \bibnamefont {Farr}},\ }\href
  {\doibase 10.3847/2041-8213/aaaa64} {\bibfield  {journal} {\bibinfo
  {journal} {The Astrophysical Journal}\ }\textbf {\bibinfo {volume} {854}},\
  \bibinfo {pages} {L9} (\bibinfo {year} {2018})}\BibitemShut {NoStop}%
\bibitem [{\citenamefont {Ryan}(1997)}]{ryan1997accuracy}%
  \BibitemOpen
  \bibfield  {author} {\bibinfo {author} {\bibfnamefont {F.~D.}\ \bibnamefont
  {Ryan}},\ }\href@noop {} {\bibfield  {journal} {\bibinfo  {journal} {Physical
  Review D}\ }\textbf {\bibinfo {volume} {56}},\ \bibinfo {pages} {1845}
  (\bibinfo {year} {1997})}\BibitemShut {NoStop}%
\bibitem [{\citenamefont {Vigeland}\ and\ \citenamefont
  {Hughes}(2010)}]{vigeland2010spacetime}%
  \BibitemOpen
  \bibfield  {author} {\bibinfo {author} {\bibfnamefont {S.~J.}\ \bibnamefont
  {Vigeland}}\ and\ \bibinfo {author} {\bibfnamefont {S.~A.}\ \bibnamefont
  {Hughes}},\ }\href@noop {} {\bibfield  {journal} {\bibinfo  {journal}
  {Physical Review D}\ }\textbf {\bibinfo {volume} {81}},\ \bibinfo {pages}
  {024030} (\bibinfo {year} {2010})}\BibitemShut {NoStop}%
\bibitem [{\citenamefont {Brink}(2008)}]{PhysRevD.78.102001}%
  \BibitemOpen
  \bibfield  {author} {\bibinfo {author} {\bibfnamefont {J.}~\bibnamefont
  {Brink}},\ }\href {\doibase 10.1103/PhysRevD.78.102001} {\bibfield  {journal}
  {\bibinfo  {journal} {Phys. Rev. D}\ }\textbf {\bibinfo {volume} {78}},\
  \bibinfo {pages} {102001} (\bibinfo {year} {2008})}\BibitemShut {NoStop}%
\bibitem [{\citenamefont {Bena}\ and\ \citenamefont
  {Mayerson}(2020{\natexlab{b}})}]{PhysRevLett.125.221602}%
  \BibitemOpen
  \bibfield  {author} {\bibinfo {author} {\bibfnamefont {I.}~\bibnamefont
  {Bena}}\ and\ \bibinfo {author} {\bibfnamefont {D.~R.}\ \bibnamefont
  {Mayerson}},\ }\href {\doibase 10.1103/PhysRevLett.125.221602} {\bibfield
  {journal} {\bibinfo  {journal} {Phys. Rev. Lett.}\ }\textbf {\bibinfo
  {volume} {125}},\ \bibinfo {pages} {221602} (\bibinfo {year}
  {2020}{\natexlab{b}})}\BibitemShut {NoStop}%
\bibitem [{\citenamefont {Nakano}\ \emph {et~al.}(2017)\citenamefont {Nakano},
  \citenamefont {Tanaka}, \citenamefont {Sago},\ and\ \citenamefont
  {Tagoshi}}]{tanaka}%
  \BibitemOpen
  \bibfield  {author} {\bibinfo {author} {\bibfnamefont {H.}~\bibnamefont
  {Nakano}}, \bibinfo {author} {\bibfnamefont {T.}~\bibnamefont {Tanaka}},
  \bibinfo {author} {\bibfnamefont {N.}~\bibnamefont {Sago}}, \ and\ \bibinfo
  {author} {\bibfnamefont {H.}~\bibnamefont {Tagoshi}},\ }\href {\doibase
  10.1093/ptep/ptx093} {\bibfield  {journal} {\bibinfo  {journal} {Progress of
  Theoretical and Experimental Physics}\ }\textbf {\bibinfo {volume} {2017}}
  (\bibinfo {year} {2017}),\ 10.1093/ptep/ptx093},\ \Eprint
  {http://arxiv.org/abs/http://oup.prod.sis.lan/ptep/article-pdf/2017/7/071E01/19176441/ptx093.pdf}
  {http://oup.prod.sis.lan/ptep/article-pdf/2017/7/071E01/19176441/ptx093.pdf}
  \BibitemShut {NoStop}%
\bibitem [{\citenamefont {Micchi}\ and\ \citenamefont
  {Chirenti}(2020)}]{micchi2020spicing}%
  \BibitemOpen
  \bibfield  {author} {\bibinfo {author} {\bibfnamefont {L.~F.~L.}\
  \bibnamefont {Micchi}}\ and\ \bibinfo {author} {\bibfnamefont
  {C.}~\bibnamefont {Chirenti}},\ }\href@noop {} {\bibfield  {journal}
  {\bibinfo  {journal} {Physical Review D}\ }\textbf {\bibinfo {volume}
  {101}},\ \bibinfo {pages} {084010} (\bibinfo {year} {2020})}\BibitemShut
  {NoStop}%
\bibitem [{\citenamefont {Wang}\ \emph {et~al.}(2020)\citenamefont {Wang},
  \citenamefont {Oshita},\ and\ \citenamefont {Afshordi}}]{wang2020echoes}%
  \BibitemOpen
  \bibfield  {author} {\bibinfo {author} {\bibfnamefont {Q.}~\bibnamefont
  {Wang}}, \bibinfo {author} {\bibfnamefont {N.}~\bibnamefont {Oshita}}, \ and\
  \bibinfo {author} {\bibfnamefont {N.}~\bibnamefont {Afshordi}},\ }\href@noop
  {} {\bibfield  {journal} {\bibinfo  {journal} {Physical Review D}\ }\textbf
  {\bibinfo {volume} {101}},\ \bibinfo {pages} {024031} (\bibinfo {year}
  {2020})}\BibitemShut {NoStop}%
\bibitem [{\citenamefont {Maggio}\ \emph
  {et~al.}(2019{\natexlab{a}})\citenamefont {Maggio}, \citenamefont {Testa},
  \citenamefont {Bhagwat},\ and\ \citenamefont {Pani}}]{Maggio}%
  \BibitemOpen
  \bibfield  {author} {\bibinfo {author} {\bibfnamefont {E.}~\bibnamefont
  {Maggio}}, \bibinfo {author} {\bibfnamefont {A.}~\bibnamefont {Testa}},
  \bibinfo {author} {\bibfnamefont {S.}~\bibnamefont {Bhagwat}}, \ and\
  \bibinfo {author} {\bibfnamefont {P.}~\bibnamefont {Pani}},\ }\href {\doibase
  10.1103/PhysRevD.100.064056} {\bibfield  {journal} {\bibinfo  {journal}
  {Phys. Rev. D}\ }\textbf {\bibinfo {volume} {100}},\ \bibinfo {pages}
  {064056} (\bibinfo {year} {2019}{\natexlab{a}})}\BibitemShut {NoStop}%
\bibitem [{\citenamefont {Micchi}\ \emph {et~al.}(2021)\citenamefont {Micchi},
  \citenamefont {Afshordi},\ and\ \citenamefont {Chirenti}}]{micchi2021loud}%
  \BibitemOpen
  \bibfield  {author} {\bibinfo {author} {\bibfnamefont {L.~F.~L.}\
  \bibnamefont {Micchi}}, \bibinfo {author} {\bibfnamefont {N.}~\bibnamefont
  {Afshordi}}, \ and\ \bibinfo {author} {\bibfnamefont {C.}~\bibnamefont
  {Chirenti}},\ }\href@noop {} {\bibfield  {journal} {\bibinfo  {journal}
  {Physical Review D}\ }\textbf {\bibinfo {volume} {103}},\ \bibinfo {pages}
  {044028} (\bibinfo {year} {2021})}\BibitemShut {NoStop}%
\bibitem [{\citenamefont {Sago}\ and\ \citenamefont
  {Tanaka}(2020)}]{sago2020gravitational}%
  \BibitemOpen
  \bibfield  {author} {\bibinfo {author} {\bibfnamefont {N.}~\bibnamefont
  {Sago}}\ and\ \bibinfo {author} {\bibfnamefont {T.}~\bibnamefont {Tanaka}},\
  }\href@noop {} {\bibfield  {journal} {\bibinfo  {journal} {arXiv preprint
  arXiv:2009.08086}\ } (\bibinfo {year} {2020})}\BibitemShut {NoStop}%
\bibitem [{\citenamefont {Maggio}\ \emph {et~al.}(2017)\citenamefont {Maggio},
  \citenamefont {Pani},\ and\ \citenamefont {Ferrari}}]{maggio2017exotic}%
  \BibitemOpen
  \bibfield  {author} {\bibinfo {author} {\bibfnamefont {E.}~\bibnamefont
  {Maggio}}, \bibinfo {author} {\bibfnamefont {P.}~\bibnamefont {Pani}}, \ and\
  \bibinfo {author} {\bibfnamefont {V.}~\bibnamefont {Ferrari}},\ }\href@noop
  {} {\bibfield  {journal} {\bibinfo  {journal} {Physical Review D}\ }\textbf
  {\bibinfo {volume} {96}},\ \bibinfo {pages} {104047} (\bibinfo {year}
  {2017})}\BibitemShut {NoStop}%
\bibitem [{\citenamefont {Maggio}\ \emph
  {et~al.}(2019{\natexlab{b}})\citenamefont {Maggio}, \citenamefont {Cardoso},
  \citenamefont {Dolan},\ and\ \citenamefont {Pani}}]{maggio2019ergoregion}%
  \BibitemOpen
  \bibfield  {author} {\bibinfo {author} {\bibfnamefont {E.}~\bibnamefont
  {Maggio}}, \bibinfo {author} {\bibfnamefont {V.}~\bibnamefont {Cardoso}},
  \bibinfo {author} {\bibfnamefont {S.~R.}\ \bibnamefont {Dolan}}, \ and\
  \bibinfo {author} {\bibfnamefont {P.}~\bibnamefont {Pani}},\ }\href@noop {}
  {\bibfield  {journal} {\bibinfo  {journal} {Physical Review D}\ }\textbf
  {\bibinfo {volume} {99}},\ \bibinfo {pages} {064007} (\bibinfo {year}
  {2019}{\natexlab{b}})}\BibitemShut {NoStop}%
\bibitem [{\citenamefont {Cunha}\ \emph {et~al.}(2017)\citenamefont {Cunha},
  \citenamefont {Berti},\ and\ \citenamefont {Herdeiro}}]{cunha2017light}%
  \BibitemOpen
  \bibfield  {author} {\bibinfo {author} {\bibfnamefont {P.~V.}\ \bibnamefont
  {Cunha}}, \bibinfo {author} {\bibfnamefont {E.}~\bibnamefont {Berti}}, \ and\
  \bibinfo {author} {\bibfnamefont {C.~A.}\ \bibnamefont {Herdeiro}},\
  }\href@noop {} {\bibfield  {journal} {\bibinfo  {journal} {Physical Review
  Letters}\ }\textbf {\bibinfo {volume} {119}},\ \bibinfo {pages} {251102}
  (\bibinfo {year} {2017})}\BibitemShut {NoStop}%
\bibitem [{\citenamefont {Addazi}\ \emph {et~al.}(2020)\citenamefont {Addazi},
  \citenamefont {Marcian{\`o}},\ and\ \citenamefont
  {Yunes}}]{addazi2020gravitational}%
  \BibitemOpen
  \bibfield  {author} {\bibinfo {author} {\bibfnamefont {A.}~\bibnamefont
  {Addazi}}, \bibinfo {author} {\bibfnamefont {A.}~\bibnamefont
  {Marcian{\`o}}}, \ and\ \bibinfo {author} {\bibfnamefont {N.}~\bibnamefont
  {Yunes}},\ }\href@noop {} {\bibfield  {journal} {\bibinfo  {journal} {The
  European Physical Journal C}\ }\textbf {\bibinfo {volume} {80}},\ \bibinfo
  {pages} {36} (\bibinfo {year} {2020})}\BibitemShut {NoStop}%
\bibitem [{\citenamefont {Chen}\ \emph {et~al.}(2019)\citenamefont {Chen},
  \citenamefont {Chen}, \citenamefont {Ma}, \citenamefont {Lo},\ and\
  \citenamefont {Sun}}]{chen2019instability}%
  \BibitemOpen
  \bibfield  {author} {\bibinfo {author} {\bibfnamefont {B.}~\bibnamefont
  {Chen}}, \bibinfo {author} {\bibfnamefont {Y.}~\bibnamefont {Chen}}, \bibinfo
  {author} {\bibfnamefont {Y.}~\bibnamefont {Ma}}, \bibinfo {author}
  {\bibfnamefont {K.-L.~R.}\ \bibnamefont {Lo}}, \ and\ \bibinfo {author}
  {\bibfnamefont {L.}~\bibnamefont {Sun}},\ }\href@noop {} {\bibfield
  {journal} {\bibinfo  {journal} {arXiv preprint arXiv:1902.08180}\ } (\bibinfo
  {year} {2019})}\BibitemShut {NoStop}%
\bibitem [{\citenamefont {Han}(2014)}]{han2014gravitational}%
  \BibitemOpen
  \bibfield  {author} {\bibinfo {author} {\bibfnamefont {W.-B.}\ \bibnamefont
  {Han}},\ }\href@noop {} {\bibfield  {journal} {\bibinfo  {journal}
  {International Journal of Modern Physics D}\ }\textbf {\bibinfo {volume}
  {23}},\ \bibinfo {pages} {1450064} (\bibinfo {year} {2014})}\BibitemShut
  {NoStop}%
\bibitem [{\citenamefont {Blackman}\ \emph {et~al.}(2015)\citenamefont
  {Blackman}, \citenamefont {Field}, \citenamefont {Galley}, \citenamefont
  {Szil{\'a}gyi}, \citenamefont {Scheel}, \citenamefont {Tiglio},\ and\
  \citenamefont {Hemberger}}]{blackman2015fast}%
  \BibitemOpen
  \bibfield  {author} {\bibinfo {author} {\bibfnamefont {J.}~\bibnamefont
  {Blackman}}, \bibinfo {author} {\bibfnamefont {S.~E.}\ \bibnamefont {Field}},
  \bibinfo {author} {\bibfnamefont {C.~R.}\ \bibnamefont {Galley}}, \bibinfo
  {author} {\bibfnamefont {B.}~\bibnamefont {Szil{\'a}gyi}}, \bibinfo {author}
  {\bibfnamefont {M.~A.}\ \bibnamefont {Scheel}}, \bibinfo {author}
  {\bibfnamefont {M.}~\bibnamefont {Tiglio}}, \ and\ \bibinfo {author}
  {\bibfnamefont {D.~A.}\ \bibnamefont {Hemberger}},\ }\href@noop {} {\bibfield
   {journal} {\bibinfo  {journal} {Physical review letters}\ }\textbf {\bibinfo
  {volume} {115}},\ \bibinfo {pages} {121102} (\bibinfo {year}
  {2015})}\BibitemShut {NoStop}%
\bibitem [{\citenamefont {Varma}\ \emph {et~al.}(2018)\citenamefont {Varma},
  \citenamefont {Stein},\ and\ \citenamefont
  {Gerosa}}]{vijay_varma_2018_1435832}%
  \BibitemOpen
  \bibfield  {author} {\bibinfo {author} {\bibfnamefont {V.}~\bibnamefont
  {Varma}}, \bibinfo {author} {\bibfnamefont {L.~C.}\ \bibnamefont {Stein}}, \
  and\ \bibinfo {author} {\bibfnamefont {D.}~\bibnamefont {Gerosa}},\ }\href
  {\doibase 10.5281/zenodo.1435832} {\enquote {\bibinfo {title}
  {{vijayvarma392/surfinBH: Surrogate Final BH properties}},}\ } (\bibinfo
  {year} {2018})\BibitemShut {NoStop}%
\bibitem [{\citenamefont {Varma}\ \emph
  {et~al.}(2019{\natexlab{a}})\citenamefont {Varma}, \citenamefont {Gerosa},
  \citenamefont {Stein}, \citenamefont {H\'ebert},\ and\ \citenamefont
  {Zhang}}]{PhysRevLett.122.011101}%
  \BibitemOpen
  \bibfield  {author} {\bibinfo {author} {\bibfnamefont {V.}~\bibnamefont
  {Varma}}, \bibinfo {author} {\bibfnamefont {D.}~\bibnamefont {Gerosa}},
  \bibinfo {author} {\bibfnamefont {L.~C.}\ \bibnamefont {Stein}}, \bibinfo
  {author} {\bibfnamefont {F.~m.~c.}\ \bibnamefont {H\'ebert}}, \ and\ \bibinfo
  {author} {\bibfnamefont {H.}~\bibnamefont {Zhang}},\ }\href {\doibase
  10.1103/PhysRevLett.122.011101} {\bibfield  {journal} {\bibinfo  {journal}
  {Phys. Rev. Lett.}\ }\textbf {\bibinfo {volume} {122}},\ \bibinfo {pages}
  {011101} (\bibinfo {year} {2019}{\natexlab{a}})}\BibitemShut {NoStop}%
\bibitem [{Note1()}]{Note1}%
  \BibitemOpen
  \bibinfo {note} {See, e.g., Sec.~III of Ref.~\cite
  {press1973perturbations}.}\BibitemShut {Stop}%
\bibitem [{\citenamefont {Hughes}(2000)}]{Hughes}%
  \BibitemOpen
  \bibfield  {author} {\bibinfo {author} {\bibfnamefont {S.~A.}\ \bibnamefont
  {Hughes}},\ }\href {\doibase 10.1103/PhysRevD.61.084004} {\bibfield
  {journal} {\bibinfo  {journal} {Phys. Rev. D}\ }\textbf {\bibinfo {volume}
  {61}},\ \bibinfo {pages} {084004} (\bibinfo {year} {2000})}\BibitemShut
  {NoStop}%
\bibitem [{\citenamefont {Han}(2010)}]{han2010prd}%
  \BibitemOpen
  \bibfield  {author} {\bibinfo {author} {\bibfnamefont {W.-B.}\ \bibnamefont
  {Han}},\ }\href {\doibase 10.1103/PhysRevD.82.084013} {\bibfield  {journal}
  {\bibinfo  {journal} {Phys. Rev. D}\ }\textbf {\bibinfo {volume} {82}},\
  \bibinfo {pages} {084013} (\bibinfo {year} {2010})}\BibitemShut {NoStop}%
\bibitem [{\citenamefont {Han}\ and\ \citenamefont {Cao}(2011)}]{han2011prd}%
  \BibitemOpen
  \bibfield  {author} {\bibinfo {author} {\bibfnamefont {W.-B.}\ \bibnamefont
  {Han}}\ and\ \bibinfo {author} {\bibfnamefont {Z.}~\bibnamefont {Cao}},\
  }\href {\doibase 10.1103/PhysRevD.84.044014} {\bibfield  {journal} {\bibinfo
  {journal} {Phys. Rev. D}\ }\textbf {\bibinfo {volume} {84}},\ \bibinfo
  {pages} {044014} (\bibinfo {year} {2011})}\BibitemShut {NoStop}%
\bibitem [{\citenamefont {Varma}\ \emph
  {et~al.}(2019{\natexlab{b}})\citenamefont {Varma}, \citenamefont {Field},
  \citenamefont {Scheel}, \citenamefont {Blackman}, \citenamefont {Gerosa},
  \citenamefont {Stein}, \citenamefont {Kidder},\ and\ \citenamefont
  {Pfeiffer}}]{PhysRevResearch.1.033015}%
  \BibitemOpen
  \bibfield  {author} {\bibinfo {author} {\bibfnamefont {V.}~\bibnamefont
  {Varma}}, \bibinfo {author} {\bibfnamefont {S.~E.}\ \bibnamefont {Field}},
  \bibinfo {author} {\bibfnamefont {M.~A.}\ \bibnamefont {Scheel}}, \bibinfo
  {author} {\bibfnamefont {J.}~\bibnamefont {Blackman}}, \bibinfo {author}
  {\bibfnamefont {D.}~\bibnamefont {Gerosa}}, \bibinfo {author} {\bibfnamefont
  {L.~C.}\ \bibnamefont {Stein}}, \bibinfo {author} {\bibfnamefont {L.~E.}\
  \bibnamefont {Kidder}}, \ and\ \bibinfo {author} {\bibfnamefont {H.~P.}\
  \bibnamefont {Pfeiffer}},\ }\href {\doibase 10.1103/PhysRevResearch.1.033015}
  {\bibfield  {journal} {\bibinfo  {journal} {Phys. Rev. Research}\ }\textbf
  {\bibinfo {volume} {1}},\ \bibinfo {pages} {033015} (\bibinfo {year}
  {2019}{\natexlab{b}})}\BibitemShut {NoStop}%
\bibitem [{\citenamefont {Sathyaprakash}\ and\ \citenamefont
  {Schutz}(2009)}]{sathyaprakash2009physics}%
  \BibitemOpen
  \bibfield  {author} {\bibinfo {author} {\bibfnamefont {B.~S.}\ \bibnamefont
  {Sathyaprakash}}\ and\ \bibinfo {author} {\bibfnamefont {B.~F.}\ \bibnamefont
  {Schutz}},\ }\href@noop {} {\bibfield  {journal} {\bibinfo  {journal} {Living
  reviews in relativity}\ }\textbf {\bibinfo {volume} {12}},\ \bibinfo {pages}
  {1} (\bibinfo {year} {2009})}\BibitemShut {NoStop}%
\bibitem [{\citenamefont {Mino}\ \emph
  {et~al.}(1997{\natexlab{a}})\citenamefont {Mino}, \citenamefont {Sasaki},
  \citenamefont {Shibata}, \citenamefont {Tagoshi},\ and\ \citenamefont
  {Tanaka}}]{mino1997black}%
  \BibitemOpen
  \bibfield  {author} {\bibinfo {author} {\bibfnamefont {Y.}~\bibnamefont
  {Mino}}, \bibinfo {author} {\bibfnamefont {M.}~\bibnamefont {Sasaki}},
  \bibinfo {author} {\bibfnamefont {M.}~\bibnamefont {Shibata}}, \bibinfo
  {author} {\bibfnamefont {H.}~\bibnamefont {Tagoshi}}, \ and\ \bibinfo
  {author} {\bibfnamefont {T.}~\bibnamefont {Tanaka}},\ }\href@noop {}
  {\bibfield  {journal} {\bibinfo  {journal} {Progress of Theoretical Physics
  Supplement}\ }\textbf {\bibinfo {volume} {128}},\ \bibinfo {pages} {1}
  (\bibinfo {year} {1997}{\natexlab{a}})}\BibitemShut {NoStop}%
\bibitem [{Note2()}]{Note2}%
  \BibitemOpen
  \bibinfo {note} {This does not lead to a diverging echo when $\omega
  \rightarrow +\infty $ since $\protect \mathcal {R}^{\protect \rm BH}_{\ell m
  \omega }$ converges to zero quickly as $\omega \rightarrow +\infty
  $.}\BibitemShut {Stop}%
\bibitem [{\citenamefont {Chen}\ \emph {et~al.}(2020)\citenamefont {Chen},
  \citenamefont {Wang},\ and\ \citenamefont {Chen}}]{chen2020tidal}%
  \BibitemOpen
  \bibfield  {author} {\bibinfo {author} {\bibfnamefont {B.}~\bibnamefont
  {Chen}}, \bibinfo {author} {\bibfnamefont {Q.}~\bibnamefont {Wang}}, \ and\
  \bibinfo {author} {\bibfnamefont {Y.}~\bibnamefont {Chen}},\ }\href@noop {}
  {\bibfield  {journal} {\bibinfo  {journal} {arXiv preprint arXiv:2012.10842}\
  } (\bibinfo {year} {2020})}\BibitemShut {NoStop}%
\bibitem [{\citenamefont {Oshita}\ \emph {et~al.}(2020)\citenamefont {Oshita},
  \citenamefont {Wang},\ and\ \citenamefont {Afshordi}}]{Oshita_2020}%
  \BibitemOpen
  \bibfield  {author} {\bibinfo {author} {\bibfnamefont {N.}~\bibnamefont
  {Oshita}}, \bibinfo {author} {\bibfnamefont {Q.}~\bibnamefont {Wang}}, \ and\
  \bibinfo {author} {\bibfnamefont {N.}~\bibnamefont {Afshordi}},\ }\href
  {\doibase 10.1088/1475-7516/2020/04/016} {\bibfield  {journal} {\bibinfo
  {journal} {Journal of Cosmology and Astroparticle Physics}\ }\textbf
  {\bibinfo {volume} {2020}},\ \bibinfo {pages} {016} (\bibinfo {year}
  {2020})}\BibitemShut {NoStop}%
\bibitem [{\citenamefont {Teukolsky}\ and\ \citenamefont
  {Press}(1974)}]{teukolsky_energy}%
  \BibitemOpen
  \bibfield  {author} {\bibinfo {author} {\bibfnamefont {S.~A.}\ \bibnamefont
  {Teukolsky}}\ and\ \bibinfo {author} {\bibfnamefont {W.}~\bibnamefont
  {Press}},\ }\href@noop {} {\bibfield  {journal} {\bibinfo  {journal} {The
  Astrophysical Journal}\ }\textbf {\bibinfo {volume} {193}},\ \bibinfo {pages}
  {443} (\bibinfo {year} {1974})}\BibitemShut {NoStop}%
\bibitem [{\citenamefont {Davis}\ \emph {et~al.}(1972)\citenamefont {Davis},
  \citenamefont {Ruffini},\ and\ \citenamefont {Tiomno}}]{davis1972pulses}%
  \BibitemOpen
  \bibfield  {author} {\bibinfo {author} {\bibfnamefont {M.}~\bibnamefont
  {Davis}}, \bibinfo {author} {\bibfnamefont {R.}~\bibnamefont {Ruffini}}, \
  and\ \bibinfo {author} {\bibfnamefont {J.}~\bibnamefont {Tiomno}},\
  }\href@noop {} {\bibfield  {journal} {\bibinfo  {journal} {Physical Review
  D}\ }\textbf {\bibinfo {volume} {5}},\ \bibinfo {pages} {2932} (\bibinfo
  {year} {1972})}\BibitemShut {NoStop}%
\bibitem [{\citenamefont {Flanagan}\ and\ \citenamefont
  {Hughes}(1998)}]{Flanagan:1997sx}%
  \BibitemOpen
  \bibfield  {author} {\bibinfo {author} {\bibfnamefont {E.~E.}\ \bibnamefont
  {Flanagan}}\ and\ \bibinfo {author} {\bibfnamefont {S.~A.}\ \bibnamefont
  {Hughes}},\ }\href {\doibase 10.1103/PhysRevD.57.4535} {\bibfield  {journal}
  {\bibinfo  {journal} {Phys. Rev. D}\ }\textbf {\bibinfo {volume} {57}},\
  \bibinfo {pages} {4535} (\bibinfo {year} {1998})},\ \Eprint
  {http://arxiv.org/abs/gr-qc/9701039} {arXiv:gr-qc/9701039} \BibitemShut
  {NoStop}%
\bibitem [{Note3()}]{Note3}%
  \BibitemOpen
  \bibinfo {note} {To account for the expansion of the Universe, one can simply
  replace the the coordinate distance $r$ with the luminosity distance
  $d_{\protect \rm L}$, and replace the total mass $M$ with the redshifted
  total mass $M(1+z)$ where $z = z(d_{\protect \rm L})$ is the redshift of the
  source.}\BibitemShut {Stop}%
\bibitem [{\citenamefont {Finn}\ and\ \citenamefont
  {Chernoff}(1993)}]{Finn:1992xs}%
  \BibitemOpen
  \bibfield  {author} {\bibinfo {author} {\bibfnamefont {L.~S.}\ \bibnamefont
  {Finn}}\ and\ \bibinfo {author} {\bibfnamefont {D.~F.}\ \bibnamefont
  {Chernoff}},\ }\href {\doibase 10.1103/PhysRevD.47.2198} {\bibfield
  {journal} {\bibinfo  {journal} {Phys. Rev. D}\ }\textbf {\bibinfo {volume}
  {47}},\ \bibinfo {pages} {2198} (\bibinfo {year} {1993})},\ \Eprint
  {http://arxiv.org/abs/gr-qc/9301003} {arXiv:gr-qc/9301003} \BibitemShut
  {NoStop}%
\bibitem [{aLI()}]{aLIGODesignNoiseCurve}%
  \BibitemOpen
  \href@noop {} {\enquote {\bibinfo {title} {{Advanced LIGO anticipated
  sensitivity curves}},}\ }\bibinfo {howpublished}
  {\url{https://dcc.ligo.org/LIGO-T1800044/public}}\BibitemShut {NoStop}%
\bibitem [{\citenamefont {Abbott}\ \emph
  {et~al.}(2017{\natexlab{b}})\citenamefont {Abbott} \emph
  {et~al.}}]{Evans:2016mbw}%
  \BibitemOpen
  \bibfield  {author} {\bibinfo {author} {\bibfnamefont {B.~P.}\ \bibnamefont
  {Abbott}} \emph {et~al.} (\bibinfo {collaboration} {LIGO Scientific}),\
  }\href {\doibase 10.1088/1361-6382/aa51f4} {\bibfield  {journal} {\bibinfo
  {journal} {Class. Quant. Grav.}\ }\textbf {\bibinfo {volume} {34}},\ \bibinfo
  {pages} {044001} (\bibinfo {year} {2017}{\natexlab{b}})},\ \Eprint
  {http://arxiv.org/abs/1607.08697} {arXiv:1607.08697 [astro-ph.IM]}
  \BibitemShut {NoStop}%
\bibitem [{\citenamefont {Mino}\ \emph
  {et~al.}(1997{\natexlab{b}})\citenamefont {Mino}, \citenamefont {Sasaki},
  \citenamefont {Shibata}, \citenamefont {Tagoshi},\ and\ \citenamefont
  {Tanaka}}]{SasakiReview}%
  \BibitemOpen
  \bibfield  {author} {\bibinfo {author} {\bibfnamefont {Y.}~\bibnamefont
  {Mino}}, \bibinfo {author} {\bibfnamefont {M.}~\bibnamefont {Sasaki}},
  \bibinfo {author} {\bibfnamefont {M.}~\bibnamefont {Shibata}}, \bibinfo
  {author} {\bibfnamefont {H.}~\bibnamefont {Tagoshi}}, \ and\ \bibinfo
  {author} {\bibfnamefont {T.}~\bibnamefont {Tanaka}},\ }\href {\doibase
  10.1143/PTPS.128.1} {\bibfield  {journal} {\bibinfo  {journal} {Progress of
  Theoretical Physics Supplement}\ }\textbf {\bibinfo {volume} {128}},\
  \bibinfo {pages} {1} (\bibinfo {year} {1997}{\natexlab{b}})},\ \Eprint
  {http://arxiv.org/abs/http://oup.prod.sis.lan/ptps/article-pdf/doi/10.1143/PTPS.128.1/5438984/128-1.pdf}
  {http://oup.prod.sis.lan/ptps/article-pdf/doi/10.1143/PTPS.128.1/5438984/128-1.pdf}
  \BibitemShut {NoStop}%
\bibitem [{\citenamefont {Press}\ and\ \citenamefont
  {Teukolsky}(1973)}]{press1973perturbations}%
  \BibitemOpen
  \bibfield  {author} {\bibinfo {author} {\bibfnamefont {W.~H.}\ \bibnamefont
  {Press}}\ and\ \bibinfo {author} {\bibfnamefont {S.~A.}\ \bibnamefont
  {Teukolsky}},\ }\href@noop {} {\bibfield  {journal} {\bibinfo  {journal} {The
  Astrophysical Journal}\ }\textbf {\bibinfo {volume} {185}},\ \bibinfo {pages}
  {649} (\bibinfo {year} {1973})}\BibitemShut {NoStop}%
\end{thebibliography}%

\end{document}